# Laser-induced plasma formation and cavitation in water: from nanoeffects to extreme states of matter

Norbert Linz[1], Sebastian Freidank[1], Xiao-Xuan Liang[1], and Alfred Vogel[1]

1) Institute of Biomedical Optics, University of Luebeck, Peter-Monnik Weg 4, 23562 Luebeck, Germany

E-mails: norbert.linz@uni-luebeck.de; alfred.vogel@uni-luebeck.de

## Abstract

We present an in-depth analysis of the energy dependence of optical breakdown in water by tightly focused laser pulses, from plasma formation to shock waves and cavitation. Laser pulses of fs to ns durations and UV to IR wavelengths are aberration-free focused through microscope objectives. Photography captures luminescent plasmas with submicrometer resolution, and bubble threshold and size are determined via probe beam scattering. The energy dependence of mechanical effects is quantified through the maximum bubble radius $R_{max}$. We find three key scenarios depicting the interaction between multiphoton and avalanche ionization, recombination, and thermal ionization from nanoeffects near threshold to extreme energy densities. They include a previously unknown scenario that emerges with single-longitudinal-mode UV ns pulses from compact lasers. It enables cost-effective creation of nanoeffects, as demonstrated on corneal tissue and glass. High-resolution colour photography revealed new insights in the spatiotemporal dynamics of plasma formation, with an interplay of breakdown waves, string formation by local instabilities of avalanche ionization, and radiative energy transport. Plasma volume data from photographs together with absorption measurements show that the average energy density of luminescent fs and ns plasmas is similar, ranging between 10 and 40 kJ/cm$^3$. However, small hot regions with up to 400 kJ/cm$^3$ are formed in ns breakdown. From the hot regions, energy is spread out via X-ray bremsstrahlung, forming a luminescent halo. Well above threshold, $R_{max}$ scales with $E^{1/3}$ across all scenarios, with 15% - 20% conversion of laser energy into bubble energy. With increasing plasma energy density, an ever-larger energy fraction is converted into shock wave energy (75% at 40 kJ/cm$^3$). We discuss guidelines for parameter selection in laser surgery and material processing in bulk media as well as for laser ablation and breakdown spectroscopy in liquids. Finally, we suggest roadmaps for future experimental and modeling work, and for broadening applications.

## 1. Introduction

Laser-induced plasma formation (optical breakdown) offers the potential of precisely tunable nonlinear energy deposition in nominally transparent dielectrics that can be used for material processing on the micro- and nanoscale [1-10]. Laser-generated plasmas in water or aqueous fluids are of major importance for laser surgery in the eye [11-17], in other transparent tissues and cells [18-27], and for microfluidics [28-35].

Tightly focused energetic laser pulses can produce fine tunable nano-and microeffects and also generate ultra-high plasma energy densities that by far exceed the energy density in explosives such as TNT. Water is an excellent model to study the process of nonlinear energy deposition from nanoeffects to extreme states of matter because it has a large bandgap of $\approx$ 9.5 eV [36,37], similar to fused silica [7,38]. While the energy density required for thermoelastic bubble formation with femtosecond (fs) pulses is $\approx$ 0.55 kJ/cm$^3$ [22], the present experiments showed that under appropriate experimental conditions the plasma energy density can reach values of 260 – 400 kJ/cm$^3$, more than 100 times larger than the vaporization enthalpy of water. This corresponds to pressure values well above 100 GPa.

The mechanical confinement of plasmas formed in bulk water results in shock wave emission and cavitation bubble generation [39-44]. Laser generation of compact plasma with tunable energy density at well-defined time and location is an excellent tool for fundamental experimental investigations of cavitation bubble dynamics and shock wave emission in free liquid and near boundaries [40,41,45-57]. Laser-produced bubbles are not disturbed by detonation gases as in underwater explosions [58,59] or by electrodes as in spark discharge [60,61], and experimental data on their dynamics are, thus, an excellent basis for model validation [40,42,43,62-68].

In this paper, we mainly explore the optical breakdown dynamics in bulk water, with references to transparent tissues and solid dielectrics. However, the underlying mechanisms are also relevant for plasma formation at linear absorbing solid targets immersed in water, such as in laser ablation in liquids (LAL) for nanoparticle generation [69-81] and in underwater laser-induced breakdown spectroscopy (LIBS) for remote analysis of target constituents [82-89]. Here, plasma formation is ignited at the target and can, well above threshold, also spread into the liquid and shield the target. The dynamics of the bubble produced during ablation influences the particle size distribution, and during consecutive ablation events bubble remnants may disturb light transmission and ablation [74,75,90-94]. Secondary laser irradiation can be used to fragment large free nanoparticles for homogenization of the size distribution and modification of their crystalline structure [95-100]. In LIBS, plasma-mediated ablation is not the primary goal but a way for obtaining a characteristic spectroscopic signal from the atomized and excited target material after the initial continuum emission from the plasma has ceased [84,88,101-103]. A mechanistic understanding of optical breakdown in bulk liquid, which is the focus of this paper, also helps to understand the rules governing plasma growth and hydrodynamic phenomena in LAL and LIBS. We will use it to derive guidelines for optimizing energy coupling into the target and minimizing shielding by plasma within the liquid with the goals of optimizing ablation yield, particle properties, and spectroscopic information.

Beginning in the 1980s, laser-induced "photodisruption" has been employed for intraocular and refractive surgery, using first nanosecond (ns) pulses and later picosecond (ps) and fs pulses focused at moderate *NA* [11-13,17,47,104,105]. That stimulated research on the breakdown mechanisms at moderate numerical aperture (*NA*) with different pulse durations [12,40,47,104,106-115]. Later it was discovered that fs pulses focused at large *NA* can produce submicrometer-sized laser effects suitable for nanosurgery on a subcellular level [20-22,24,116-119] and for creating tiny voids in glass and crystalline solids [1,3,120-122]. Although laser-induced bubble formation was often employed for investigating cavitation phenomena [46,48,51,53,56,68], the process of nonlinear energy deposition itself received little attention as long as the generated bubbles had an approximately spherical shape [52,54]. Due to the difficulty of time-resolved studies on a nano- and microscale, details of the optical breakdown dynamics at large *NA* and for energies well above the breakdown threshold have experimentally been studied merely for a few individual laser parameters [19,40,110,123,124]. Attention was largely focused on the parameter dependence of the breakdown threshold and laser effects close to this threshold [36,37,112,125,126]. Furthermore, it was often restricted to ultrashort pulse durations because researchers thought that nanosecond breakdown was too stochastic and disruptive to allow for a generation of reproducible nanoeffects with precisely tunable magnitude [114,120,125,127-129].

Although optical breakdown with ultrashort laser pulses has received a lot of attention in the last three decades, various principal questions remain unresolved. In the lower range of plasma energy densities, there is a quest for compact and cost-effective devices enabling precisely adjustable nonlinear energy deposition for nanosurgery and material processing. A key question in this context is whether ns breakdown is really intrinsically stochastic and only fs laser effects are deterministic and precisely adjustable. We will show that ns breakdown is not per se stochastic, delineate the conditions for deterministic energy deposition with cost-effective ns pulses and demonstrate the prerequisites for producing precisely tunable nanoeffects.

An open question of great practical interest is how the partitioning of absorbed laser energy into vaporization, bubble and shock wave energy depends on plasma energy density because this partitioning governs the disruptiveness of laser effects. Of prime interest is also the question, which factors determine the upper limits of global and local plasma energy density in fs and ns breakdown and how these limits are linked to the spatiotemporal dynamics of energy deposition. Finding answers requires a precise determination of plasma size, inner structure and absorptivity.

We address the above questions through a systematic experimental exploration of the energy dependence of optical breakdown phenomena as a function of pulse duration $\tau_L$ (fs to ns), laser wavelength $\lambda_L$ (UV to IR) and focusing *NA* (0.3 to 0.9). In this parameter space, we cover the range from the bubble threshold in water to extremely high plasma energy densities.

Compared to solid dielectrics, the 'self-healing' properties of water largely facilitate parametric studies. Bubble formation provides a straightforward and readily detectable breakdown criterion [37,117], and the maximum bubble radius $R_{max}$ is a convenient measure for the magnitude of the laser effect. This way, a large amount of data can be acquired without the need for a continuous movement of the target. Tight focusing of the laser beam largely avoids nonlinear beam propagation effects [130,131], and the use of water-immersion microscope objectives guarantees aberration-free focusing [117]. A probe beam scattering technique [117] is employed to rapidly collect $R_{max}(E_L)$ data for a large number of irradiation parameters. These features are utilized to establish scaling laws for the energy dependence of nonlinear energy deposition in the ($\tau_L$, $\lambda_L$, *NA*) parameter space.

Orthogonal adjustment of pump- and probe microscope objectives enables us to photograph plasma luminescence in side view with sub-micrometer resolution. With tight focusing, plasma luminescence can be recorded even for fs plasmas. For ns breakdown at visible wavelengths, scattered pump laser light is also recorded on the plasma colour photographs. This makes it possible to distinguish between regions of primary energy deposition and plasma inflation by radiative energy transport. Evaluation of the plasma size from the time-integrated photographs together with absorption data from transmission measurements provides values for the average plasma energy density. These data sets are then used to establish an energy balance between vaporization, shock wave and bubble energy as a function of plasma energy density. Analysis of the plasma structure and the spectral information contained in the colour photographs in conjunction with the concept of a moving breakdown front in the bulk of water [108,132-136] provides information on the breakdown dynamics at super-threshold pulse energies.

Through the systematic experimental investigations in a large parameter space, we discovered a new regime of fine-adjustable deterministic energy deposition with temporally smooth UV ns pulses from microchip lasers. This enables material nanoprocessing and precise surgery with cost-effective compact laser sources, which is demonstrated on corneal dissection and refractive index modifications in glass. We explain this finding through the interplay of strong-field ionization (SI), avalanche ionization (AI), recombination and thermal ionization (TI). Here SI provides seed electrons, inhibition of AI by recombination at low electron densities results in nanoeffects, and at higher pulse energies AI combined with TI at larger electron densities leads to a second breakdown step exhibiting the well-known luminescent plasma formation and strong mechanical effects.

Previous work has shown that the plasma energy density in bulk water is limited by the movement of an optical breakdown wave during the laser pulse [108,133,135], and for ns breakdown additionally by an enlargement of the plasma size through radiative energy transport from the regions of primary energy deposition [108]. Consequently, the energy density in plasmas produced by energetic ns laser pulses focused at $NA = 0.25$ was found to be limited to about 40 kJ/cm$^3$ [40], which is smaller than in the plasma skin layer produced by ultrashort laser pulses at target surfaces [6,137-139]. In this paper, we determine the average plasma energy density achievable by tight focusing ($0.3 \leq \mathrm{NA} \leq 0.8$) of laser pulses in large ($\tau_L$, $\lambda_L$, $E_L$) parameter space and explore local inhomogeneities and 'hot spots' through high-resolution photography of plasma luminescence. Since regions of primary energy deposition scatter the incoming pump laser light, they can be distinguished from the broadband plasma luminescence. Photographs reveal a fascinating structural complexity, with high-density strings formed during primary energy deposition, which are a consequence of an intrinsic spatial instability of avalanche ionization that is proportional to irradiance *and* local electron density. Energy redistribution by soft X-rays emitted from the strings then inflates the plasma into a larger region with lower average energy density that emits the diffuse luminescence known from previous research.

Our findings pose challenges and provides new impulses for future experimental investigations of the extreme states of matter in laser-induced breakdown as well as for optical breakdown modeling. The spatiotemporal interplay of nonlinear beam propagation and plasma formation in breakdown by ultrashort laser pulses is well covered by existing models [130,131,136,140,141]. However, modeling of the interplay between SI, AI, TI, recombination during nanosecond pulses is still in its infancy, and we are not aware of any optical breakdown models covering the intrinsic instability of AI and the role of bremsstrahlung –mediated energy redistribution. Future experimental studies resolving the dynamics of plasma and shock wave formation during ns laser pulses with picosecond temporal and sub-micrometer spatial resolution can provide a quantitative basis for model validation.

The broad experimental coverage of the optical breakdown dynamics in the ($\tau_L$, $\lambda_L$, $NA$, $E_L$) parameter space provides a solid basis for numerical simulations. However, detailed simulations go beyond the scope of this paper, especially since theoretical models for some aspects of the observed breakdown dynamic still need to be developed. Instead, we present in section 2 a conceptual heuristic framework on the interplay of SI, AI, TI and recombination as a function of irradiance and pulse energy. This framework guides the experiments described in section 3 and is needed to understand the results presented in section 4. In section 5, we first interpret the experimental findings and place them in the context of previous work, and then we use the new insights to derive consequences for laser surgery and material processing in bulk dielectric media, laser ablation of absorbing targets in liquids for nanoparticle generation, fragmentation of the laser-produced nanoparticles, and laser-induced breakdown spectroscopy. Finally, section 6 formulates challenges and a roadmap for future experimental exploration and advanced modelling of plasma formation and plasma-induced hydrodynamic events, which open avenues for novel applications beyond laser surgery, LAL and LIBS. Furthermore, it expands the view towards low-density plasma phenomena below the cavitation bubble threshold and their applications.

Altogether, we establish a comprehensive picture of nonlinear energy deposition, which provides a deeper understanding of the underlying mechanisms, opens new avenues for nanoprocessing with cost-effective compact devices, facilitates parameter selection for material processing and laser surgery, and outlines perspectives for future research in high-energy density physics.

## 2. Theoretical considerations and experimental rationale

Optical breakdown is a sequence of nonlinear energy deposition via plasma formation associated with rapid heating and pressure rise in the plasma followed by the emission of an acoustic transient or shock wave and a phase transition with cavitation bubble formation [40,42]. The mechanisms underlying nonlinear energy deposition in the ($\tau_L$, $\lambda_L$, $E_L$) parameter space are very complex – even for focusing at large NA, where nonlinear beam propagation plays little role. In this section, we present an overview of the expected breakdown behavior based on generic equations and heuristic arguments that motivate the experimental rationale and provide a framework for the interpretation of the measurement results.

### 2.1. Pathways of nonlinear energy deposition and thermalization

The general sequence of nonlinear energy deposition in large-bandgap dielectrics at high irradiance is depicted in **Fig. 1**. Plasma formation is initiated by strong-field ionization (SI) via multiphoton ionization (MPI) or tunneling across the band gap between valence band (VB) and conduction band (CB). Seed electron generation is followed by avalanche ionization (AI) that consists of a sequence of inverse bremsstrahlung absorption (IBA) events of single photons through which the kinetic energy of the CB electrons is raised until it suffices to free another VB electron by impact ionization. Conservation of energy and momentum during IBA requires that the photon interacts with the electron during a collision with a heavy particle (atom or ion). Moreover, it requires a kinetic energy of the impacting electron $\geq (3/2)\,\tilde{\Delta}$ [37,142,143], where $\tilde{\Delta}$ is the ionization potential of the dielectric. This potential is given by the sum of band gap energy $E_{gap}$ and the oscillation (quiver) energy $E_{osc}$ of CB electrons in the strong electromagnetic field [37,144]. Avalanche ionization initiated by strong-field ionization is the backbone of nonlinear energy deposition into large band gap materials such as water because it leads to a rapid multiplication of the number of free electrons [6,36,37,143].

The energy deposited into the electronic system is thermalized via collisional electron-phonon coupling and electron-hole recombination. The resulting volumetric energy density $U$ is given by the time-integrated product of CB electron density, $\rho_c$, and the average energy $\varepsilon_{avg}$ carried by CB electrons, where $\varepsilon_{avg}$ is the sum of $\tilde{\Delta}$ and the average kinetic energy of the electrons, $\varepsilon_{kin.avg}$. For fs breakdown, only one set of free electrons is produced during the laser pulse, and their energy is thermalized afterwards [22,37,115]. Here, $U$ is simply the product ($\rho_c \times \varepsilon_{avg}$) but for ns breakdown, it can be much larger. When a sufficiently high thermodynamic temperature, $T_{equil}$, of the breakdown region is reached, VB electrons can be lifted into the CB via thermal ionization. For sufficiently long pulse durations, recombination and thermal ionization create feedback loops acting at different time scales through which the CB-electron density, $\rho_c$, and $T_{equil}$ act back on the energy deposition process [**Fig. 1(b)**]. The respective feedback strength is given by the dependence of avalanche-ionization and recombination rates on free electron density, and by the temperature dependence of thermal ionization. In the following, we will see how the different time constants of primary energy deposition processes and feedback loops result in different features of the breakdown dynamics depending on laser pulse duration and wavelength.

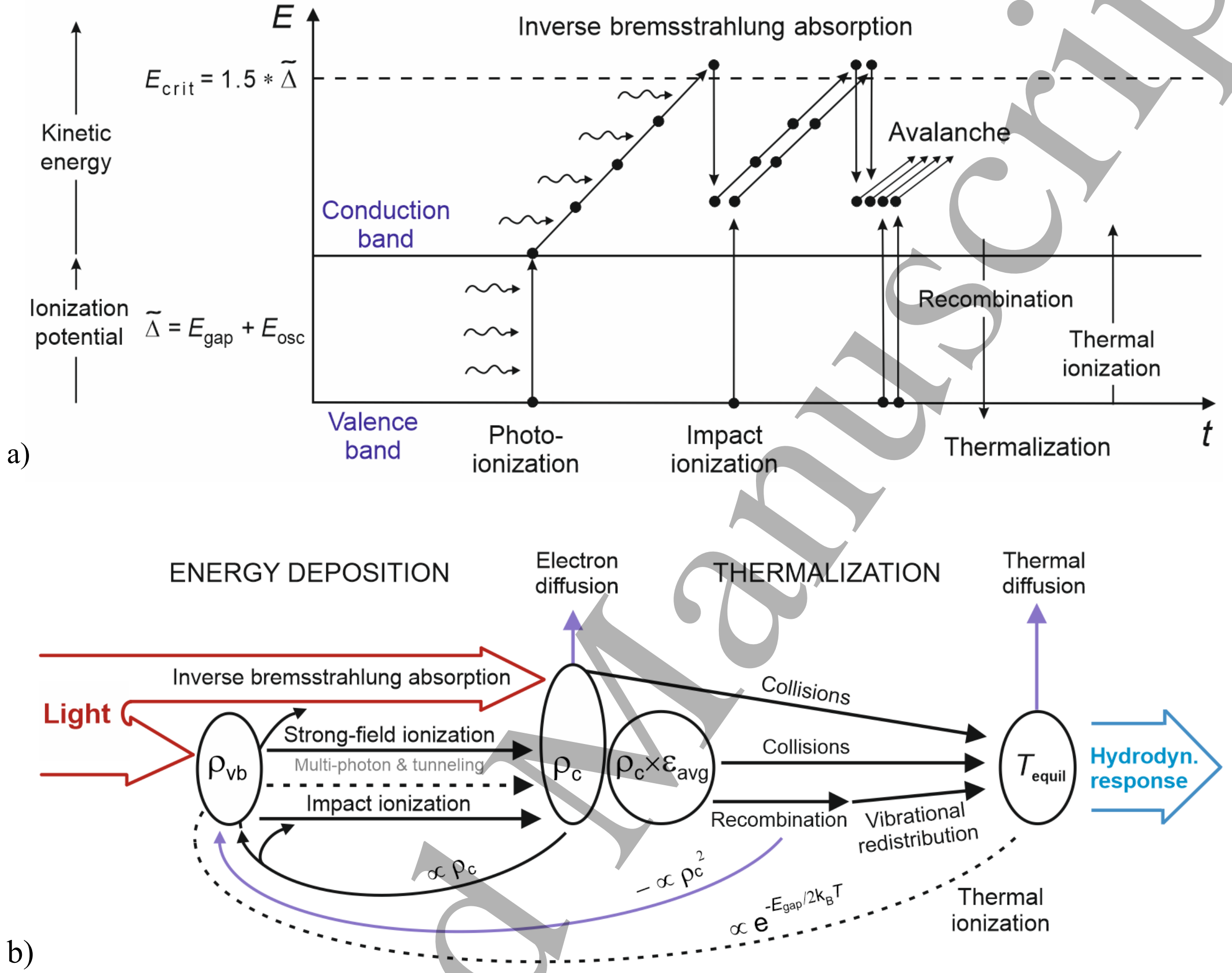

**Figure 1** Pathways of nonlinear energy deposition and thermalization during optical breakdown in large-bandgap dielectrics. (a) Visualization of the pathways within the band structure of the dielectric, (b) Visualization with feedback loops between the individual mechanisms. The energy flow from the laser light into the electronic system of the dielectric is followed by thermalization and the hydrodynamic response of the medium. These pathways include feedback loops acting at different time scales through which the free-electron density $\rho_c$ and temperature $T_{equil}$ produced by nonlinear absorption and thermalization act back on energy deposition and ionization. When the thermodynamic temperature of the medium is sufficiently high, nonlinear absorption is complemented by thermal ionization. Pathways illustrating the coupling of incident laser light into VB and CB electrons are marked in red, and pathways indicating CB electron losses and heat dissipation are marked in violet. While the feedback of free-electron density on the avalanche ionization rate occurs on a fs time scale, the recombination time ranges from < 1ps to several ps (depending on $\rho_c$), and the thermalization time as a whole amounts to ≈20 ps. Recombination, thermalization, and thermal ionization play no role during fs breakdown but become increasingly important for longer pulse durations. The hydrodynamic response (shock wave emission and cavitation) sets in when the energy of the free electrons is thermalized.

*Strong-field ionization* provides seed electrons quasi-instantaneously. Seed electron generation by MPI dominates from UV to near-IR and at low irradiance values, whereas tunneling becomes relevant at longer wavelengths [6,22,145-147]. The dependence of strong-field ionization on irradiance, $I$, becomes complicated, when tunneling gets involved but when multiphoton ionization dominates, it is simply proportional to $I^k$, where $k$ is the order of the multiphoton process [113,144]. With increasing wavelength, simultaneous absorption of more and more photons is required to cross the band gap, and the multiphoton ionization rate $\eta_{MPI}$ decreases.

For bubble formation at $\tau_L \geq 100$ fs, most electrons are produced by *avalanche ionization* [37,143]. This process is subject to time constraints because impact ionization can occur only after conduction band (CB) electrons have gained sufficient energy through inverse bremsstrahlung absorption. Free electrons can absorb photons only during collisions with heavy particles, which occur in intervals of ≈ 1 fs [37,113,115,148]. Since several subsequent absorption processes are needed to collect sufficient energy for impact ionization, each doubling sequence of free electron density takes several femtoseconds and the feedback of free-electron density on the avalanche ionization rate has a time-lag [113,146]. Nevertheless, since it occurs on a fs time scale, this feedback loop plays a significant role already in fs breakdown. The energy distribution of CB electrons is initially nonstationary and its evolution during this time has to be described by a multi-rate equation [143,146,149]. However, for sufficiently long pulse durations the energy distribution evolves asymptotically into a stationary distribution, whereby the transition time depends on bandgap, wavelength, and irradiance. For water, this time amounts to about 100 fs [37,149]. Once the stationary regime has been reached, avalanche ionization is proportional to the CB electron density and approximately proportional to $I$ [37,113,150]: $(\mathrm{d}\rho_c/\mathrm{d}t) = \eta_{AI} \times \rho_c\, I$. The wavelength dependence of AI is contained in the avalanche ionization rate, $\eta_{AI}$. With increasing wavelength, the photon energy decreases and more and more inverse bremsstrahlung events are needed to provide the energy required for impact ionization but the rate constant for individual IBA events is approximately proportional to $\lambda_L^2$ [37,38,113]. Both relations together yield a relatively weak overall dependence of the AI rate $\eta_{AI}$ on wavelength [37].

The pulse duration dependence of plasma-mediated ablation and bubble formation is determined by an interplay of SFI and AI. Time constraints of AI play a significant role for the pulse duration dependence of optical breakdown thresholds in the fs range. For pulses between 100 fs and 5fs, Lenzner et al. reported a relatively slow drop of the radiant exposure ablation threshold $F_{th}$ with decreasing $\tau_L$, corresponding to a fast increase of the breakdown irradiance $I_{th}$ [126]. This behavior reflects the growing time constraints of AI with decreasing $\tau_L$, which goes along with an increase of the relative importance of SFI. However, Lenzner et al. did not take the time-lag of AI into account and attributed the resulting mismatch between theoretical predictions and experimental observations to deficiencies of the Keldysh model for SFI.

Time constraints of AI were first discussed by Kennedy [113] and are included in all models that explicitly treat the gain of kinetic CB electron energy as a succession of IBA events separated by a finite collision time. Coupling of laser energy into CB electrons during collisions and collisional losses are key elements in the description of AI [6,125,142,151-153]. Stuart et al. showed that an approximate description of these processes by a rate equation based on the Drude model yielded good results with less numerical expense than full models based on solving kinetic equations [125]. Noack et al. adapted the rate equation approach to breakdown in water and considered the finite time needed to reach the impact ionization energy [115]. Simulations considering SFI initiation, AI, and recombination led to good agreement with experimental data for pulse durations between 100 fs and 76 ns. Vogel et al. then combined numerical simulations of plasma formation in water with a model of thermoeleastic bubble formation [22,117]. Rethfeld introduced a multirate equation (MRE) model, which describes the temporal evolution of the CB electron energy distribution and ionization avalanche rate during the laser pulse. As already mentioned, the MRE approach is particularly important for short pulse durations below 100 fs and at low irradiance [146,153]. Christensen & Balling modified the approach to make it fully consistent with the law of energy conservation [143], and Liang et al. adapted the MRE model to breakdown in water considering the complex band structure of water [149], which contains an intermediate energy level for solvated electrons that facilitates breakdown initiation. Linz et al. modeled the wavelength dependence of ns and fs breakdown in water taking this band structure into account [36,37]. Model fits to experimental $I_{th}(\lambda)$ data for IR ns breakdown provided evidence that AI is initiated by multiphoton ionization rather than through background electrons or impurities. Fits to experimental results for fs breakdown at 250 fs pulse duration showed a dominant role of AI over SFI in the entire investigated wavelength range from 335 nm to 1085 nm.

The time scale for *recombination* is longer than the time constants relevant for avalanche ionization. In liquid water, recombination proceeds mainly as non-radiative interaction of CB electrons with $H_3O^+_{aq}$ ions and as electron attachment to neutral OH fragments [37,154,155], while in plasmas with large electron density and dissolved band structure electron-ion recombination dominates [156]. At irradiance values leading to optical breakdown with cavitation bubble formation, cross recombination processes between electrons and holes or electrons and ions from independent ionization events dominates by far over geminate recombination [157]. Under these circumstances, recombination is proportional to the square of free-carrier density because two types of free-carriers are involved in each event [113,158]. The recombination time ranges from ≈ 1ps to several ps, depending on $\rho_c$ [37]. Therefore, recombination influences the breakdown dynamics in ps and ns breakdown but is negligible for fs pulses.

*Thermalization of the absorbed laser energy* involves energy transfer from free electrons to heavy particles occurring through elastic and inelastic collisions [159-161] as well as through recombination followed by energy dissipation via vibrational relaxation [162-164]. Thermalization through inelastic collisions with vibrational excitation of water molecules and dissipation through the hydrogen bond network is fast and dominates for low-density plasmas such as the electron cascade produced by ionizing radiation and occurs on a time scale below 1 ps [165-168]. Experimental investigations yielded plasma life times ≤ 300 fs for fs plasmas with low electron densities ≤ 3 x $10^{18}$ $cm^{-3}$ [169,170], two orders below the bubble threshold [37]. In low-density plasma, hydrating CB electrons interact with intact water molecules and are then solvated in traps below the CB [171]. By contrast, in high-density plasmas well above the breakdown threshold, water molecules are largely dissociated and the band structure is disturbed [156,172,173]. In this regime, electron collisions are mainly elastic, and longer thermalization times, $\tau_{therm}$, in the order of 20-30 ps have been observed [123,174]. Here, thermalization proceeds mainly via electron-ion recombination. The recombination rate for high-density plasmas was determined by analyzing the decay of plasma scattering [123,174] and luminescence [158]. Before thermalization, the electron temperature exceeds the ion or lattice temperature, respectively, and during the thermalization process both temperatures equilibrate and a thermodynamic temperature evolves [152,156,175-178]. The thermodynamic temperature then governs the onset of hydrodynamic phenomena such as shock wave emission and bubble formation [7,153,179].

Heating of the focal volume during the laser pulse causes a gradual change of the electronic energy distribution from Fermi-Dirac statistics towards a Boltzmann distribution [159,180,181]. The change of the energy distribution can significantly contribute to the CB electron density, when high temperatures are reached and the high-energy tail of the Boltzmann distribution reaches into the CB. We denote this temperature-related process as "*thermal ionization*." Since it refers to the electron density distribution in thermal equilibrium between electron and heavy particle populations, it differs significantly from strong-field and avalanche ionization, which create a thermal imbalance by depositing energy into the electronic system. As long as the band structure is still intact, the temperature-related density of electrons in the conduction band, $\rho_{therm}$, can be assessed using the theory of electric conductivity of semiconductors [180,181]. When the band structure is dissolved at high temperatures, the ionization degree is described by the Saha equation [182-184]. Both approaches predict a proportionality $\rho_{therm} \propto exp(-E_{gap}/2k_B T)$, where $E_{gap}$ is the band gap energy, and $k_B$ the Boltzmann constant [159,180-183]. The exponential increase of $\rho_{therm}$ with $T$ suggests a rapid onset of thermal ionization beyond a critical temperature $T_{cr}$, which will become relevant at times $t > \tau_{therm}$.

With intact band structure, ionization will cease, when $\rho_c$ reaches such a large value that the highest molecular orbital of the valence band is depleted [6,143,156,185]. This level is the $1b_1$ level, which contains a pair of bound electrons [185,186]. Since the molecular density of liquid water is $3.34 \times 10^{22}$ cm$^{-3}$, full depletion of the $1b_1$ level corresponds to a free electron density of $\rho_c = 6.68 \times 10^{22}$ cm$^{-3}$ [113]. However, electron densities that fully deplete the $1b_1$ level are likely associated with a dissolution of the band structure of water and dissociation of water molecules leading to the development of an electron-ion plasma. In any case, once a very high ionization degree and electron density are reached, most absorbed laser energy goes into inverse bremsstrahlung absorption with little further impact ionization, which results in rapid heating of the free electrons to high levels of kinetic energy [125].

*Electron and heat diffusion* out of the focal volume inhibit the breakdown process both during surface ablation of transparent dielectrics and plasma formation in bulk media but on very different time scales. During surface ablation, a thin plasma skin layer with nanometer thickness evolves, and electron and heat diffusion become relevant already on a time scale of a few picoseconds [125,137,139,177,178,187]. The heat diffusion time is proportional to the square of the thickness of the absorbing layer, and the heat penetration depth is proportional to $(D \cdot t)^{1/2}$, where $D$ is the heat diffusion coefficient and $t$ the time [187,188]. Hence, the heated zone spreads by a factor of $\tau_L^{1/2}$ during the laser pulse, which must be compensated by an increase of radiant exposure (fluence) by the same factor to achieve the ablation temperature. This leads to a proportionality of breakdown fluence to $\tau_L^{1/2}$ both for dielectrics and metals [112,125,177,187]. By contrast, for breakdown in bulk dielectrics, no skin layer evolves [133,134]. Even with tight focusing, focal dimensions are in the range of hundreds of nanometers to several micrometers, and diffusion becomes relevant only on a time scale of hundreds of nanoseconds [22,156,188,189].

Therefore, electron diffusion and heat conduction can be neglected. Here the breakdown threshold for pulse durations > 10 ps depends mainly on the irradiance at which multiphoton-generated seed electrons for AI are available [22].

The above considerations reveal that essentially four interaction mechanisms acting on different time scales govern the evolution of CB electron density in bulk water. These are strong-field ionization SI, avalanche ionization AI, recombination, and thermal ionization TI. While fs breakdown can be understood by analyzing only the interplay of SFI and AI, for longer pulse durations also recombination and thermal ionization must be considered. This brings in two new feedback loops (Fig. 1), which complicates the interplay of the basic mechanisms. For getting a general understanding of the breakdown behavior in the entire pulse duration range from fs to ns, it is useful to keep the description of the individual mechanisms as simple as possible. Therefore, we describe their dependencies on irradiance $I$, CB electron density $\rho_c$ and temperature $T$ by a set of generic equations:

$$\left(\frac{d\rho_c}{dt}\right)_{MPI} = \eta_{MPI}\, I^k\,, \tag{1}$$

$$\left(\frac{d\rho_c}{dt}\right)_{AI} = \eta_{AI}\, \rho_c\, I\,, \tag{2}$$

$$\left(\frac{d\rho_c}{dt}\right)_{rec} = -\eta_{rec} \times \rho_c^2\,, \tag{3}$$

$$\rho_{c,therm} \propto e^{-E_{gap}/2k_BT}\,. \tag{4}$$

For the sake of simplicity, we consider here only multiphoton ionization as representative for strong-field ionization and use the simple AI rate equation for the stationary avalanche ionization regime. For the sake of generality, we assume a simple band gap without energy levels between valence and conduction band. The actual band structure is more complex in water, where an intermediate state at the energy level of solvated electrons facilitates plasma initiation [36,37,149], and in fused silica, where self-trapping of excitons impairs avalanche ionization [190,191]. Furthermore, we assume a constant band gap during plasma formation although the band structure deteriorates once high electron densities and temperatures are reached in the focal volume, and the rate constants for all ionization processes change.

The relative strength of MPI and AI depends on laser wavelength and pulse duration, as discussed further above for fs breakdown. The strength of recombination and thermal ionization compared to MPI and AI depends mainly on pulse duration and will be discussed in detail in the next section.

Recombination comes into play for pulse durations longer than the thermalization time of CB electrons, and thermal ionization plays a role when recombination and collisional dissipation have lasted long enough to heat the focus to temperatures ≥ 3000 K [159]. It can be seen from Fig. 1 that recombination and thermal ionization counteract each other – while recombination inhibits the ionization process for long laser pulses, it is enhanced by thermal ionization. A consideration of their interplay reveals interesting features that have been overlooked in the past.

While consideration of details is required to explore and quantify the shades of optical breakdown dynamics, the simple form of Eqs. (1) – (4) helps in obtaining a qualitative understanding of the great picture in the entire range of pulse durations from fs to ns. It enables to anticipate and characterize general scenarios for the irradiance and energy dependence of the breakdown behavior in the ($\tau_L$, $\lambda_L$) parameter space. The next section describes three qualitatively different scenarios covering the range from low-density to high-density plasma formation. Two scenarios are well known from previous research, and a third one is predicted by analyzing the interplay of nonlinear energy deposition, recombination, and thermalization pathways. The detailed features of these scenarios and their borders in the ($\tau_L$, $\lambda_L$) space will then be explored experimentally.

## 2.2. Breakdown scenarios from precisely tunable energy deposition to "big bang"

*Scenario 1: Continuous tunability such as observed in fs breakdown*. At ultrashort pulse durations, the breakdown dynamics is governed solely by strong-field and avalanche ionization. Recombination and thermalization can be neglected since they occur on a ps time scale after the laser pulse. Strong-field-created seed electrons are abundant but the value of $\rho_c$ at the end of the laser pulse depends mainly on the avalanche ionization rate, as discussed in the previous section. Since the avalanche ionization rate is limited by time constraints, $\rho_c$ increases only gradually with growing irradiance [Eq. (2)]. Therefore, the onset of breakdown is smooth, and the amount of deposited energy – and the bubble size - can be continuously tuned by varying $E_L$.

*Scenario 2: "Big bang" such as observed in IR ns breakdown*. With increasing laser pulse duration, more and more doubling sequences of the ionization avalanche can occur during a pulse. Therefore, the irradiance threshold for bubble formation could largely decrease for long pulses if seed electrons for avalanche ionization were readily available. However, seed electron generation by multiphoton ionization requires a minimum irradiance. This hurdle is particularly large for long wavelengths, where high-order multiphoton processes are required for crossing the bandgap. Thus, for IR ns pulses seed electron generation by multiphoton ionization is the critical hurdle for the occurrence of breakdown [36].

Once this hurdle is overcome, the ionization avalanche proceeds very fast. In conjunction with thermal ionization, it can largely overshoot the threshold condition and produce bright plasma luminescence and a large bubble already at the onset of breakdown. In this "big bang" scenario, tunability of energy deposition is restricted to the high-density plasma regime.

*Scenario 3: Stepwise tunability in ps and ns breakdown.* At short wavelengths, multiphoton ionization always provides sufficient seed electrons for AI, even at nanosecond pulse durations. However, although the ionization avalanche has an easy start, it is considerably slower than in IR breakdown because the avalanche ionization rate is smaller for shorter wavelengths as it is approximately proportional to $\lambda^2$ [37]. The slow ionization avalanche will thus be inhibited by recombination because the recombination rate is proportional to $\rho_c^2$ [Eq. (3)], while the avalanche ionization rate exhibits a linear dependence on $\rho_c$ [Eq. (2)]. In the lower irradiance range near threshold, the growth of the free electron density is limited and a dynamic equilibrium between avalanche ionization and recombination evolves in which laser energy is continuously deposited into the electronic system and thermalized simultaneously. The focal temperature reached at the end of the laser pulse depends on the equilibrium level of $\rho_c$, which is determined by the focal irradiance. The above description implies that ns pulses at UV wavelengths and possibly also at short visible (VIS) wavelengths fulfill essential prerequisites for precisely tunable energy deposition and the creation of nano- and micro effects – a feature, which traditionally has been ascribed to ultrashort laser pulses and scenario 1.

With increasing irradiance (or $E_L$, respectively), the ionization avalanche produces an ever higher free electron density before it is inhibited by recombination. Thus, the focal temperature reached at the end of the pulse increases gradually as long as recombination can hold the ionization process. Beyond a critical temperature, the breakdown dynamics will change dramatically due to the exponential dependence of thermal ionization on temperature [Eq. (4)]. Together with avalanche ionization, thermal ionization now overcomes the inhibiting action of recombination, and $\rho_c$ can suddenly shoot up to very high ionization levels. Consequently, brightly luminescent plasma with high energy density is formed. The existence of a second breakdown threshold at the upper end of the nano/micro regime is a key difference to scenario 1 with continuous tunability. Thus, scenario 3 is an intermediate between the other better-known scenarios, sharing features of scenario 1 at low irradiance and of scenario 2 at high irradiance. The abrupt transition between the tunable low- and high-density plasma regimes goes along with a sudden, stepwise increase of the vaporized liquid volume, plasma pressure and bubble size.

The tunable energy deposition and creation of nanoeffects in the low-energy part of scenario 3 opens interesting perspectives for the use of compact diode-pumped UV solid state lasers in applications for which to date fs lasers are employed. Some evidence for the possibility of creating nano- and microeffects with tightly focused UV-A ns lasers has been reported previously [192,193]. However, this possibility has not yet been systematically explored because of the widespread view that only fs pulses allow for reproducible (“deterministic”) energy deposition whereas ns breakdown exhibits “stochastic” character and is associated with vigorous laser effects [22,127,129,194,195]. In the present paper, we investigate the prerequisites for deterministic breakdown behavior and explore, in which wavelength range nanoeffects by non-luminescent ns plasmas can be produced.

### 2.3. Deterministic vs. stochastic breakdown behavior

For ultrashort laser pulses, where seed electrons are abundant, breakdown dynamics and bubble threshold show little dependence on impurities as long as pulse duration and energy remain stable. In these regions, the breakdown dynamics is, therefore, “inherently” deterministic. By contrast, when MPI initiation is the critical hurdle for the occurrence of breakdown, the breakdown dynamics depends more sensitively on small variations of the initial conditions. Here, a deterministic behavior can be expected only for highly reproducible laser pulses with smooth temporal profile and good beam quality, and for largely impurity-free media. While the temporal profile of mode-locked laser pulses is fairly stable, many solid state ns lasers are run in longitudinal multimode operation in which the pulse shape exhibits intensity spikes arising from statistical interference of the longitudinal modes that affect the multiphoton excitation rate [Eq. (1)]. Pulse-to-pulse fluctuations of the spiking behavior with intensity peaks of varying height at different times during the pulse introduce strong fluctuations in seed electron generation that result in a stochastic breakdown behavior.

Impurities can also influence the seed electron generation - either by localized heating of tiny particles followed by thermionic electron emission or by providing intermediate energy levels in the water band gap that facilitate multiphoton excitation. Contaminations of low concentration in a nominally ‘pure’ medium differ from biomolecules in aqueous biological media that provide numerous centers of reduced excitation energy at high concentration [196-198]. Both thermionic emission of seed electrons and/or reduced excitation energy for MPI at impurities in biological tissues can lower $I_{th}$ in IR ps and ns breakdown [109,199]. By contrast, in UV breakdown, seed electrons are readily available, and significant changes of $I_{th}$ can here only arise when the impurity concentration is so high that it significantly widens the MPI-channel [37], or when the seed electrons become available at fluctuating times during a laser pulse exhibiting spikes.

Impurity-related fluctuations of ns breakdown in bulk transparent dielectrics have been reported in studies in which the laser beam was only weakly focused at *NA*s below 0.1 [109,200,201]. Under such circumstances, the probability of hitting upon particulate impurities is much larger than for the large *NA*s investigated in the present study because the focal volume $V$ increases rapidly with decreasing $NA$: $V \propto 1/NA^4$. For sufficiently large $NA$, plasmas are compact [54,108,202] and likely smaller than the average distance between impurities. Therefore, impurities will hardly matter, and deterministic behavior can be expected not just for ultrashort laser pulses but also for ns breakdown by laser pulses with smooth temporal shape.

In our study, the influence of the temporal laser pulse shape is addressed by comparing the breakdown behavior of longitudinal single- and multimode laser pulses. Optimum reproducibility of all other irradiation parameters is ensured by using laser beams of high quality and diffraction limited focusing conditions. The potential influence of impurities is explored by varying the focusing strength between $0.3 < NA < 0.9$, corresponding to an 81-fold variation of the focal volume.

## 2.4. Scaling of laser-induced effects with pulse energy

Laser effects in transparent dielectrics may range from free-electron mediated chemical changes [22,198,203,204] through a phase transition up to vigorous shock wave emission and cavity formation [22,40,114,121,123,176,205]. In the present paper, we use the onset of a phase transition (bubble formation) in water as benchmark for the optical breakdown threshold and employ the maximum bubble radius $R_{max}$ as convenient measure for the magnitude of the laser-produced effects above threshold [117]. This approach, which was introduced by Vogel et al. [117], facilitates a detailed exploration of scaling laws in a large ($\tau_L$, $\lambda_L$, $E_L$) parameter space because the $R_{max}$ ($E_L$) scaling can be fast and precisely determined through pump-probe measurements. Furthermore, the bubble radius provides a uniform measurement parameter over a very large range of plasma energy densities, whereas for a similar energy range in solids the damage type changes from colour center formation [206] through refractive index modifications [207] to cavity and crack formation [208,209].

## 2.5. Characterization of energy partitioning as a function of plasma energy density

The $R_{max}$ ($E_L$) curves are a convenient measure for assessing the magnitude of laser-induced effects in water and soft tissue. However, additional information is required to fully characterize the hydrodynamic effects that besides bubble formation also include shock wave emission. The volumetric energy density $U_{\text{Plasma}} = E_{\text{abs}} / V_{\text{Plasma}}$ of the laser-produced plasma is a key parameter governing the partitioning of absorbed laser energy in fractions going into vaporization, bubble formation, and shock wave emission.

The lower end of the plasma energy density scale in our investigations is given by the threshold for homogeneous bubble nucleation, which for ns pulses equals the kinetic spinodal limit and amounts to $U_{\text{th}} \approx 1.2$ kJ cm$^{-3}$ [188]. For laser pulse durations shorter than the acoustic transit time through the plasma, energy deposition is stress-confined and thermoelastic tensile stress reduces $U_{\text{th}}$ to $\approx 0.6$ kJ cm$^{-3}$ [22,210]. Close to threshold, a large percentage of the absorbed energy goes into vaporization of liquid within the plasma volume, whereas in high-density plasmas most energy will be converted into the mechanical energy of shock wave and cavitation bubble produced upon plasma expansion [40,41,58]. The upper end of the plasma energy density in bulk dielectrics and the corresponding temperatures and pressures are of great interest to the scientific community [114,120,121,211,212] but there is still a need for systematic measurements of the energy density and partitioning covering a large range in ($\tau_{\text{L}}$, $\lambda_{\text{L}}$) parameter space.

An experimental determination of $U_{\text{plasma}}$ requires measurements of the absorbed energy fraction of the incident laser light, $E_{\text{abs}}$, and of the plasma volume, $V_{\text{plasma}}$. Although high-density ns plasmas emit bright radiation delineating the plasma shape [108], the luminescence from fs plasmas in bulk media is very weak, especially when the laser pulses are focused at small or moderate *NA*. That made it impossible to use plasma luminescence as indicator for fs breakdown and led researchers to introduce bubble formation as alternative criterion [110,113]. Thus far, data on the size of fs-laser produced plasmas rely on estimates [120,121,211,212], numerical modeling of the energy deposition [130,136,213-215], shadowgraphs of the plasma or the laser-induced phase transition [110,114,170,216-218], shadowgraphs combined with photoacoustic imaging [218,219], third harmonic imaging [220], or on the damage morphology in dielectric solids [208,221]. Evaluation of the damage morphology is useful for well-delineated low-density plasma effects but becomes problematic for high-density plasmas when cracks or cavities impair a precise localization of the plasma border. It is not at all possible for liquid breakdown because no permanent trace is left when the laser-induced bubble has vanished. A way for rapid in-situ determination of plasma size in water is offered by Favre's observation that white-light emission of fs plasmas becomes detectable when the laser pulses are focused at large *NA* [222]. In their study, water droplets were illuminated by a collimated laser beam, and light reflected at the droplet wall was focused back into the droplets. However, that goes along with strong spherical aberrations producing a caustic.

In this paper, we take side-view photographs of the plasma luminescence produced by laser pulses focused through an aberration-free microscope objective. Recording of the plasma luminescence becomes possible by integration over the luminescence from a series of pulses.

Photographic determination of $V_{plasma}$ is also a prerequisite for establishing an energy balance for the absorbed laser energy, $E_{abs}$. Knowledge of the plasma volume enables to determine the fraction of absorbed laser energy going into vaporization, $E_V$, when the luminescent region is completely vaporized. The bubble energy $E_B$ is readily obtained from the maximum bubble radius (see section III.B). The absorbed energy can be determined by measuring the plasma transmission if most of the incident light is absorbed. This is approximately true for luminescent plasmas, where we found that backscattering and reflection amounts to less than 2% of the incident laser energy [223]. The shock wave energy can then be obtained by $E_{SW} = E_{abs} - E_V - E_B$ if other energy fractions play no significant role. Energy losses by plasma radiation leaving the luminescent region are neglected because very little energy leaves the water cell as plasma radiation. Literature values for blackbody temperatures of ns plasmas range from 10000 K for 4 mJ pulses [224] to 16000 K at much larger pulse energies ≥ 100 mJ [225,226]. Vogel et al. estimated the energy content of blackbody radiation from a luminescent ns plasma with $T_{blackbody}$ = 10000 K and found that it amounts to less than 0.1 % of the absorbed laser energy [114]. However, although little radiation energy leaves the plasma, we will see in section IV.C. that in nanosecond breakdown very hot regions are produced within the region of primary energy deposition, which leads to radiative redistribution of a significant energy amount *within* the luminescent region. Work done against surface tension and to overcome liquid viscosity must be considered for nano- and microbubble dynamics close to the breakdown threshold [42] but plays no role for millimeter-sized bubbles [63,66,227] and little role for the smallest bubbles produced by luminescent plasmas, which still have a size larger than 10 μm [42]. We showed previously, that neglecting this part of the energy balance as well as losses by plasma radiation, an indirect assessment of $E_{SW}$ from $E_{abs}$, $E_v$, and $E_B$ becomes possible, which avoids difficult measurements of shock wave amplitude and duration that would be needed for a direct determination of $E_{SW}$ [40,114,228,229].

Knowledge of the plasma energy density provides also an independent starting point for determining plasma temperature, $T_{plasma}$, and pressure, $p_{plasma}$, different from spectroscopic approaches because these physical quantities are linked to mass density $\rho_0$ and energy density $U$ (or internal energy, respectively) through the equation of state (EOS) of water [230]. The term "plasma temperature" here refers to the thermodynamic temperature after thermalization of electron energies, which is completed after 20-30 ps.

In fs breakdown, a thermodynamic temperature and state are established only after the laser pulse, when the thermalization of CB electron energy is completed. However, since hydrodynamic phenomena such as shock wave emission and bubble formation set in on a time scale of tens to about 100 ps, when sound propagation and particle movement become relevant compared to the source size, use of the thermodynamic temperature for characterizing the plasma with regard to hydrodynamic phenomena seems justified also for fs breakdown.

With ultrashort laser pulses, energy deposition is isochoric and no significant plasma expansion occurs during the laser pulse. Therefore, the mass density at the end of the laser pulse is known and $U_{\text{plasma}}$ data can be used to determine $T_{\text{plasma}}$ and $p_{\text{plasma}}$. Since thermal expansion takes place at sound velocity, the pulse duration $\tau_L$ must be shorter than the acoustic transit time through the plasma volume to fulfill this condition [188]. Therefore, it is often also denoted as ‘stress confinement’ condition. We will see in section IV.D that stress confinement applies not only for fs breakdown but also for luminescent ns plasmas.

## 3. Experimental methods

### 3.1. Experimental setup

The setup for the investigation of optical breakdown in water is presented in Fig. 2. Laser pulses of different durations, wavelengths, and temporal pulse shapes were focused through microscope objectives of various *NA*s into a cuvette containing double-distilled water (Braun, aqua ad iniectabilia) that had been filtered with a 0.22 µm microfilter (Millipore). Water-immersion objectives (Leica HCX APO L U-V-I) built into the wall of the cuvette provided diffraction-limited focusing conditions for wavelengths between 340 nm and 1100 nm. We investigated three *NA*s: $NA = 0.8$ (40x objective), $NA = 0.5$ (20x objective), and $NA = 0.3$ (10x objective). The use of large *NA*s avoided nonlinear beam propagation effects for ultrashort pulse durations [3,117,140] and stimulated Brillouin scattering for ns pulses, respectively [231-233].

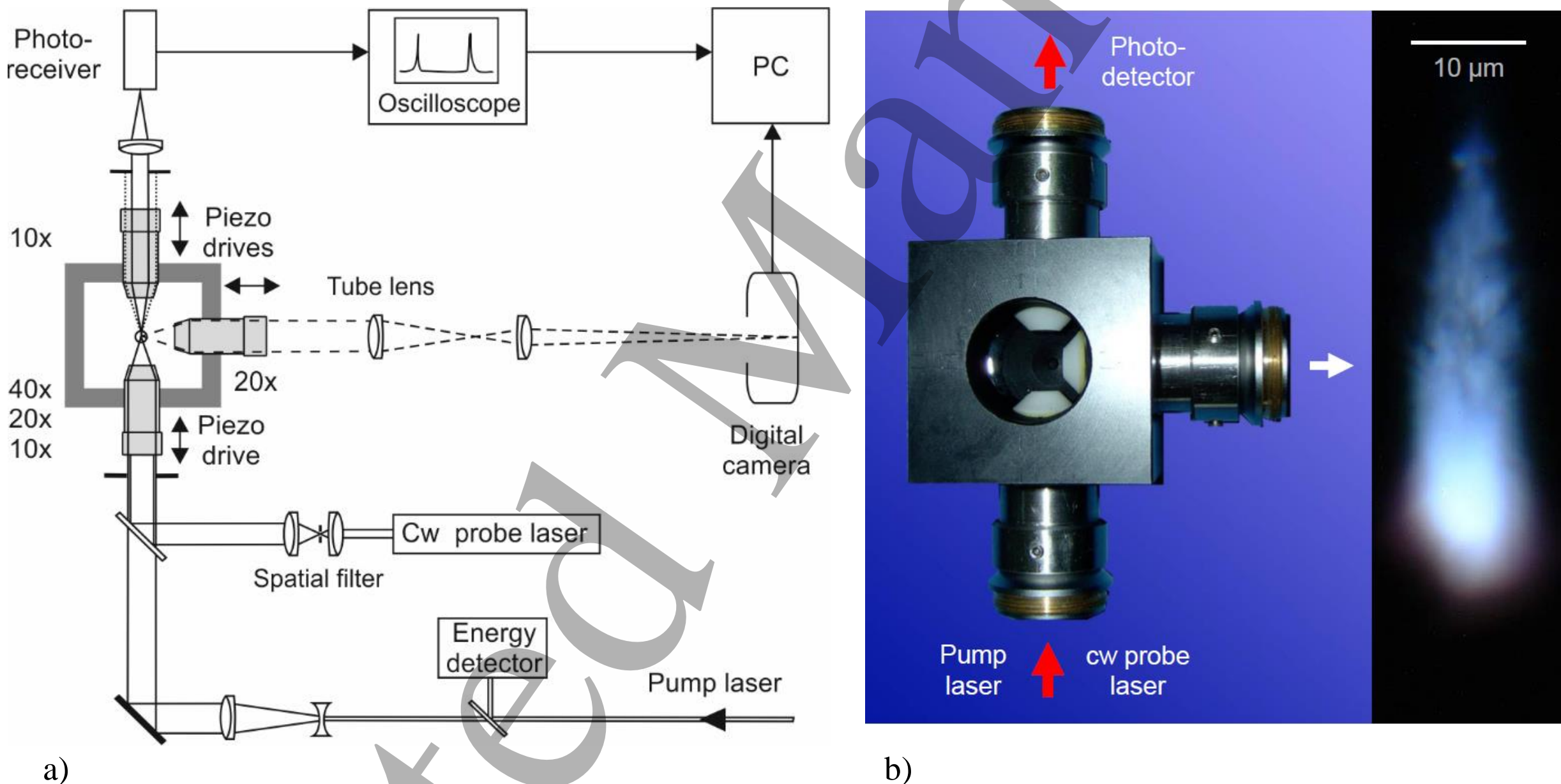

**Figure 2** a) Experimental setup for investigating the energy dependence of femtosecond to nanosecond laser-induced breakdown in water at various pulse durations, wavelengths, *NA*s, and laser operation modes. b) Photograph of the water cell with inbuilt water immersion objective microscope objectives, and an example of a high-resolution plasma photograph taken through the 20x, $NA = 0.5$.

Orthogonal confocal adjustment of the objectives with long working distance (3.3 mm, 3.5 mm, and 3.6 mm, respectively, for the 40x, 20x, and 10x objectives) enabled high-resolution plasma photography in side view. To achieve confocality, the cylindrical ceramic cladding behind the conical front part was removed and replaced by a thinner stainless steel cover. Openings in the water cell made of polyoxymethylene (POM) were exactly fitted to the diameter of the objectives to be watertight without further sealing. The *z*- position of the objectives could be controlled by piezo-driven devices.

The pump laser beam producing the breakdown events was expanded to overfill the rear entrance pupil of the focusing objective. This created an approximately uniform irradiance distribution corresponding to an Airy pattern in the focal plane. Single pulses were selected from the laser pulse train using a mechanical shutter. The pulse energy was adjusted by a combination of rotatable λ/2 plate and thin film polarizer and measured by diverting part of the incident laser pulse onto a calibrated energy meter (Ophir PD10-pJ, or Ophir PD 10).

The laser systems used in the experiments, together with the respective values of pulse duration, wavelength, and approximate values of the beam quality parameter $M^2$ are listed in Table 1.

| Laser | Pulse duration (FWHM) | Wavelength (nm) | Beam quality parameter $M^2$ |
|---|---|---|---|
| High Q Laser Production Yb:glass IC-1045-30-fs | 280 fs<br>306 fs<br>350 fs | 347<br>520<br>1040 | 1.3 |
| TEEM Photonics PNV-001525-140 | 0.56 ns | 355 | 1.2 |
| CryLas FTSS 355-50 with IR option, and FTSS 355/532-50 | 0.93 ns<br>0.95 ns<br>1.01 ns | 355<br>532<br>1064 | 1.4 |
| Spectra Physics Nd:YAG Quanta Ray (instable resonator) | 6.8 ns<br>8.8 ns<br>11.2 ns | 355<br>532<br>1064 | 1.6 |

**Table 1** Features of laser systems used in the experiments.

The Quanta Ray nanosecond Nd:YAG laser could either be run in multi-longitudinal-mode (mlm) operation or produce seeded, single-longitudinal-mode (slm) nanosecond pulses. The Teem Photonics and CryLas lasers emit amplified microchip ns laser pulses that operate intrinsically in single longitudinal mode due to the short resonator length. Occasional mode hops due to thermal drift of resonator length slightly shift the laser frequency but do not affect the single-mode emission itself. The High-Q Yb:glass fs laser system consists of a mode-locked (ml) oscillator and a 1-kHz regenerative amplifier. Mode-locked and slm emission go along with smooth temporal pulse shapes, while mlm emission exhibits stochastic longitudinal mode beating, which results in a spiky temporal intensity profile.

Nanosecond and picosecond pulse shapes were detected by a fast photodiode (ANTEL AR-S1) with a rise time < 100 ps and recorded by means of a 6 GHz oscilloscope (Tektronix DPO 70604). A Gaussian function was fitted to the experimental data to obtain the FWHM time. Femtosecond pulse durations were obtained from a $sech^2$-fit to an autocorrelation trace (APE pulse check).

### 3.2. Imaging of plasma luminescence and scattering

Open-shutter photography in side view was used to record time-integrated images of the plasma region. The Leica HCX APO L U-V-I 20x, $NA = 0.5$ microscope objective was used for imaging, when the pump pulses were focused through a 40x or 10x objective, and the 40x, $NA = 0.8$ objective was employed, when the pump pulses were focused through the 20x objective. The image produced by the microscope objective and its tube lens was further magnified using an enlarger lens (Nikon, Nikkor 63mm/1:2.8) corrected for 8x magnification. The total magnification was 150 times with 20 x objective and 312 times with 40x objective. Photographs were recorded by a digital camera (Canon EOS 5D, 4368 × 2912 pixels) connected to a PC. In most cases, time-integrated images of single breakdown events were taken but for imaging weak plasma luminescence exposures were integrated over 70-100 laser pulses at ISO 3200. The spectral sensitivity of the chip reaches from 410 nm to 690 nm. All images were recorded using the Auto White Balance function of the camera. Because of the use of water immersion objectives and the large image magnification, the spatial resolution is diffraction limited (0.6 µm for the 20x, $NA = 0.5$ objective and 500 nm wavelength).

On photographs of optical breakdown events produced by 532-nm laser pulses, scattered laser light is visible besides the plasma luminescence. It demarcates the region of primary energy deposition, whereas the plasma luminescence portrays the energy distribution and plasma volume at the end of the laser pulse. A comparison of the extent of both regions provides information on possible energy transport processes during plasma formation. To reveal size and structure of the light scattering region, 532 nm laser radiation was partially blocked by a long-pass colour glass filter (Schott OG 550 with 0.28% transmission at 532 nm) or a dielectric laser mirror acting as a notch filter (Laser Optics 45°, HR at 532 nm, HT around 400 nm and 700 nm, with 0.03% transmission at 532 nm). Scattered laser light could not be detected when UV or IR wavelengths were used, because they are outside the sensitivity range of the Canon EOS 5D camera chip.

### 3.3. Determination of breakdown threshold values and threshold sharpness

The onset of bubble formation serves as threshold criterion for optical breakdown. It is detected by a sensitive probe beam scattering technique that also enables to determine the maximum bubble size [117]. For this purpose, a spatially filtered single-frequency cw probe laser beam (CrystaLaser, 660 nm, 40 mW) was adjusted collinear and confocal with the pulsed laser beam. Transmitted probe light was collected by a 10x water immersion objective and imaged onto an AC-coupled amplified photoreceiver (FEMTO, 25 kHz –200 MHz bandwidth), which was protected from the respective pump laser irradiation by appropriate blocking filters.

Breakdown energy thresholds ($E_{th}$) were determined by counting how frequently bubble formation occurred as the energy was increased from sub-threshold to super-threshold values. Data were binned into small energy intervals (n ≥ 15) with > 20 events per interval, and fitted using the Gaussian error function. $E_{th}$ corresponds to 50 % breakdown probability, and the sharpness $S$ of the breakdown threshold is defined as $S = E_{th}/\Delta E_L$, where $\Delta E_L$ is the energy difference between 10 % and 90 % breakdown probability.

The threshold irradiance $I_{th}$ was calculated using the equation

$$I_{th} = \frac{E_{th}}{\tau_L \, \pi \, (M^2 d / 2)^2} \times 3.73. \tag{5}$$

Here $\tau_L$ denotes the laser pulse duration, $M^2$ is the beam quality parameter as listed in Table 1, and $d$ is the diffraction-limited diameter of the Airy pattern arising from focusing a beam with top-hat profile of wavelength $\lambda_L$, which is given by $d = 1.22\ \lambda_L/NA$. The factor 3.73 relates the average irradiance values within the pulse duration and focal spot diameter to the respective peak values, which determine the onset of optical breakdown phenomena.

### 3.4. Determination of bubble size and energy

For large bubbles, the maximum bubble radius, $R_{max}$, is related to the period of the first bubble oscillation, $T_{osc}$, by the Rayleigh equation [234]

$$R_{max} = \frac{T_{osc}}{1.83} \sqrt{\frac{p_{stat} - p_v}{\rho_{liq}}} \,, \tag{6}$$

where $p_{stat} = 100$ kPa is the hydrostatic pressure, $\rho_{liq} = 998$ kg/m$^3$ the liquid density, and $p_v$ the equilibrium vapor pressure inside the bubble (2330 Pa at 20° C). However, for bubble sizes below a few micrometers Eq. (6) must be corrected for the influence of surface tension and viscosity. Surface tension produces a pressure scaling inversely proportional to the bubble radius that adds to the hydrostatic pressure, and viscosity also becomes ever more important with decreasing bubble size [63]. Therefore, we determined $R_{max}$ using the Gilmore model of cavitation bubble dynamics, which considers both surface tension and viscosity [42,63] and compared the outcome with the result of Eq. (6). The $R_{max}$ value obtained with the Gilmore model is larger than the value predicted using Eq. (6) by a factor $f = \left[ R_{\mathrm{max,Gilmore}}(T_{\mathrm{osc}}) / R_{\mathrm{max,Rayleigh}}(T_{\mathrm{osc}}) \right]$. For convenience, we provide a fitted function for the dependence of this ratio on $T_{osc}$ that facilitates the $R_{max}$ determination from measured oscillation times:

$$f(T_{osc}) = 0.2697 \times e^{\frac{-T_{osc}}{658.31\mu s}} + 0.5163 \times e^{\frac{-T_{osc}}{140.98\mu s}} + 0.6651 \times e^{\frac{-T_{osc}}{35.071\mu s}} + 0.0709 \times e^{\frac{-T_{osc}}{5410.1\mu s}} + 1. \quad (7)$$

When the influence of surface tension and viscosity on $R_{max}(T_{osc})$ is considered, the light scattering technique provides reliable results even for bubbles with less than 10 ns oscillation time and 100 nm radius. $T_{osc}$ could be measured with better than ± 1 ns accuracy, corresponding to ± 10 nm for $R_{max}$.

The bubble energy ($E_B$) is given by

$$E_B = \frac{4}{3}\pi R_{\max}^3 \left( p_{stat} - p_v + \frac{2\sigma}{R_{\max}} \right), \quad (8)$$

where $\sigma = 0.073$ N/m is the surface tension of water at room temperature.

### 3.5. Determination of absorbed energy and its partitioning

The absorbed laser energy was obtained from measurements of the plasma transmittance $T_{tra}$

$$E_{abs} = E_L\,(1 - T_{tra}). \quad (9)$$

This approach works well for dense plasmas, where reflection and backscattering are small [218,223,235]. However, close to the bubble threshold, scattering may be more important than absorption, and Eq. (9) should not be used.

For plasma transmission measurements, a 63x water immersion objective ($NA = 0.9$) was used to collect all transmitted and some forward-scattered laser light onto a calibrated energy meter. Calibration accounted for light losses by reflections at optical surfaces and by absorption in the microscope objective and in water.

For luminescent plasmas, we investigated the partitioning of $E_{abs}$ into fractions going into vaporization of the liquid within the plasma volume, shock wave emission, and bubble formation. The plasma volume $V_{plasma}$ was evaluated from the photographs in side-view, and the vaporization energy calculated as

$$E_V = \rho_{liqu} V_{plasma} [c_p (T_2 - T_1) + \Delta H_{vap}], \quad (10)$$

with $c_p$ = 4.187 kJ K$^{-1}$ kg$^{-1}$, $T_2$ = 100°C, $T_1$ = 20°C, and $\Delta H_{vap}$ = 2256 kJ kg$^{-1}$.

The bubble energy was obtained from Eq. (8), and the shock wave energy estimated as

$$E_{SW} = E_{abs} - E_V - E_B. \quad (11)$$

### 3.6. Determination of plasma energy density, electron density, pressure, and temperature

The *average plasma energy density* is given by the ratio of absorbed laser energy and plasma volume

$$U_{avg} = \frac{E_{abs}}{V_{plasma}}. \quad (12)$$

The $E_{abs}$ and $V_{plasma}$ data had to be determined in separate measurement series because the 63x $NA = 0.9$ objective used for the transmission measurements could not be combined in confocal arrangement with the 20x $NA = 0.5$ objective for plasma photography. To determine the dependence of $U_{avg}$ on pulse energy, functions were fitted through the $V_{plasma}$ ($E_L$) and $E_{abs}$($E_L$) data (Jandel Scientific, TableCurve2D), and $U_{avg}$($E_L$) curves were calculated by dividing the fit functions.

For fs breakdown, the average *free electron density* $\rho_c$ can be deduced from $U_{avg}$ because here only one set of free electrons is produced, as recombination during the pulse plays no significant role. The average energy of CB electrons is given by the sum of effective ionization energy and kinetic energy $\varepsilon_{avg} = \tilde{\Delta} + \varepsilon_{kin,avg}$. It was shown to be constant for irradiance values between optical breakdown threshold and full ionization of the 1b$_1$ level [149] and can be approximated by $\varepsilon_{avg} = (9/4)\tilde{\Delta}$ [22,37]. This yields

$$\rho_c = \frac{U_{avg}}{(9/4)\tilde{\Delta}}. \quad (13)$$

In order to assess laser plasma coupling, it is of interest to compare $\rho_c$ with the critical electron density $\rho_{crit}$ at which the plasma frequency $\omega_p = \sqrt{\rho_c e^2 / m_e \varepsilon_0}$ equals the laser frequency $\omega_L$ [6,115,151]. Here $e$ denotes the electron charge, $m_e$ its mass, and $\varepsilon_0$ is the vacuum dielectric permittivity. Above $\rho_{crit}$, the electron movement can follow the electric field oscillations, and both reflectivity and absorptivity increase [151,236]. The *critical free-electron density* is given by

$$\rho_{crit} = \omega_L^2 \frac{m_e \varepsilon_0}{e^2}. \quad (14)$$

The energy density data are also employed for calculating *plasma pressure and temperature* from EOS data, assuming isochoric energy deposition. The absorbing volume remains constant during energy deposition if neither thermal nor hydrodynamics expansion occur. Since thermal expansion proceeds with sound velocity, thermoelastic stress cannot relax, and isochoric energy deposition is stress-confined. The criterion for stress confinement is that an acoustic transient cannot completely traverse the absorbing volume during the laser pulse duration [22,188]. For 10-ns laser pulses and a sound speed of 1500 m/s, this the case for plasma diameters ≥ 15 μm. Since this criterion holds for all luminescent ns plasmas observed in this paper, we can use EOS data to estimate pressure and temperature of luminescent plasmas for the entire range of investigated pulse durations from femto- to nanoseconds.

Figure 3 illustrates the dependence of pressure and temperature on internal energy given by three different EOS. The IAPWS-95 formulation is a generally accepted reference for temperatures up to 1273 K and pressures up to 1000 MPa (10 kbar) [230]. It can be extrapolated also to higher internal energies but this extrapolation does not explicitly consider the dissociation of water molecules occurring at high temperatures. Dissociation is included in the SESAME table 7150 [237] and in the quantum molecular dynamics (QMD) calculations for high-density water by Mattsson and Desjarlais [172,173].

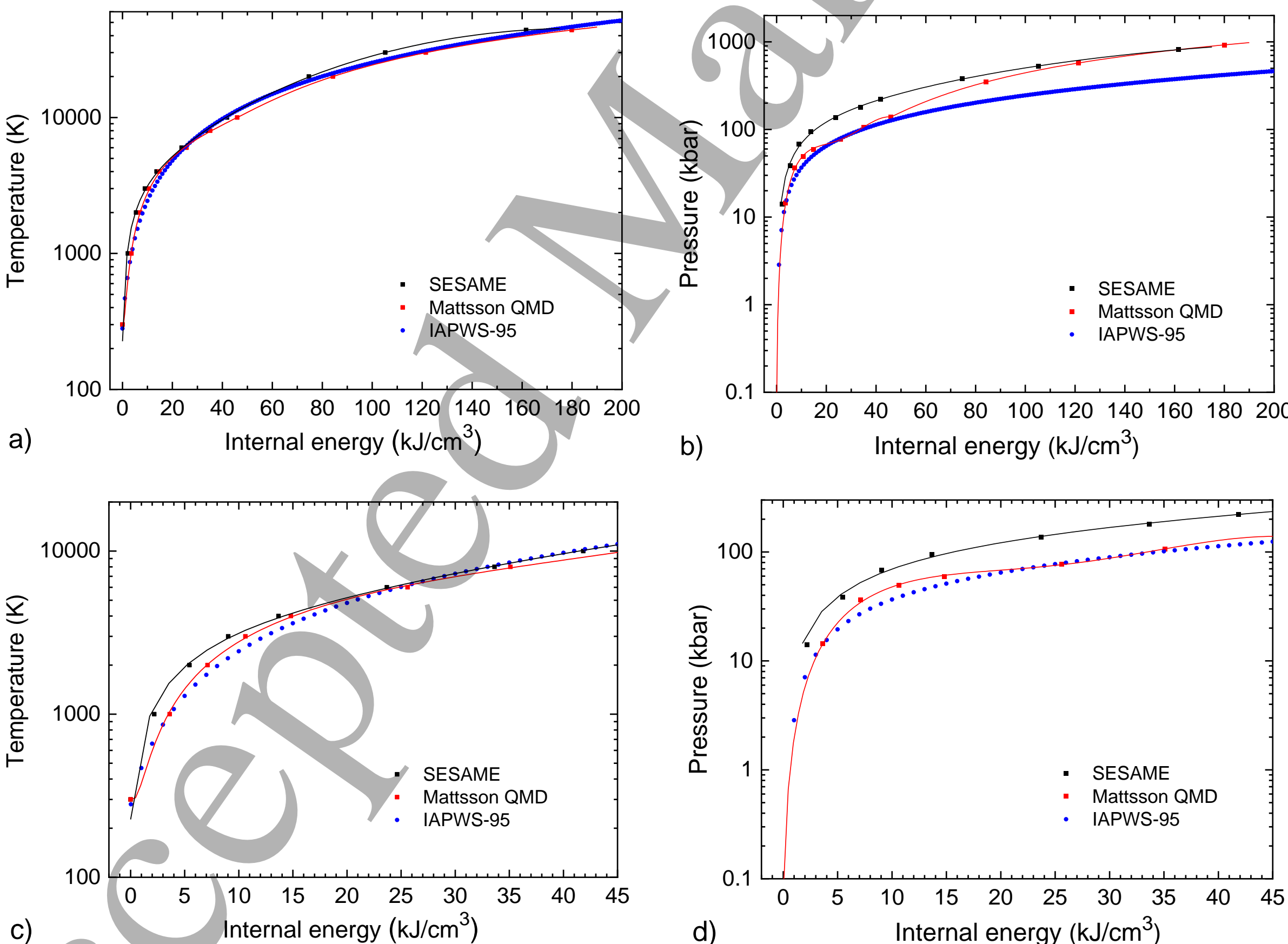

**Figure 3** Water temperature and pressure as a function of internal energy $U$ after isochoric energy deposition into liquid water ($\rho_{water}$ = 998 kg/m$^3$) at room temperature. $T(U)$ and $p(U)$ plots are shown for the full available data range in (a) and (b), and as zoom for internal energies up to 45 kJ/cm$^3$ in (c) and (d), respectively. Each graph contains values based on the IAPWS-95 formulation of the EOS of water [230], data from SESAME table 7150

[237], and QMD calculations from Refs. [172,173]. Differences between the EOS data are mainly related to the way in which water dissociation is considered (see text).

For internal energies around 40 $kJcm^{-3}$, the SESAME pressure is significantly higher than the QMD predictions. According to Ref. [173], this is due to an overestimation of the degree of dissociation in the SESAME table. For higher energy densities, where water is largely dissociated, the agreement with the molecular dynamics calculations is again very good. The extrapolated IAPWS-95 formulation agrees well with the QMD simulations for $U \leq 50$ $kJcm^{-3}$, but the predicted pressure values are too low for $U > 50$ $kJcm^{-3}$ when water is largely dissociated, which is not considered in IAPWS-95. Since experimental $U_{avg}$ data obtained in the present paper stay below 50 $kJcm^{-3}$ and because relatively few data points are available from the QMD simulations, we use the IAPWS-95 formulation for the interpretation of these data.

### 3.7. Demonstration of micro-material-processing capability of slm UV nanosecond pulses

Scenario 3 of section II.B suggests that the energy deposition by UV ns pulses with smooth temporal shape features a tunable nanoregime and an also tunable but more disruptive micro-regime, with a distinct threshold between them. The existence of both regimes and their potential for material processing in transparent dielectrics is demonstrated by producing dissections in cornea such as needed for flap creation in LASIK refractive surgery [13,17,24,105], and by creating micro-patterns and refractive index changes in borosilicate crown glass (BK7). For this purpose, single-longitudinal mode (slm) laser pulses with 355 nm wavelength and 0.56 ns duration were focused into the samples through a Zeiss LD Plan-Neofluar 63x/0.75corr objective, with the correction ring adjusted to the desired focusing depth. Samples were moved laterally using a computer-controlled translation stage programmed in LabView, and the focal depth was adjusted using a piezoelectric objective positioner (PI P-725.4CD PIFOC).

For the experiments on cornea, porcine eyes were obtained from a slaughterhouse and stored in physiological saline at 4°C until use, no longer than 4h after enucleation. Specimens with 9 mm diameter were extracted using a trephine and placed in a sample holder that was capped with a microscope cover glass. The specimens were immersed in physiological saline for index matching purposes. To mimic conditions in refractive LASIK surgery, the numerical aperture was reduced to 0.38 by inserting a mask into the objective's rear aperture.

# 4. Results

## 4.1. Breakdown thresholds and threshold behavior

Table 2 summarizes the experimental results on thresholds for bubble formation and bright plasma luminescence and on the threshold sharpness for different pulse durations, wavelengths, and modes of laser operation. Figure 4(a) displays the pulse duration and wavelength dependence of bubble thresholds, and Fig. 4(b) illustrates the threshold behavior for slm and mlm laser operation.

| λ (nm) | $\tau_L$ (FWHM) | Laser mode | *NA* | *d* (μm) | Breakdown criterion | $E_{th}$ (nJ) | $I_{th}$ ($10^{11}$ W/cm$^2$) | *S* |
|---|---|---|---|---|---|---|---|---|
| 1040 | 350 fs | ml | 0.8 | 1.59 | Bubble | 25.0 | 80.0 | 30.5 |
| 520 | 306 fs | ml | 0.8 | 0.79 | Bubble | 4.8 | 70.0 | 34.3 |
| 347 | 280 fs | ml | 0.8 | 0.53 | Bubble | 3.6 | 131.5 | 66.4 |
| 355 | 0.56 ns | slm | 0.3 | 1.44 | Bubble | 830 | 2.40 | 47.9 |
| 355 | 0.56 ns | slm | 0.3 | 1.44 | BPL | 2,950 | | 43.2 |
| 355 | 0.56 ns | slm | 0.5 | 0.87 | Bubble | 330 | 2.62 | 47.3 |
| 355 | 0.56 ns | slm | 0.5 | 0.87 | BPL | 1,050 | | 133.4 |
| 355 | 0.56 ns | slm | 0.8 | 0.54 | Bubble | 140 | 2.83 | 55.8 |
| 355 | 0.56 ns | slm | 0.8 | 0.54 | BPL | 380 | | 96.8 |
| 1064 | 1.02 ns | slm | 0.8 | 1.62 | Bubble = BPL | 4,210 | 3.83 | 23.2 |
| 532 | 0.95 ns | slm | 0.8 | 0.81 | Bubble | 840 | 3.30 | 49.5 |
| 532 | 0.95 ns | slm | 0.8 | 0.81 | BPL | 1,200 | | 48.5 |
| 355 | 0.93 ns | slm | 0.8 | 0.54 | Bubble | 270 | 2.45 | 63.1 |
| 355 | 0.93 ns | slm | 0.8 | 0.54 | BPL | 770 | | 66.5 |
| 1064 | 11.2 ns | mlm | 0.8 | 1.62 | Bubble = BPL | 25,000 | 1.59 | 2.7 |
| 1064 | 11.2 ns | slm | 0.8 | 1.62 | Bubble = BPL | 80,800 | 5.15 | 24.9 |
| 532 | 8.8 ns | mlm | 0.8 | 0.81 | Bubble | 2,000 | 0.65 | 1.1 |
| 532 | 8.8 ns | slm | 0.8 | 0.81 | Bubble | 7,500 | 2.43 | 4.8 |
| 532 | 8.8 ns | slm | 0.8 | 0.81 | BPL | 84,400 | | 7.4 |
| 355 | 6.8 ns | mlm | 0.8 | 0.54 | Bubble | 600 | 0.56 | 0.7 |
| 355 | 6.8 ns | slm | 0.8 | 0.54 | Bubble | 1,400 | 1.31 | 2.1 |
| 355 | 6.8 ns | slm | 0.8 | 0.54 | BPL | 42,400 | | 9.1 |

**Table 2** Thresholds for bubble formation and bright plasma luminescence (BPL), and threshold sharpness *S* for various pulse durations $\tau_L$, wavelengths $\lambda_L$, and modes of laser operation. Modes are: ml = mode-locked (sech$^2$ pulse shape); slm = single-longitudinal-mode (Gaussian pulse shape), mlm = multi-longitudinal-mode (statistically varying temporal profile as shown in Fig. 5a). Irradiance thresholds $I_{th}$ were calculated from the diffraction-limited diameter of the focal Airy pattern, *d*, and $E_{th}$ using Eq. (5). When the BPL threshold is larger than the bubble threshold, only the energy threshold is given. Here, the plasma is significantly larger than the focal spot and irradiance values referring to the focal spot diameter would be erroneously high.

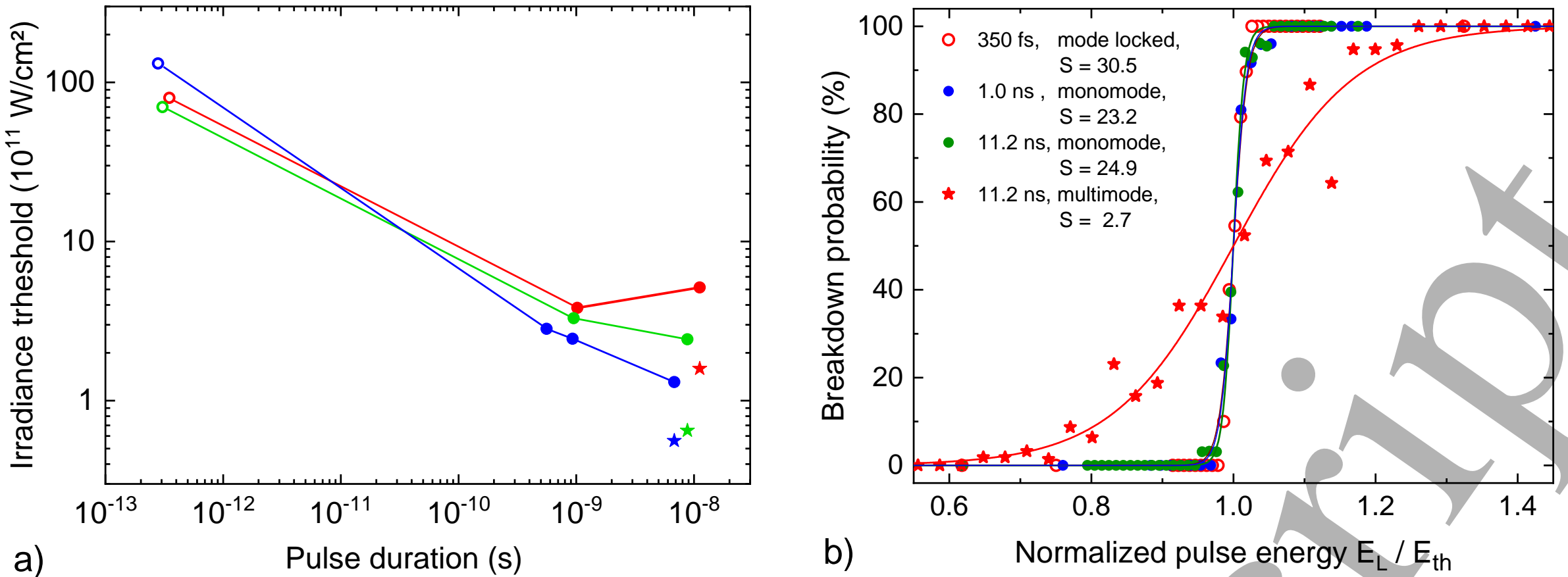

**Figure 4** (a) Graphical illustration of the pulse duration and wavelength dependence of bubble irradiance thresholds based on the data in Table 2. The wavelengths are indicated by the colour of the symbols and the mode of laser operation by their shape: ○ mode-locked, ● single-longitudinal mode (slm), ★ multi-longitudinal mode (mlm). To guide the eye, straight lines connect data points for ml and slm operation. (b) Dependence of threshold behavior on the mode of laser operation. Probability curves for bubble formation in water by IR laser pulses focused at $NA = 0.8$ are shown for mode-locked fs pulses (350 fs, 1040 nm), slm ns pulses (1.0 ns and 11.2 ns, 1064 nm) and mlm ns pulses (11.2 ns, 1064 nm). Threshold sharpness $S$ is high for the fs and slm ns laser pulses with smooth temporal profile but low for mlm ns pulses exhibiting irregular intensity spikes (Fig. 5a).

In Fig. 4(a), we see a pronounced decrease of the threshold irradiance for bubble formation with increasing pulse duration and a weak dependence on wavelength, both in agreement with previous work [36,37,108,114,115,125]. For mlm ns pulses exhibiting intensity spikes (see Fig. 5a), the threshold irradiance averaged over the spikes is lower than for temporally smooth slm pulses. For visible and IR wavelengths, slm irradiance threshold values change little between 1 ns and ≈10 ns, as predicted in [22].

Figure 4(b) shows that the threshold sharpness depends strongly on the mode of laser operation, which influences the reproducibility of the temporal pulse shape. Bubble thresholds are equally sharp ($S \approx 25$) for slm ns pulses with smooth temporal shape as for fs laser pulses. By contrast, for ns multimode pulses exhibiting statistically fluctuating intensity spikes, the threshold sharpness is reduced to $S = 2.7$. Generally, the threshold sharpness is 3-9 times higher for ns breakdown by slm pulses than for the corresponding mlm case – both for the bubble and the BPL thresholds. The threshold variations observed with mode-locked and slm laser pulses resemble the pulse-to-pulse energy fluctuations of the respective laser systems. Thus, with appropriate mode control, ns and fs optical breakdown thresholds are equally "deterministic".

### 4.2. Energy dependence of cavitation bubble size

Figure 5 shows the dependence of maximum cavitation bubble radius on laser pulse energy for different pulse durations and wavelengths. Again, ns breakdown is highly irregular for multimode pulses [Figs. 5(a) and (b)] but well reproducible for slm pulses [Fig 5(c)].

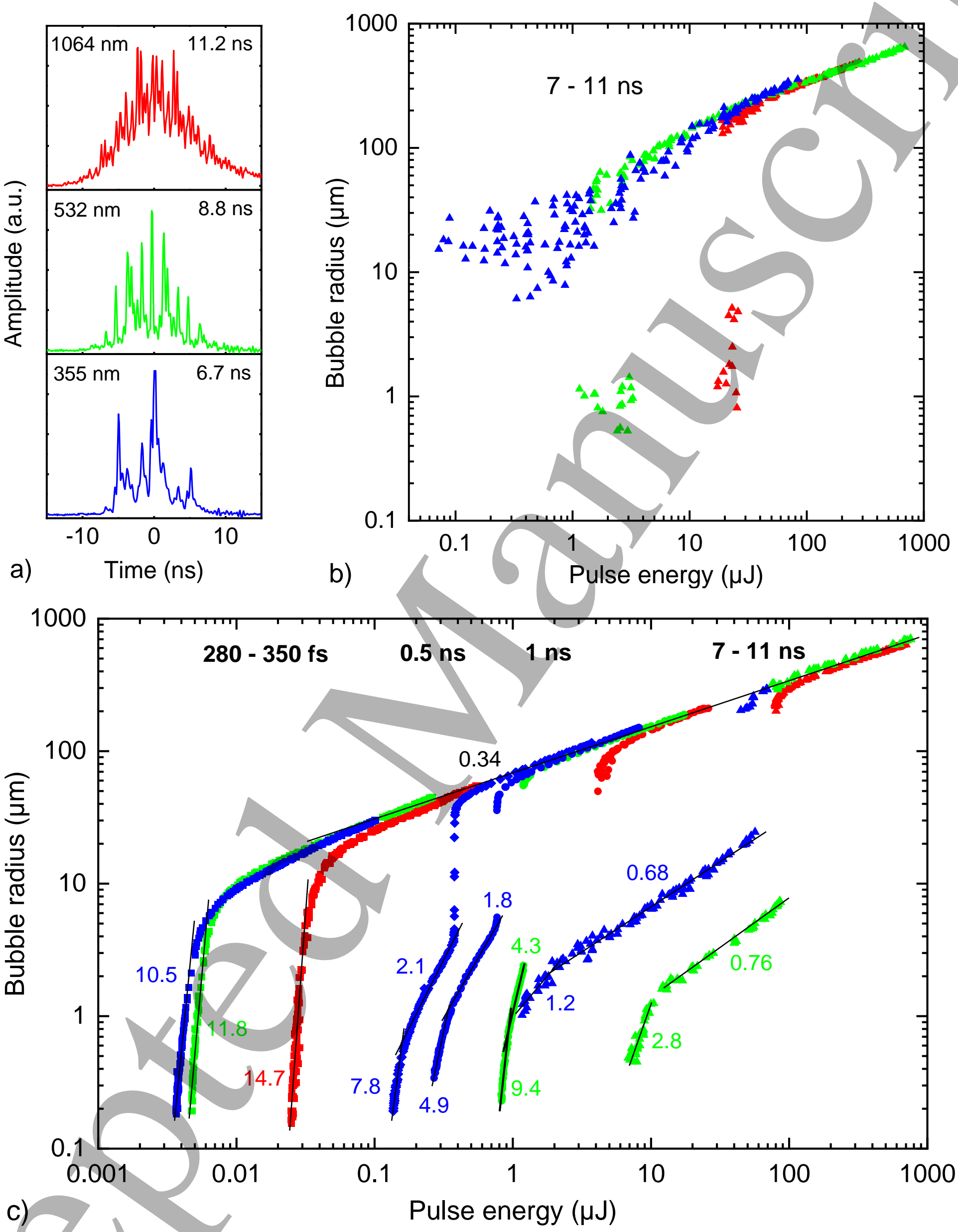

**Figure 5** Maximum cavitation bubble radius as a function of laser pulse energy for different pulse durations, wavelengths, and laser modes; *NA* = 0.8. The colours denote the wavelengths: blue = 347 nm for fs pulses, and 355 nm for ns pulses; green = 520 nm and 532 nm, red = 1040 nm and 1064 nm. Examples for pulse shapes of longitudinally multimode laser pulses of 7-11 ns duration are given in (a), and the $R_{max}(E_L)$ data for these irradiation conditions are presented in (b). All ns data in (c) were obtained using longitudinally single mode (Gaussian) laser pulses. The fs data refer to mode-locked pulses of $sech^2$ pulse shape. In (c), the $R_{max}(E_L)$ curves are piecewise approximated by straight lines, with slopes indicated. The $R_{max}(E_L)$ data refer to breakdown scenario 1 for fs pulses, to scenario 2 for IR ns pulses, and to scenario 3 for UV and VIS pulses with duration ≥ 0.5 ns.

The $R_{max}(E_L)$ dependencies in Fig. 5(c) reflect the three scenarios described in section 2.2. In fs breakdown, $R_{max}$ increases continuously with $E_L$ from 200 nm to large bubble sizes, initially with steep and later with smaller slope (scenario 1). By contrast, IR ns pulses generate luminescent plasmas and large bubbles with 50 to 200 µm radius already at threshold (scenario 2). In the third scenario, applicable to UV and VIS ns pulses with smooth temporal shape, the $R_{max}(E_L)$ curves feature an abrupt increase in bubble size that correlates with the onset of bright plasma luminescence. We use the terms "small-bubble regime" and "BPL regime" to distinguish the regions before and after the step. At the bubble threshold, nano-bubbles with a radius as small as 200 nm can be produced, just like in fs breakdown. In the small-bubble regime, the $R_{max}(E_L)$ curve increases more slowly than for fs breakdown close to threshold but in the BPL regime it exhibits the same trend as in the other scenarios. Figure 6 shows the same pattern of stepwise tunability in UV ns breakdown for a range of numerical apertures $0.3 < NA < 0.8$. The sharpness of the BPL threshold decreases slightly with decreasing $NA$ from $S = 96.8$ at $NA = 0.8$ to $S = 43.2$ at $NA = 0.3$. At the same time, the relative jump in bubble radius becomes smaller; the ratio $R_{max>BPL}/R_{max<BPL}$ drops from a value of 10 at NA = 0.8 to a value of 4 at $NA = 0.3$. Here $R_{max<BPL}$ and $R_{max>BPL}$ denote the radii of the expanded bubble just below and above the BPL threshold.

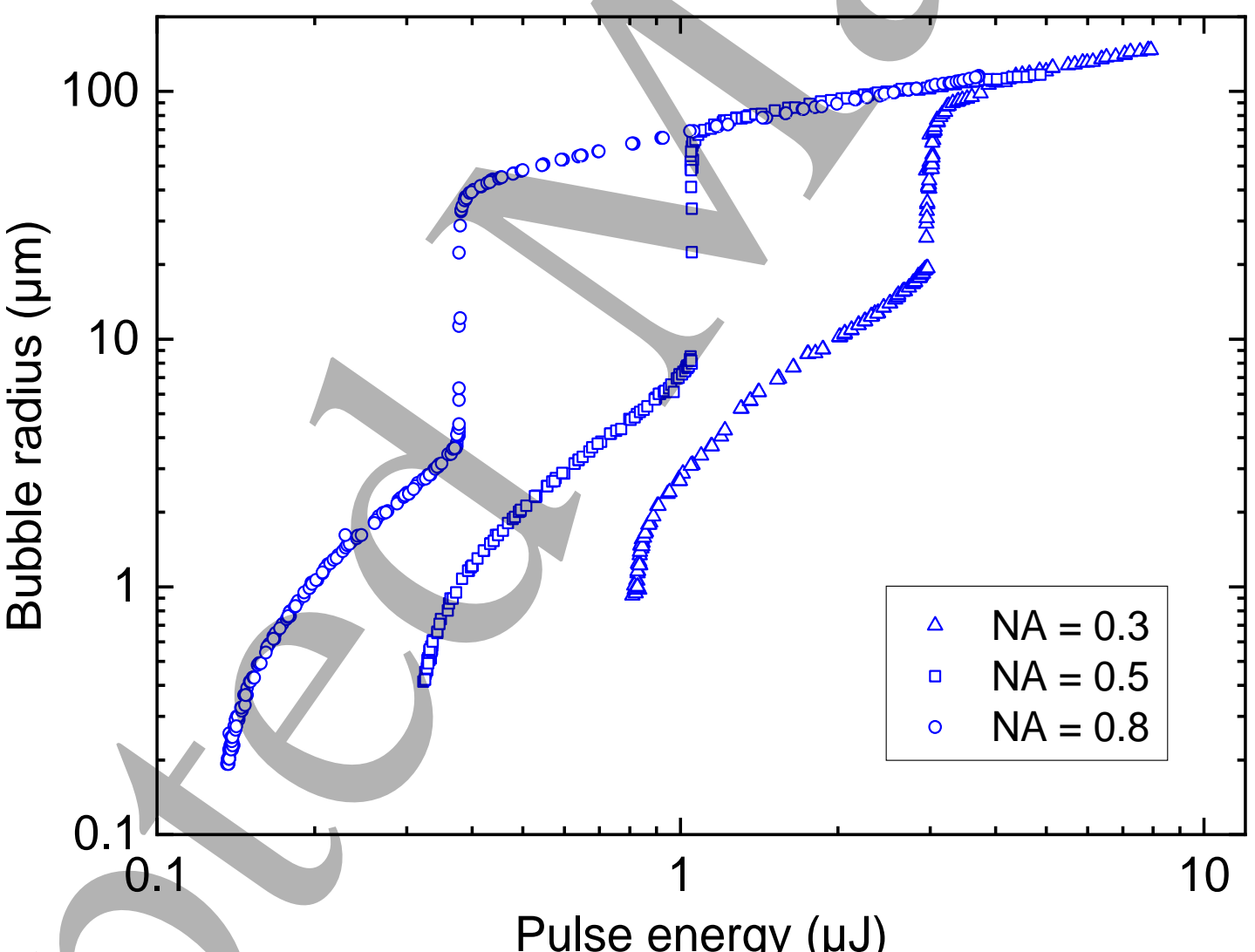

**Figure 6** $R_{max}(E_L)$ curves for slm UV laser pulses (355 nm) of 0.56 ns duration at different *NA*s.

The width of the small-bubble regime for UV and VIS ns pulses increases with laser pulse duration [Fig. 5(c)]. For 560-ps and 1-ns pulses, it comprises an energy range of $\approx 3\times E_{thB}$, while for 8-ns pulses the range is $\approx 20\times E_{thB}$. The nanoregime is broader for ns pulses than for fs breakdown. We use the normalized energy range with bubble sizes between 200 nm and 3 µm, $(E_{L3\mu m}-E_{L200nm})/E_{L200nm}$, as a measure for the width of the nanoregime. For UV wavelengths, this ratio has a value of 0.37 for 280-fs pulses and increases by a factor of four to 1.46 for 560-ps pulses. For 6.8-ns pulses at 355 nm and 8.8-ns pulses at 532 nm, the small-bubble regime is even larger but the minimum bubble size is not quite as small as with fs and sub-ns pulses.

Although the scaling laws for plasma-mediated energy deposition represented by the $R_{max}(E_L)$ curves are complex in the vicinity of the bubble threshold, a similar energy scaling is observed for all pulse durations and wavelengths at pulse energies well above threshold. The straight line with slope 1/3 in the log-log plot represents a dependence $R_{max} \propto (E_L)^{1/3}$, which according to Eq. (8) corresponds to $E_B \propto E_L$.

### 4.3. Plasma luminescence and scattering

Figure 7 presents time-integrated photographs of the shape and inner structure of plasma luminescence with 0.6 µm spatial resolution. Luminescence of fs laser induced plasmas could not be seen with the naked eye or after single photographic exposures but only after integration over 70 breakdown events. Weak luminescence appeared at about $3\times E_{thB}$ and gradually increased with $E_L$ (Fig. 7a). Luminescence from fs plasmas stayed confined to the laser cone angle, whereas the luminescence of BPL ns plasmas extended for all wavelengths up to 20 µm beyond the laser focus and outside the cone angle of the laser beam (Figs. 7 (b) – (e)).

Luminescence from UV ns breakdown is shown in Fig 7b. In the small-bubble regime [left column of (b)], no luminescence is observed and the plasma becomes visible only through scattering of the incident laser light. It appears green because the camera chip is not sensitive at UV wavelengths but can detect 2$^{nd}$ harmonic radiation that leaked through the wavelength separator after third-harmonic generation. The abrupt onset of bright plasma luminescence at $E_L$ = 45 µJ (right column of Fig 7b) coincides with the sharp increase of the bubble radius seen in Figs. 5 and 6. At the BPL threshold, thermal ionization becomes sufficiently strong to overcome the inhibiting action of recombination, as described in section II.B. The sharpness of this threshold is similar to that for bubble formation (Table 2).

The photographs of plasmas produced by 532-nm ns pulses show scattered pump laser light emanating from a region within the cone angle of the laser beam, which is surrounded by a luminescent halo [Figs. 7 (c), (e), and (f)]. The scattering region defines the region of primary energy deposition by the incident laser light. It consists of branched strings with ≈ 1 µm diameter that become brighter in upstream direction. Oversaturation from scattered pump laser light is avoided by blocking filters. The dielectric filter used in Fig. 7(e) and (f) had 0.03 % transmission at 532 nm, about 1/10 of the transmission of the colour glass filter used in Fig. 7(c). This revealed the string-like substructure of the primary-energy-deposition region, which cannot be resolved in Fig. 7(c).

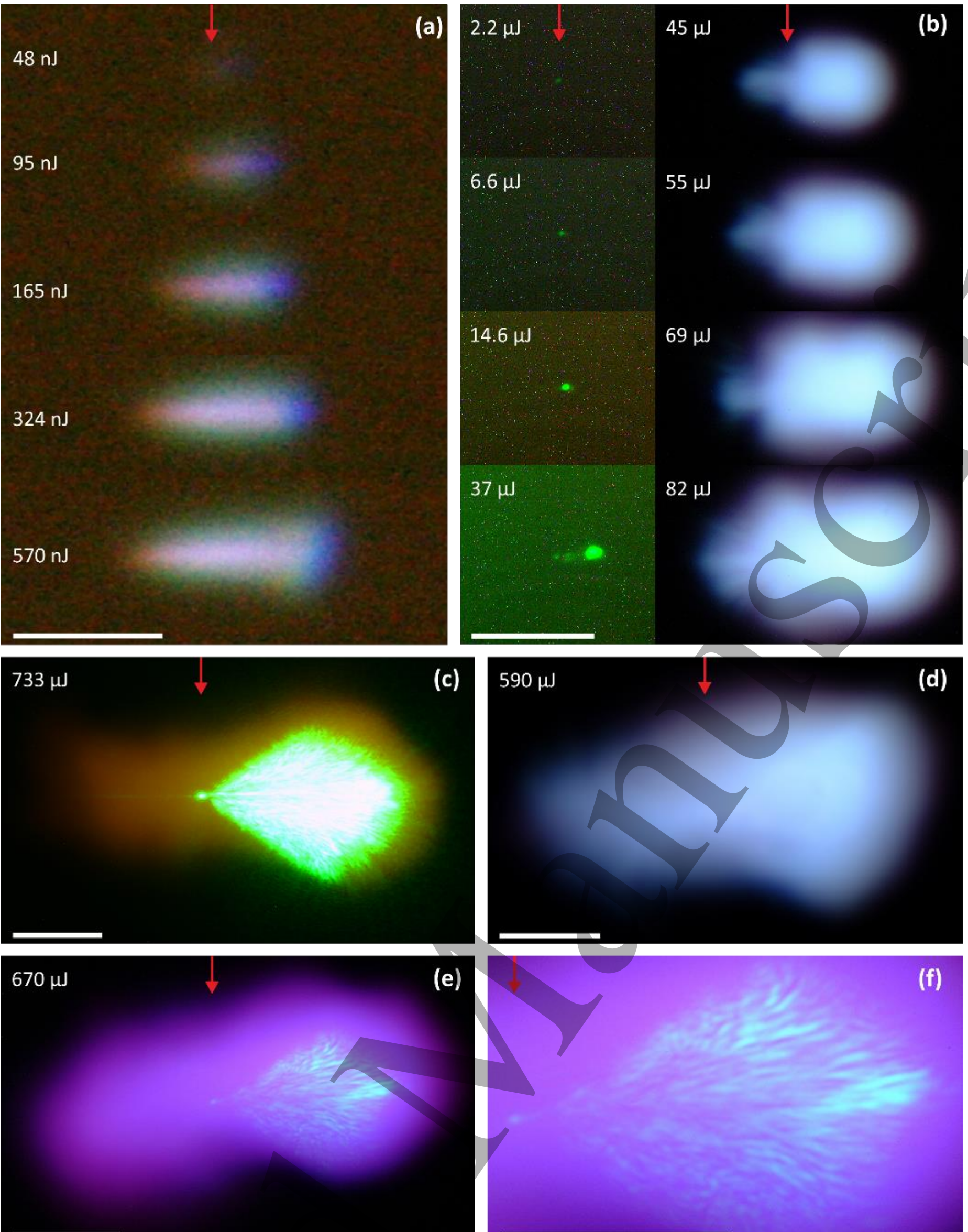

**Figure 7** Side-view photographs of plasmas in water produced at *NA* = 0.8 by temporally smooth laser pulses of different duration and wavelength: 350-fs, 1040 nm in (a), 6.8-ns 355 nm in (b), 8.8 ns, 532nm in (c), (e) and (f), and 11.2 ns, 1064 nm in (d). The corresponding pulse energies are given in each frame. Laser light is incident from the right, and the position of the beam waist is marked by an arrow. Luminescence is integrated over 70 pulses in (a) and over 100 pulses in the left column of (b), both at ISO 3200. All other photographs are from single laser pulses. In (b), the laser cone angle is depicted in the picture taken at $E_L$ = 82 µJ. In (c), (e), and (f), the camera detected both plasma luminescence and scattered laser light from the 532-nm ns pulse, whereby (f) shows an enlarged view of the central region of (e). Scattered laser radiation was in (c) partially blocked by a long-pass colour glass filter, and in (e), (f) by a dielectric notch filter. The scattered light demarcates the region of primary energy deposition within the laser cone angle. Redistribution of the plasma energy then causes the luminescence emanating from a larger volume that partly lies outside the region reached by the laser beam. This phenomenon is visible also in (b) and (d). In (d) the plasma core can be identified by its stronger luminescence. Length of scale: 5 µm in (a), and 20 µm in (b) - (f). In (e) and (f) one can see bright strings within the plasma core that have diameters in the order of 1 µm. These strings are indicative for strong scattering of the incoming laser light and, thus, demarcate zones of particularly intense primary energy deposition. The strings are only visible at 532 nm pump laser wavelength because the CMOS camera chip does not detect the light at 355 nm and 1064 nm.

A separation between a bright plasma core within the laser cone angle and a surrounding diffusely luminescent halo is vaguely perceptible also for ns breakdown at 1064 nm (Fig. 7 (d) and 355 nm (Fig. 7 (b), right column). However, the core is here not demarcated by scattered laser light because UV and IR wavelengths cannot be detected by the digital camera employed in our experiments. The brighter appearance of the core is rather due to an increase of plasma luminescence caused by a high volumetric energy density.

Generally, the extent of the luminescent halo beyond the plasma core is largest in forward direction. The existence of a halo outside the laser cone angle at large *NA* is indicative for a massive energy transport during breakdown. This transport is most likely mediated by energetic photons produced as bremsstrahlung from hot electrons in the string-like regions of primary energy deposition [238-240] as will be discussed in detail in section V.D.

Plasma features change for ns breakdown at moderate *NA*, as seen in Fig. 8 with photographs of plasmas produced by 6.8-ns, 355-nm pulses focused at *NA* = 0.3. Here the luminescent plasma region extends from the beam waist upstream towards the incoming laser beam, and it is largely confined within the laser cone angle. A bright spot demarcating the region of highest plasma energy density is located in the upstream part of each plasma, while the regions closer to the beam waist luminesce only weakly and exhibit a mottled substructure with fairly sharp borders. By contrast, the region around the hot spots is blurred and surrounded by a halo with colour transition from a whitish appearance in the hot spot to a reddish hue in the outer region. The hot spot and the halo are particularly pronounced for pulse energies between ≈ 250 µJ and ≈ 750 µJ. The reddish outer rim reaches partially into the region outside the laser cone angle and arises from ongoing luminescence during initial plasma expansion, when adiabatic cooling red-shifts the emission spectrum. For estimating the maximum energy density in the hot region within the cone angle, we assume that the volume marked by the red circle in the plasma produced with $E_L$ = 574 µJ contains all energy deposited during the second half of the laser pulse. This way, we get that 287 µJ are deposited into a volume of 1100 µm$^3$, corresponding to an ellipsoid with half axes of 7.5 µm and 6 µm, which leads to a local energy density of 260 kJcm$^{-3}$.

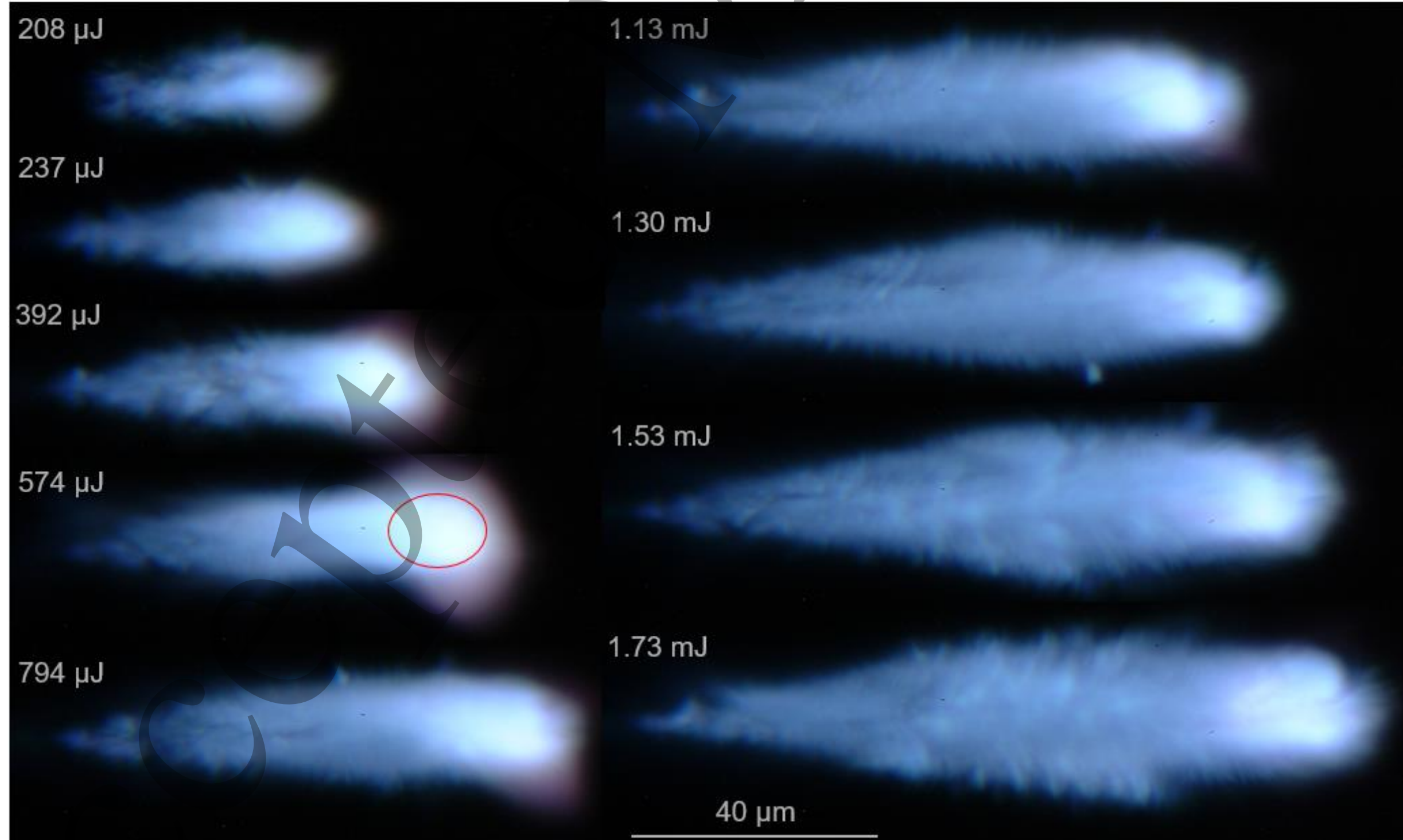

**Figure 8** Side-view photographs of plasmas in water produced at *NA* = 0.3 by temporally smooth laser pulses of 6.8-ns duration, 355 nm wavelength and different pulse energies. The bubble threshold is $E_{th}$ = 44 μJ, and the BPL threshold (50% breakdown probability) is 238 μJ. Laser light is incident from the right.

### 4.4. Energy density, electron density, pressure, and temperature of luminescent plasmas

Figure 9 shows the procedure of determining the average plasma energy density in dependence of laser pulse energy on the examples of IR fs and UV ns pulses focused at $NA$ = 0.8.

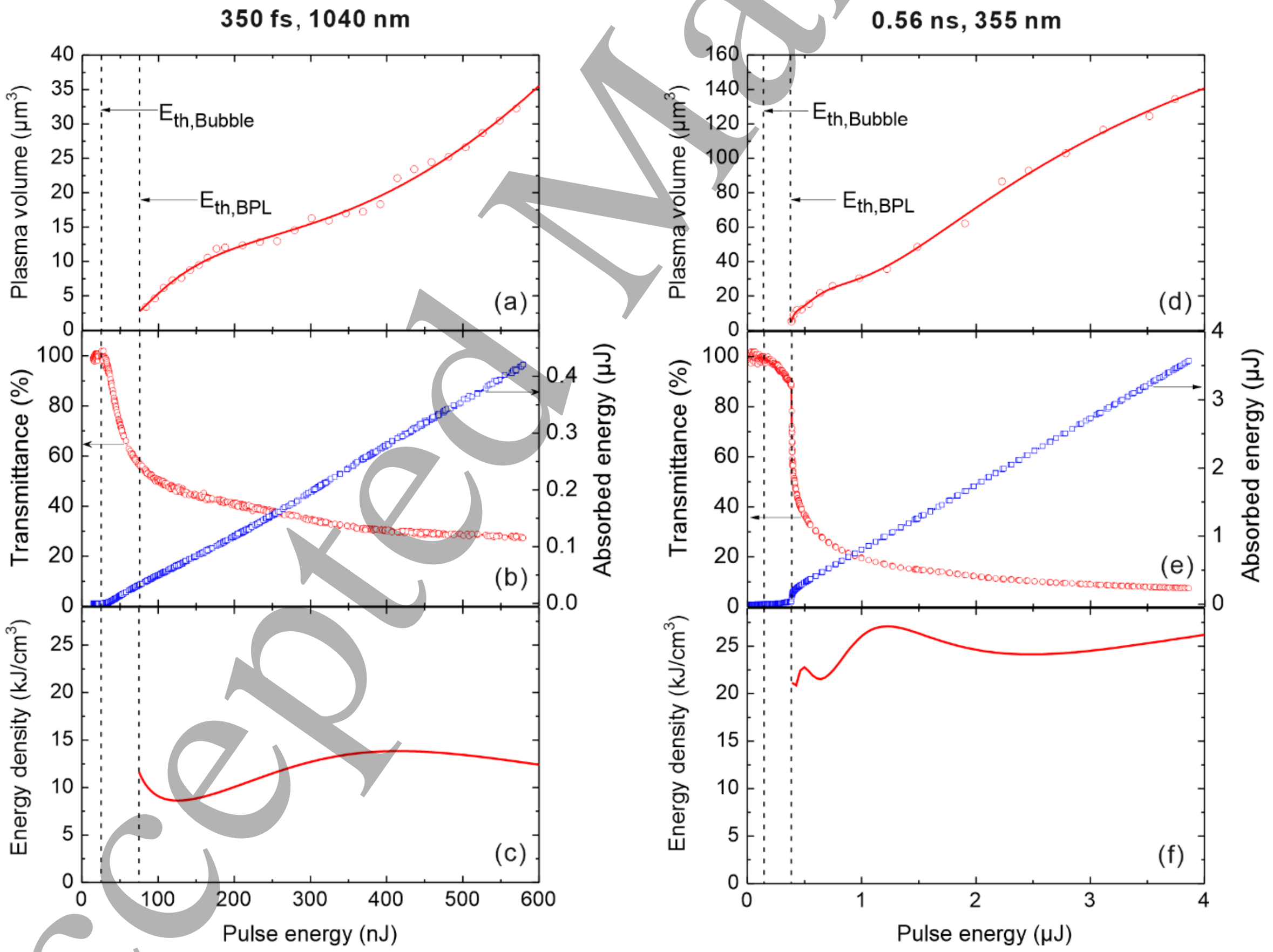

**Figure 9** Determination of average plasma energy density $U_{avg} = E_{abs}/V_{plasma}$ for 1040-nm, 350-fs pulses (a-c), and 355-nm, 0.56-ns pulses (d-f) focused at $NA$ = 0.8. The plasma volume $V_{Plasma}$ ($E_L$) in (a) and (d) was determined from photographs, and the absorbed energy $E_{abs}$($E_L$) in (c) and (e) was obtained from plasma transmittance $T_{tra}$($E_L$)

as $E_{abs} = E_L(1\text{-}T_{tra})$. The $U_{avg}$ ($E_L$) curves in (c) and (f) were then obtained by dividing a fit curve through the $E_{abs}(E_L)$ data by the fit curve through the $V_{Plasma}(E_L)$ data.

The $U_{avg}(E_L)$ curves of Figs. 9 (c) and (f) refer to the high-density plasma regime because only luminescent plasmas could be evaluated. For the examples shown, $U_{avg}$ varies little with pulse energy, and similar observations were also made for other wavelengths and pulse durations. Therefore, we determined mean values, $\bar{U}_{avg}$, to facilitate the comparison of plasma energy densities in different regions of the ($\tau_L$, $\lambda_L$) space. The $E_L$ range used for averaging starts for fs pulses at the onset of plasma luminescence ($3 \times E_{thB}$) and for ns pulses at $1.5 \times E_{thBPL}$. Its upper end is always given by the available laser pulse energy.

The $\bar{U}_{avg}$ values were used to derive plasma pressure and temperature with the help of the $p(U)$ and $T(U)$ diagrams of Fig. 3. This approach requires stress confined energy deposition, which applies for all pulse durations including ns breakdown because luminescent ns plasmas are much larger than the focal volume, as seen in Figs. 7 and 8. The movement of breakdown wave in upstream direction during the laser pulse further enhances stress confinement as it reduces the local energy deposition time below the laser pulse duration.

For fs plasmas, where only one set of free electrons is produced during breakdown, we used $\bar{U}_{avg}$ to determine the average electron density using Eq. (13) and compare it to the critical free-electron density $\rho_{crit}$ at which the plasma frequency reaches the frequency of the laser light [Eq. (14)]. The resulting values for $\bar{U}_{avg}$, $\bar{P}_{avg}$, $\bar{T}_{avg}$, $\bar{\rho}_{c,avg}$ and $\rho_{crit}$ are summarized in **Table 3**.

Plasma energy densities in luminescent plasmas are similar for fs and ns breakdown but $\bar{U}_{avg}$ varies with wavelength between ≈ 10 kJcm$^{-3}$ and ≈ 40 kJcm$^{-3}$ for both pulse durations. These values are 10-100 times larger than the energy density at the bubble threshold, which amounts to ≈ 1.2 kJ cm$^{-3}$ for ns pulses [188] and ≈ 0.6 kJ cm$^{-3}$ for fs breakdown [22]. Using the EOS date in Fig. 3, we see that the average plasma temperatures in the BPL regime range between 2800 and 10,100 K, and the corresponding pressures lie between 42 and 116 kbar. The $\rho_{c,avg}/\rho_{crit}$ ratio increases with wavelength from 0.37 at 347 nm through 2.94 at 520 nm to 4.82 at 1040 nm wavelength. Thus, for visible and IR wavelengths, the average free-electron density in luminescent fs plasmas produced at $NA = 0.8$ is supercritical.

| $\lambda_L$ (nm) | $\tau_L$ (FWHM) | Laser mode | *NA* | Energy range | $\bar{U}_{avg}$ (kJ/cm$^3$) | $\bar{P}_{avg}$ (kbar) | $\bar{T}_{avg}$ (K) | $\bar{\rho}_{c,avg}$ ($10^{21}$ cm$^{-3}$) | $\rho_{crit}$ ($10^{21}$ cm$^{-3}$) |
|---|---|---|---|---|---|---|---|---|---|
| 1040 | 350 fs | ml | 0.8 | (3-20)×$E_{thB}$ | 11.8 | 42 | 2850 | 4.96 | 1.03 |
| 520 | 306 fs | ml | 0.8 | (3-30)×$E_{thB}$ | 41.3 | 116 | 10090 | 12.1 | 4.12 |
| 347 | 280 fs | ml | 0.8 | (3-30)×$E_{thB}$ | 17.0 | 57 | 4090 | 3.45 | 9.28 |
| 355 | 0.56 ns | slm | 0.3 | (1.5-2.5)×$E_{thBPL}$ | 32.5 | 95 | 7880 | | |
| 355 | 0.56 ns | slm | 0.5 | (1.5-5.0)×$E_{thBPL}$ | 39.2 | 111 | 9560 | | |
| 355 | 0.56 ns | slm | 0.8 | (1.5-10)×$E_{thBPL}$ | 24.9 | 77 | 6000 | | |
| 1064 | 1.02 ns | slm | 0.8 | (1.5-7.0)×$E_{thBPL}$ | 35.8 | 103 | 8710 | | |
| 532 | 0.95 ns | slm | 0.8 | (1.5-14)×$E_{thBPL}$ | 38.3 | 109 | 9340 | | |
| 355 | 0.93 ns | slm | 0.8 | (1.5-12)×$E_{thBPL}$ | 36.4 | 105 | 8860 | | |
| 1064 | 11.2 ns | mlm | 0.8 | (1.5-10)×$E_{thBPL}$ | 12.0 | 43 | 2900 | | |
| 1064 | 11.2 ns | slm | 0.8 | (1.5-7.0)×$E_{thBPL}$ | 23.6 | 74 | 5690 | | |
| 532 | 8.8 ns | mlm | 0.8 | (5.0-250)×$E_{thB}$ | 8.5 | 32 | 2090 | | |
| 532 | 8.8 ns | slm | 0.8 | (1.5-7.0)×$E_{thBPL}$ | 12.5 | 44 | 3020 | | |
| 355 | 6.8 ns | mlm | 0.8 | (20-100)×$E_{thB}$ | 12.5 | 44 | 3020 | | |
| 355 | 6.8 ns | slm | 0.8 | (1.5-2.3)×$E_{thBPL}$ | 9.7 | 36 | 2360 | | |

**Table 3** Mean values of the average energy density, $\bar{U}_{avg}$, of luminescent plasmas for different pulse durations, wavelengths, and *NA*s, together with the corresponding values of plasma pressure, $\bar{P}_{avg}$, and temperature, $\bar{T}_{avg}$. For fs breakdown, also the free-electron density $\rho_{c,avg}$ corresponding to $\bar{U}_{avg}$ is listed, together with the critical free-electron density $\rho_{crit}$. The maximum number density of electrons that can be ionized in single ionization of the $1b_1$ valence band level is $6.68 \times 10^{22}$ cm$^{-3}$ [113].

Peak values for plasma energy density, pressure, and temperature are, of course, larger than the average values. The difference can be very pronounced for ns plasmas, which exhibit an inhomogeneous substructure (Figs. 7 and 8). It must be considered when average plasma temperatures derived from $\bar{U}_{avg}$ are compared to spectroscopically determined values that are closer to peak temperatures (see section 5.4).

## 4.5. Energy partitioning

For laser surgery and material processing, it is of interest to know, which fraction of the incident energy is absorbed by the plasma. However, close to the bubble threshold, $E_{abs}$ cannot well be assessed through transmission measurements because scattering is here stronger than absorption (section 3.5).

Alternatively, bubble size and energy can be used to characterize energy deposition. This approach is convenient but provides only a lower estimate for the efficiency of energy deposition and its transformation into mechanical energy because the shock wave energy is not considered. The conversion efficiency $E_B/E_L$ as a function of incident laser energy is shown in Fig. 10. Close to threshold, $E_B/E_L$ is very small for fs pulses (< 0.001%) and even smaller for ns pulses (< 0.0001%). In the BPL region well above threshold, it converges to a value $E_B/E_L \approx 20\%$ for all laser parameters.

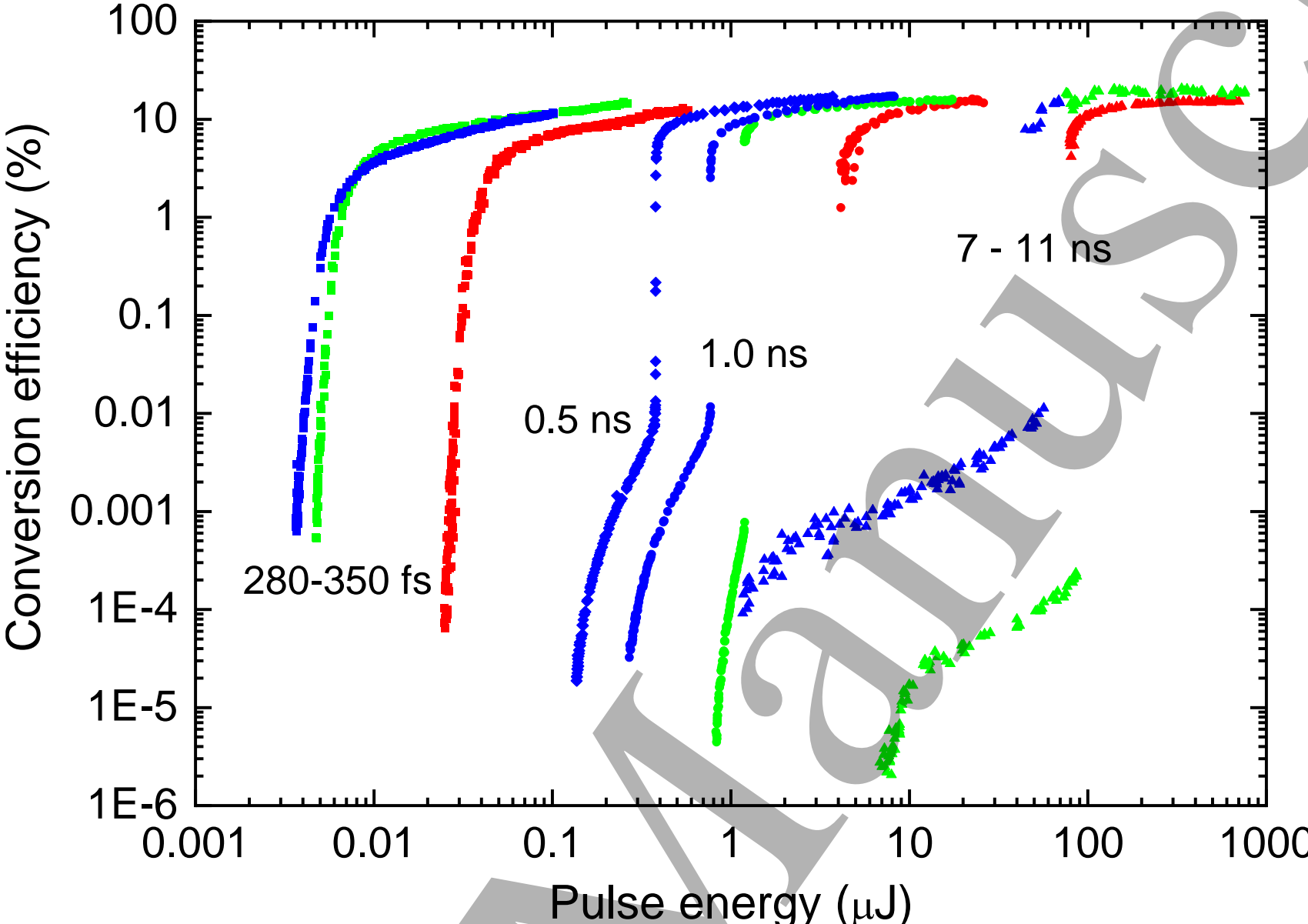

**Figure 10** Conversion of laser energy $E_L$ into bubble energy $E_B$ for data of Fig. 5c.

Let us now look at energy partitioning in the luminescent plasma regime well above the bubble threshold, where $\bar{U}_{avg}$ can be determined as shown in Fig. 9. For this regime, we explored the dependence of energy partitioning on plasma energy density based on the data listed in Table 3. Figures 11(a)-(c) show the dependence of conversion ratios $E_V/E_{abs}$, $E_B/E_{abs}$ and $E_{SW}/E_{abs}$ on $\bar{U}_{avg}$. In Fig. 11(d), averaged values from these data are combined in one graph to present the complete energy balance as a function of $\bar{U}_{avg}$. With increasing plasma energy density, the energy fraction required for vaporization of the liquid in the plasma volume drops since the absolute value of the specific vaporization energy is constant. Therefore, an ever-larger fraction is transformed into mechanical energy, most of it into shock wave energy. Interestingly, the conversion into bubble energy ($E_B/E_{abs}$) remains approximately constant between 15% and 20% when $\bar{U}_{avg}$ increases from 8.5 kJcm$^{-3}$ to 41 kJcm$^{-3}$. However, the fraction going into shock wave energy increases from $\approx 55\%$ to $\approx 75\%$, while the vaporization energy drops from $\approx 30\%$ to little more than 5%.

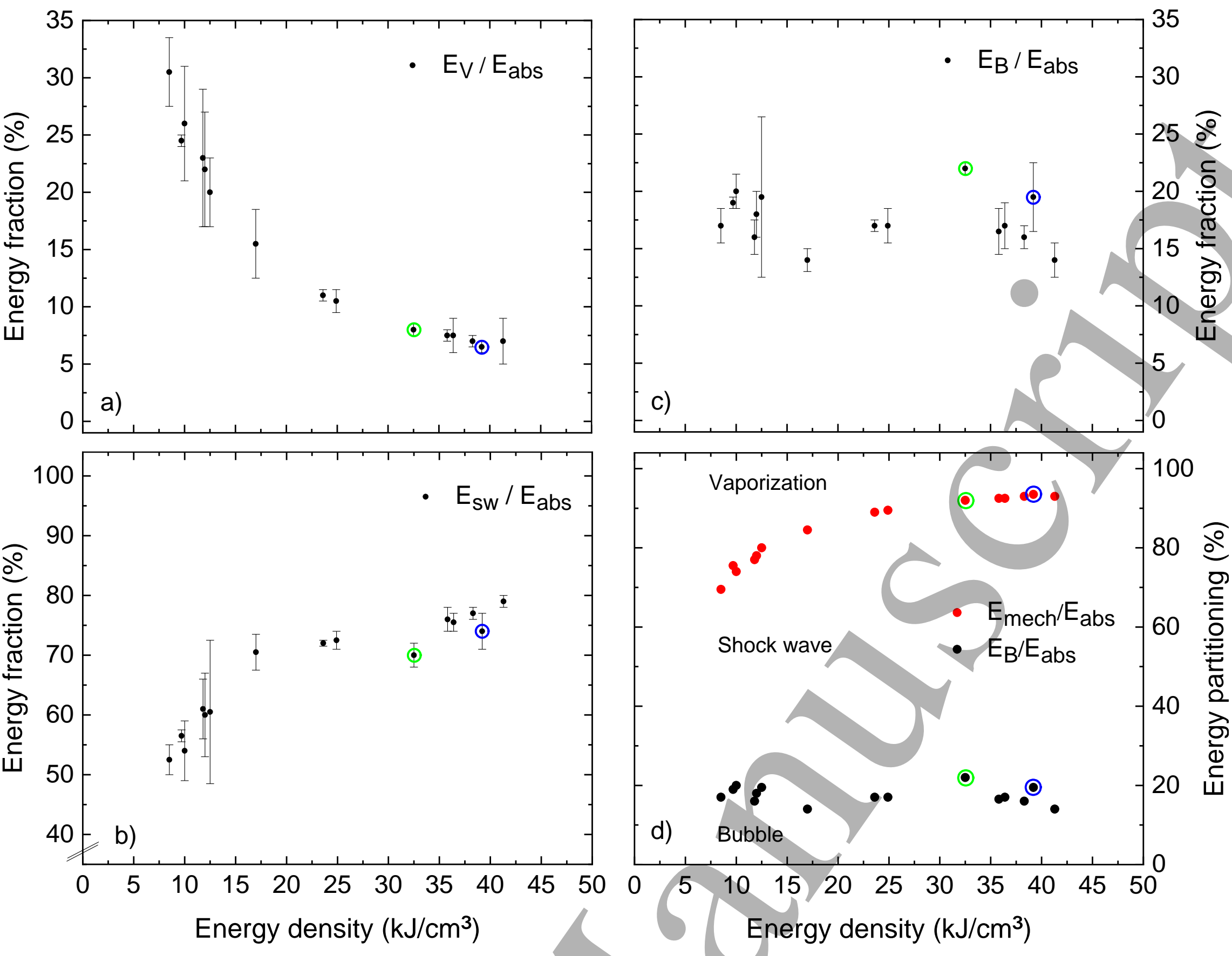

**Figure 11** Energy partitioning of absorbed laser energy as a function of average plasma energy density compiled from the results obtained with all laser systems. The individual graphs show the fractions going (a) into vaporization of the liquid within the plasma volume, (b) into shock wave emission, and (c) into bubble formation. All data are mean values averaged over the investigated range of laser pulse energies. The bars denote the variations of $E_V/E_{abs}$, $E_B/E_{abs}$ and $E_{SW}/E_{abs}$ around the average values. The complete energy balance is shown in (d), where the fractions going into bubble energy, $E_B/E_{abs}$, and mechanical energy, $(E_{SW} + E_B)/E_{abs}$ are plotted as a function of energy density. Here, $(E_{mech} - E_B)/E_{abs}$ stands for the part going into shock wave energy and $(E_{abs} - E_{mech})/E_{abs}$ represents the fraction needed for vaporization. Most data refer to $NA = 0.8$; values for $NA = 0.3$ and $NA = 0.5$ are marked with green and blue circles, respectively.

### 4.6. Material processing using longitudinally single-mode UV ns laser pulses

The energy dependence of cavitation bubble size presented in figures 5 and 6 confirms the possibility of creating fine adjustable nanoeffects by slm UV ns pulses, with stepwise transition to a more disruptive microregime. Figure 12 now demonstrates that the slm UV ns pulses can be utilized for micro-material processing such as corneal refractive surgery, micro-patterning inside glass, and the generation of refractive index modifications.

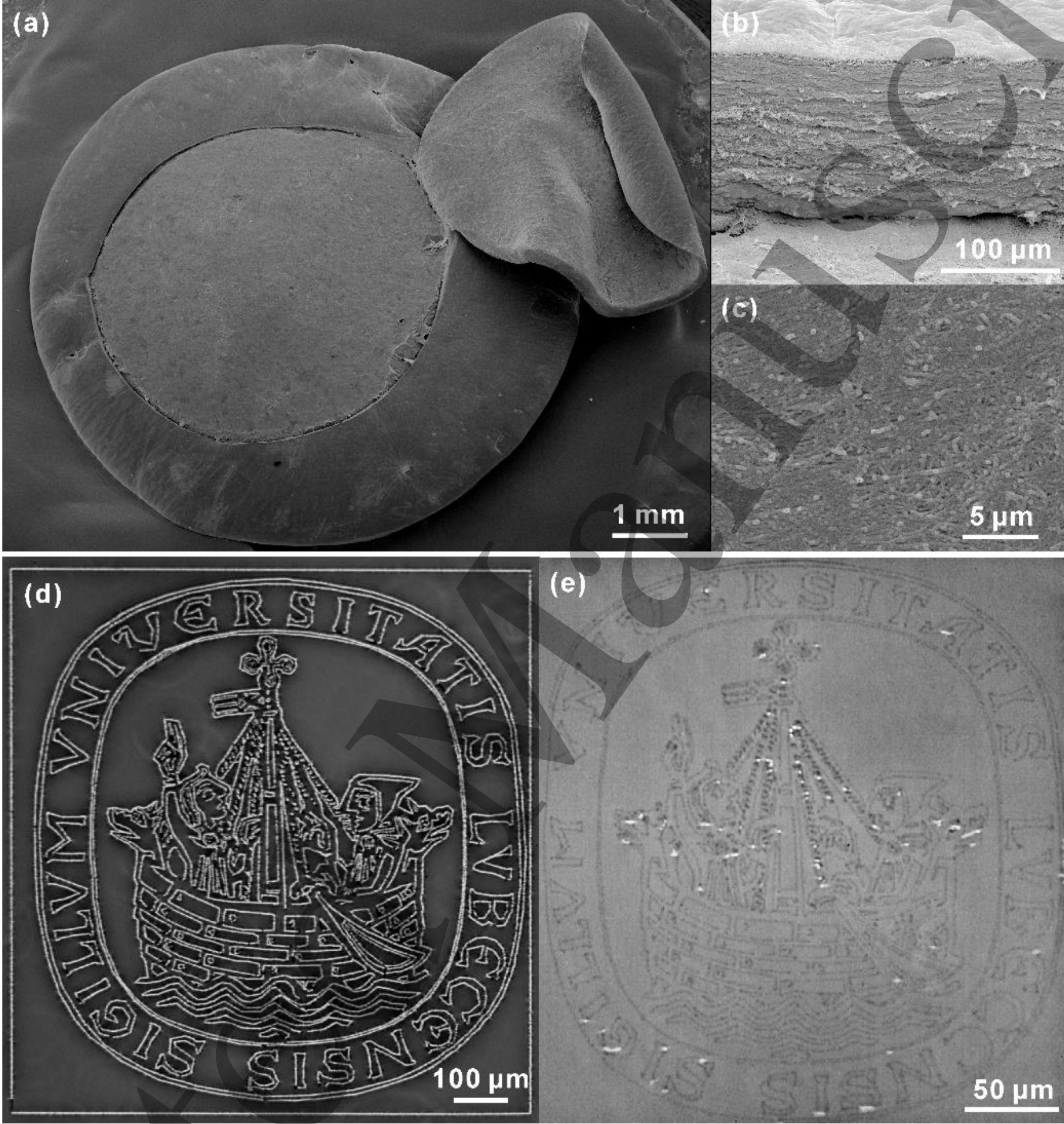

**Figure 12** Micro-material processing by means of 0.56-ns, 355-nm single longitudinal mode laser pulses applied with 1 kHz pulse repetition rate. (a) Scanning electron micrograph of a 6-mm LASIK flap in porcine cornea produced with 6 µm separation between focal spots. (b) Side cut of the flap. Due to the high cutting precision, individual corneal lamellae can be identified. (c) Enlarged view of the flap bed. (d) Logo of the University of Luebeck written within a microscope glass slide (height 1 mm). (e) Phase contrast image of the logo written at lower pulse energy, height 300 µm. The laser pulses have produced small, light scattering cavities in (d) and refractive index changes in (e). Bright spots indicate glitches, where a void was created. Laser pulses were focused at $NA = 0.28$ in (a) and (c), and at $NA = 0.75$ in (b), (d), and (e). Laser pulse energies were 1.0 µJ in (a) and (c), 2 µJ in (b), 0.44 µJ in (d), and 0.27 µJ in (e).

In state-of-the-art “Femto-LASIK”, a flap is cut into the anterior corneal stroma using fs laser pulses of 1030-1040 nm wavelength. The pulses are focused in a raster pattern into the desired dissection plane to produce the cut, and the flap is then lifted for subsequent excimer laser ablation [13,105,241]. Figs. 12(a) to (c) demonstrate that UV-A ns pulses can fulfill the same task as fs laser pulses. The cutting precision is even better because, at equal *NA*, focal diameter and length at 355 nm are only 1/3 of the values for IR Femto-LASIK [17].

Figs. 12 (d), and (e) show patterns in borosilicate glass produced at 1 kHz pulse repetition rate. At 0.44 µJ pulse energy, the expanding plasma creates tiny light-scattering cavities (Fig. 12d), while at an energy of 0.27 µJ mainly refractive index changes are produced (Fig. 12e). The size of individual laser effects as determined by observation with a phase contrast microscope is < 1µm, right at the optical resolution limit. The energy range in which refractive index changes can be formed without disrupting the glass corresponds to the small bubble regime in water (Figs. 5 and 6), while cavity formation occurs in the BPL regime. In this analogy, the transition from refractive index formation to cavity formation in glass resembles the abrupt increase of the effects size in water at the upper end of the small-bubble regime. The bright spots visible in Fig 12(e) are glitches arising from localized transitions into the BPL regime.

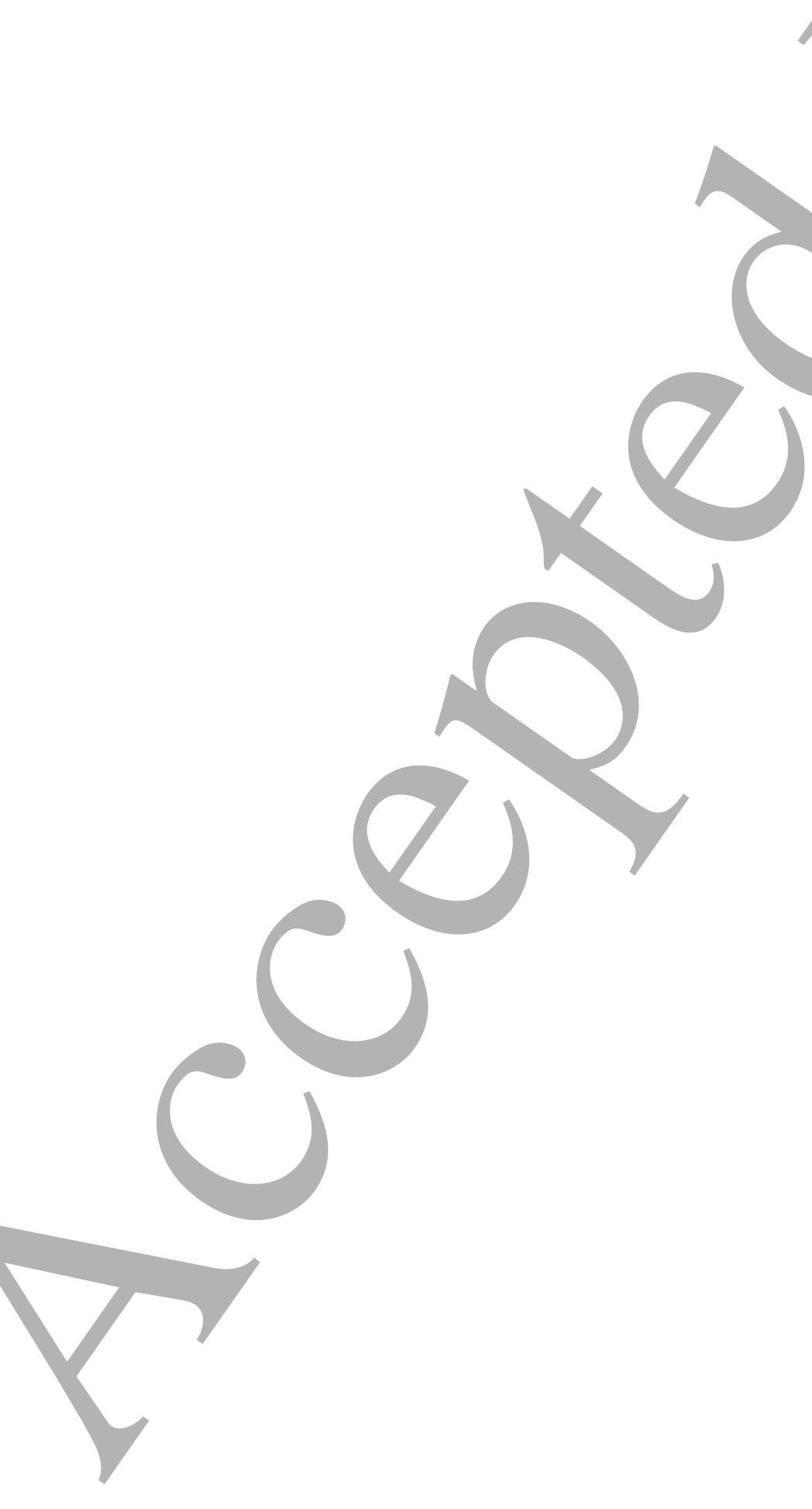

# 5. Discussion

## 5.1. Deterministic vs stochastic breakdown behavior

Our experimental investigations showed that breakdown thresholds are much sharper for femtosecond and single-longitudinal-mode nanosecond laser pulses with smooth temporal shape than for ns multimode pulses exhibiting intensity spikes from longitudinal mode beating [Figure 4(b) and Table 2]. Spikes in ns pulses facilitate multiphoton seed electron generation, which lowers the threshold [Fig 4(a)], and the pulse-to-pulse variations of spiking behavior introduce a stochastic component [Fig. 5(b)]. Similar correlations between laser mode and breakdown behavior have been reported previously on breakdown in bulk $SiO_2$ and silicate glasses [231,232]. At sample surfaces, threshold fluctuations were observed even with slm pulses [127,232] but these fluctuations were later found to be due to remnant impurities from the surface polish and were absent in pure $SiO_2$ [231]. A correlation between laser mode and breakdown behavior is also found in the $R_{max}(E_L)$ curves in Fig. 5. They are smooth for slm pulses but the bubble sizes exhibit strong statistical variations for multimode laser operation. Thus, with appropriate laser mode control also nanosecond breakdown in transparent dielectrics is a reproducible, “deterministic” event as long the impurity content is low.

Highly reproducible breakdown behavior in distilled water for tight focusing was observed at numerical apertures down to $NA = 0.3$ (Fig. 6) because under these conditions, hardly any impurities are present in the focal volume. The density of particulate impurities is $\approx 2.5\times10^4$ cm$^{-3}$ in distilled and $\approx 1.2\times10^6$ cm$^{-3}$ in tap water [201]. For $NA = 0.3$ and $\lambda = 355$ nm, one obtains a very small probability $P \approx 10^{-4}$ % of finding a particle within the focal volume in distilled water, and for tap water the probability is $P \approx 5\times10^{-3}$ %. It is still only about 9% for $NA = 0.1$ but it rises quickly, when the $NA$ is further reduced, which explains the previously reported stochastic behavior for small focusing angles [109,200,201]. Thus, the conditions for reproducible breakdown behavior are reproducible laser pulse shape, low impurity density in the medium, and tight focusing.

For biological tissues, cells and cell culture media, stronger statistical threshold fluctuations have been observed [193,242,243]. This applies especially to IR ns breakdown, where the seed electrons for avalanche ionization are the critical hurdle for the onset of breakdown. Since the cross section for multiphoton ionization depends on the excitation energy, biomolecules can lower $I_{th}$ if they possess intermediate energy levels between valence and conduction band of water that facilitate multiphoton-generation of seed electrons. In UV ns breakdown, where seed electrons are always readily available, changes of $I_{th}$ will arise only when the width of the MPI-channel is considerably enlarged, which requires a relatively high concentration of biomolecules. By contrast, with IR wavelengths, a much lower concentration may already affect $I_{th}$.

Local variations of the optical properties do not only affect the breakdown threshold but also lower the threshold sharpness (see section 2.3). In a similar way, it is expected that for UV-A and VIS ns pulses the transition between the small-bubble and luminescent-plasma regime is not as sharp in cells with inhomogeneous optical properties as in water. Experimental $R_{max}(E_L)$ data for different aqueous biological media showed that within a certain energy range, both small and much larger bubbles can be formed and the magnitude of the laser effect varies strongly from shot to shot [193,242]. Nevertheless, small bubbles could reproducibly be produced in murine intestine by slm 355-nm, 0.5-ns pulses with energies close to the bubble threshold [242]. Reliable low-density plasma formation by 0.5-ns, 532-nm slm pulses was also achieved in cell culture media [193].

Fluctuations of the breakdown threshold in biological media have also been reported for fs pulses, with dependencies on concentration of biomolecules, local optical properties, and staining [116,243,244]. However, sudden jumps to high-density plasma formation are here not observed because the $R_{max}(E_L)$ dependence is continuous [Fig. 5(c)]. Nevertheless, threshold fluctuations may result in fairly large variations of the effect size due to the steep slope of the $R_{max}(E_L)$ curves in the small-bubble regime.

## 5.2. Regimes and scaling laws of nonlinear energy deposition

The $R_{max}(E_L)$ curves of Fig 5(c) cover a large tuning range of nonlinear energy deposition in water. The volume of laser-produced bubbles in the ($\tau_L$, $\lambda_L$, $E_L$) parameter space spans 11 orders of magnitude. With fs pulses and UV/VIS ns pulses, the bubble volume can be tuned by 8 orders of magnitude, when the pulse energy is varied by a factor of only 10-100. The specific shapes of the $R_{max}(E_L)$ curves in different regions of the ($\tau_L$, $\lambda_L$, $E_L$) parameter space reflect the three breakdown scenarios introduced in section II. While pronounced differences between the scenarios are observed close to threshold, the energy dependence well above threshold is always the same. In high-density plasmas, most of the incoming laser energy is absorbed, and a constant fraction of the absorbed energy goes into bubble energy [Fig. 10(c)], which leads to $R_{max} \propto (E_L)^{1/3}$ [58,104]. In the following, we discuss the characteristic features of the different scenarios in the energy range between bubble threshold and high-density plasma regime.

***Scenario 1*** relates to fs breakdown at any wavelength. Here seed electrons from photoionization are readily available because of the high irradiance needed to complete the breakdown process. However, time constraints limit the possible number of doubling sequences during a femtosecond pulse and the strength of the ionization avalanche [37,113,115,146]. Therefore, the onset of breakdown is smooth and its endpoint can be precisely tuned by varying the laser irradiance.

Close to threshold, only a tiny fraction of the energy of the fs pulse is absorbed (Fig. 10). The $R_{max}$ ($E_L$) scaling is here determined by nonlinear absorption through the interplay of MPI and AI, and by the growth of the plasma volume with increasing $E_L$. If nonlinear absorption was dominated by multiphoton effects, the slope of the $R_{max}(E_L)$ curves in the double-logarithmic graph should be similar to the cube root of the order of the multiphoton process needed for ionization, which is 3 at $\lambda = 347$ nm and 8 at $\lambda = 1040$ nm. Thus, close to threshold it should be m = $\sqrt[3]{3} = 1.44$ at $\lambda = 347$ nm, and m = $\sqrt[3]{8} = 2$ at $\lambda = 1040$ nm if the plasma volume was constant, and a bit steeper, when the growth of the plasma volume with $E_L$ is considered. However, the actual slope is much steeper and amounts to $m = 10.5$ at $\lambda = 347$ nm, and $m = 14.7$ at $\lambda = 1040$ nm. The large slope steepness far beyond the value expected from MPI alone indicates a prominent contribution of avalanche ionization, in accordance with the results of our previous study on the wavelength dependence of fs breakdown [37]. Well above threshold, most of the incident laser energy is absorbed and the specific energy dependencies of MPI and AI cease to play a role. Here, the slope of the $R_{max}(E_L)$ curve gradually converges towards $m = 1/3$.

***Scenario 2*** describes the "big-bang" region at long pulse durations and wavelengths, where luminescent plasmas and large bubbles are produced already at threshold. Here, seed electron generation by photoionization is the critical hurdle for the occurrence of breakdown at IR wavelengths. Because of the high order of multiphoton processes, this requires a high irradiance. At the same time, avalanche ionization is powerful because its rate increases roughly proportional to $\lambda^2$ and many doubling sequences can occur during a ns pulse [113,115]. Due to the combination of high irradiance and AI rate, the onset of breakdown is abrupt and the ionization avalanche can progress very rapidly [22,231]. Already at threshold, about 50% of the incident energy is absorbed [223] and temperatures > 5000 K as well as pressures >7000 MPa are reached (Table 3), which are associated with bright plasma luminescence [Fig. 7(d)]. Correspondingly, the laser induced effects are highly disruptive [40,110,114,194,228]. For energies larger than $2\times E_{th}$, the scaling law can be approximated by $R_{max} \propto (E_L)^{1/3}$ [Fig. 5(c)], while close to threshold $R_{max} \propto (E_L - E_{th})^{1/3}$ provides better results [12].

***Scenario 3*** applies to slm UV and VIS ns pulses with smooth temporal shape. Here, multiphoton-generated seed electrons are readily available because of the large photon energies. Like in fs breakdown, this leads to a gradual onset of plasma formation, which makes it possible to create nanoeffects. The avalanche ionization rate is comparatively small at short wavelengths and the "slow" avalanche is, furthermore, inhibited by recombination. This results in a smaller conversion efficiency into bubble energy at threshold (Fig. 10) than for fs breakdown. Moreover, the slope of the $R_{max}(E_L)$ curves is smaller and the energy range in which nanoeffects and small microeffects can be produced, is broader. Close to threshold, the slope of the $R_{max}(E_L)$ curves first decreases because the growth of free electron density with increasing laser pulse energy is inhibited by recombination $\propto \rho_c^2$ [Eq. (3)]. However, this trend is counteracted by the exponential increase of $\rho_{therm}$ with focal temperature [Eq. (4)]. When thermal ionization overcomes recombination, an abrupt transition to luminescent plasmas and large bubbles occurs [Figs. 5(c) and 7(b)]. In the high-density-plasma regime above the step in the $R_{max}(E_L)$ curve, the scaling law quickly converges to $R_{max} \propto (E_L)^{1/3}$, as in scenarios 1 and 2.

Let us now look at the transitions between scenarios 1 – 3. The "big bang" scenario 2, which characterizes breakdown at long wavelengths and pulse durations, merges into scenario 1, when the pulse duration is reduced below the thermalization time of CB electrons and thermal ionization ceases to play a role. At $\lambda$ = 1064 nm, this occurs around $\tau_L$ = 30 ps [114,223]. With decreasing wavelength, seed electrons for AI become readily available, and scenario 2 changes into scenario 3. For $\tau_L$ = 2 ns, and $\lambda \geq 720$ nm, the onset of breakdown is still abrupt [36], whereas a smooth onset with nanobubble formation is observed for ns breakdown at $\lambda$ = 532 nm. In the short wavelength range, scenario 1 with continuous tunability changes into scenario 3 exhibiting a step in the $R_{max}(E_L)$ curves, when the pulse duration becomes longer than recombination and thermalization times for CB electrons. Since the thermalization time is in the order of 20-30 ps [123,174], a step is expected for $\tau_L$ > 30 ps. The precise borders of the individual regions still need to be explored in future.

Figure 6 shows that slm UV ns pulses can produce nanoeffects at $NA \geq 0.3$, i.e. within a large range of focusing angles. The energy difference between bubble and BPL thresholds increases significantly, when the pulse duration is prolonged from ≈ 0.5 ns to several ns [Fig. 5(c)]. Because of the lower threshold irradiance at longer pulse durations (Table 2), the AI rate decreases and the inhibiting role of recombination becomes stronger. This leads to a broader tuning range of nano- and microeffects.

Historically, attention has been paid mostly to scenarios 1 and 2. Our discovery of scenario 3 for UV/VIS nanosecond breakdown became possible by a combination of several factors: Diffraction limited focusing at large *NA* eliminates disturbances by aberrations or stimulated Brillouin scattering. Use of the bubble criterion for breakdown in conjunction with a sensitive measurement technique enables the detection of nanobubbles produced by non-luminescent plasma. Use of slm laser pulses eliminates statistical fluctuations of $R_{max}$ and makes it possible to record smooth $R_{max}(E_L)$ curves. The shape of these curves can be understood by considering the interplay between multiphoton ionization and avalanche ionization with recombination and thermal ionization. Recently, Agrez et al. reported also separate thresholds for the formation of nanobubbles and shock wave emission plus large bubbles by tightly focused 515-nm, 60-ps pulses but without explaining the underlying dynamics of plasma formation [245].

The integral hydrodynamic response of optical breakdown reflected in the scaling law for cavitation bubble formation is similar for fs and ns plasmas but differs in the small-bubble region close to threshold. We shall see in the next two sections that the spatio-temporal characteristics of plasma formation itself, the resulting plasma substructure and the origin of plasma luminescence differ strongly also in the high-density plasma region.

### 5.3. Plasma structure, energy density and pressure in femtosecond breakdown

In fs breakdown, no plasma luminescence is observed near the bubble threshold. Only for energies $\geq 3\times E_{thB}$, we could photographically detect faint luminescence by integrating over many breakdown events (Fig. 7a). To the best of our knowledge, this is the first recording of plasma luminescence produced by nanojoule fs laser pulses focused into water. Previously, experiments were mostly performed with weakly or moderately focused laser beams [110,246] that are subject to an interplay of self-focusing, nonlinear absorption and plasma defocusing at super-threshold pulse energies. This interplay limits the free electron density in the relatively long plasma region and strongly reduces the intensity of plasma radiation [135,140,214,217,247]. Luminescence was reported only in one study, where water droplets were irradiated by a powerful collimated fs laser beam and back reflection at the droplet walls caused localized energy deposition within the droplets [222].

For laser-induced breakdown in bulk media, the plasma energy density is limited by the movement of the breakdown front during the laser pulse [108,132-136,248] (Fig. 13). This feature distinguishes it from breakdown at material surfaces during which a thin plasma skin layer evolves [6,137-139]. In the skin layer, high electron temperatures are produced, which results in X-ray emission [249-251]. During breakdown in bulk media by weakly focused ultrashort laser pulses, the breakdown front moves with the pulse because the plasma region is considerably longer than the geometrical length of the laser pulse (Fig. 13b). As a consequence, the pulse is progressively attenuated by plasma formation while approaching the focus [135,248,252,253]. The details of this process are complex because linear focusing interacts with filamentation caused by the interplay of nonlinear self-focusing and plasma defocusing [214,218,252,254] and scattering by the plasma adds to the energy depletion [136]. With tight focusing, the plasma region is shorter than the geometrical laser pulse length. In this case, breakdown starts at the beam waist and the breakdown front moves upstream towards the incoming laser pulse [108,133,134] (Fig. 13c). This always happens in ns and ps breakdown but applies also for fs pulses that are focused at large *NA*. For focusing at moderate *NA*, when the spatial extent of a fs laser pulse resembles the plasma length, no clear upstream or downstream movement of a breakdown front is observed [135,136]. The "front" is blurred by the smooth onset of plasma formation in fs breakdown but the concept of a moving breakdown wave is very useful for identifying general trends in plasma dynamics. We will see in the following that upstream movement of the breakdown front produces much larger plasma energy densities than a downstream movement accompanied by nonlinear beam propagation and energy depletion.

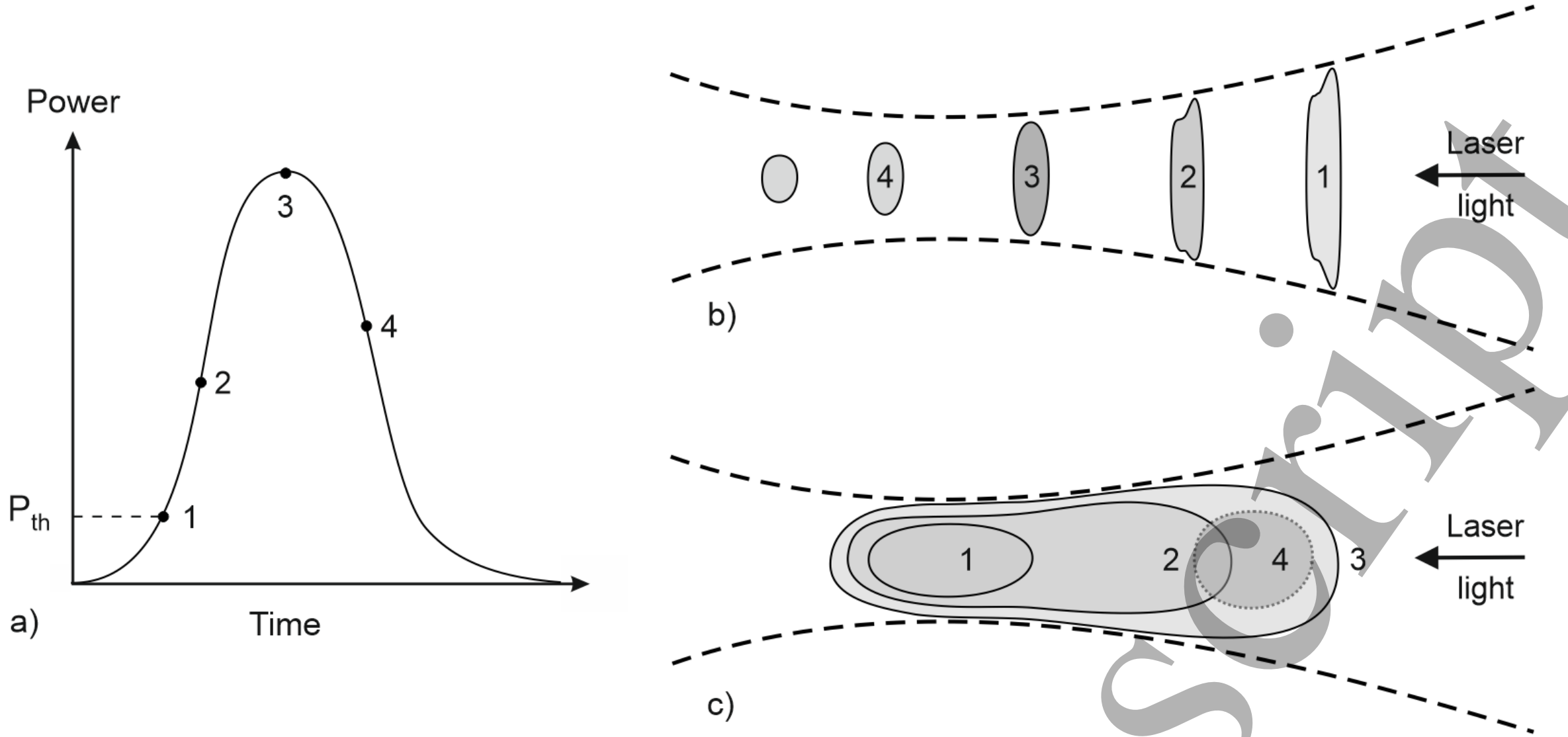

**Figure 13** Optical breakdown dynamics in dependence on pulse duration and numerical aperture. a) Time evolution of the laser power during a superthreshold laser pulse. b) Dynamics for weakly focused ultrashort pulses with pulse length shorter than the plasma length. Plasma formation starts in the upstream region as soon as the irradiance of the focused pulse becomes large enough to exceed the breakdown threshold (1). While the pulse moves downstream and forms new plasma (2-3), its energy is ever-more depleted. At the same time, the ongoing focusing of the laser beam maintains a large irradiance of the travelling pulse such that plasma formation can continue up to the beam waist. In this process, linear focusing interacts with nonlinear focusing and with defocusing by the laser-induced plasma, which may result in filamentation (4). Nevertheless, electron and energy density are largest near the nonlinear focus. c) Breakdown dynamics for laser pulses with a spatial pulse length longer than the plasma length, which applies for long pulse durations and for fs pulses focused at large *NA*. Breakdown starts at the beam waist (1) and the plasma front moves upstream towards the incoming laser pulse (2), with the maximum extent reached at time 3. During the second half of the pulse, the irradiance is too small for further extending the breakdown region, and the laser energy is absorbed in the already existing plasma. This creates a hot spot in its upstream part (4), as seen in the plasma photographs in Fig. 8. Note that the drawings in (b) and (c) are not to scale; the actual focusing angle is much smaller and the plasma length much larger in (b) than in (c).

For plasmas produced at $NA$ = 0.8, the photographs of fs breakdown in Fig. 7a show no signs of nonlinear beam propagation and filamentation, and the plasma is only ≈ 5 µm long even for $E_L$ = 500 nJ, twenty times above threshold. Hence, the plasma region is much shorter than the length of the laser pulse, which in water is ≈ 80 µm for a 350-fs pulse. Under these conditions, breakdown starts at the beam waist and the breakdown front moves towards the incoming laser pulse during the first half of the laser pulse, while the laser power increases (Fig 13 c) [102]. The velocity of the breakdown wave is fastest within the beam waist and slows down towards the far-field and towards the peak of the laser pulse. A quantitative description of the upstream movement of the breakdown wave has been given by Docchio et al. [133] and slightly modified by Vogel et al. to consider the finite length of the beam waist [108]. Both researchers derived relations for the dependence of plasma length on dimensionless laser pulse energy, $l_{plasma}(E_L/E_{th})$ at a given $NA$. These approaches portray the change of the location of iso-intensity lines characterizing the breakdown threshold during the laser pulse but do not consider nonlinear beam propagation, the lowering of the breakdown threshold by UV and X-ray radiation from plasma formed earlier during the pulse, and the change of direction of wave propagation in fs breakdown. The simple model yields reasonable agreement with experimental data for pulse durations of tens of picoseconds but are insufficient for fs breakdown at small and moderate NAs, where nonlinear beam propagation plays a role and for ns breakdown, where radiative energy transport comes into play [108]. For ultrashort laser pulses, spatiotemporal breakdown models are already available that describe the interplay of plasma formation with nonlinear beam propagation by Kerr-lens-induced self-focusing and plasma defocusing [102,130,131,136,141,156,252,255]. Models considering the influence of radiative energy transport during ns breakdown on plasma size are not yet available.

The average speed of the breakdown wave is given by the plasma length on the photographs divided by ($\tau_L/2$). For the plasma formed at $E_L$ =570 nJ, the average velocity of the breakdown front is 28570 km/s, which is 13% of the speed of light in water. The plasma growth ceases at the peak of the laser pulse, and during the second half of the pulse the light energy is absorbed in already existing plasma [108,133,134,188]. That results in a small plasma volume and high average energy densities between 11.8 and 41.3 kJcm$^{-3}$ (Table 2). These values are 4.6-16 times larger than the vaporization enthalpy of water.

For fs plasmas, in which only one set of free electrons is generated, the free-electron density can be derived from the average energy density (Table 3). It peaks at 520 nm, where it reaches a value of 0.12×10$^{22}$ cm$^{-3}$. This corresponds to 18% of full ionization of the 1b$_1$ valence band level and to a supercritical plasma state ($\rho_c/\rho_{crit}$ = 2.9). At 347 nm, $\rho_c$ reaches 5.2 % of full ionization and the plasma remains subcritical ($\rho_c/\rho_{crit}$ = 0.37), while at 1030 nm, the ionization degree is 7.4 % and the plasma is highly supercritical ($\rho_c/\rho_{crit}$ = 4.8). The complex wavelength dependence reflects the interplay of two counteracting trends: the rate of avalanche ionization increases strongly with $\lambda_L$ [37] while $\rho_{crit}$ drops with $1/\lambda_L^2$ [Eq. (14)].

The upstream movement of the breakdown wave introduces a spatial inhomogeneity, as can be seen by the colour change in the photographs of IR fs plasma in Fig. 7(a). The change from reddish to bluish hue in upstream direction indicates a shift towards higher average photon energies in the plasma radiation. The spectral composition of the plasma radiation depends on the time-averaged kinetic energy distribution of the CB electrons during emission (as will be discussed in section 5.5.), and a shorter center wavelength corresponds to a larger time-averaged mean energy. The reddish hue in the beam waist region can be explained by shielding from plasma generated during the upstream movement of the breakdown wave, while the bluish hue arises from the longer-lasting energy deposition in the upstream part during the second half of the laser pulse. A colour shift of plasma emission during a movement of the breakdown wave by only 5 µm implicates a sub-micrometer penetration depth of the incident laser light in the plasma. This is consistent with the supercritical electron density at 1030 nm resulting in a very high plasma absorptivity [125,151,236,256]. When $\rho_c$ exceeds $\rho_{crit}$, both plasma reflectivity and absorptivity increase [151,256], which augments laser-plasma coupling. Above $\rho_{crit}$, about 30% of the pump light are absorbed within a layer of less than 1 µm thickness, and the rest is reflected or backscattered [236,257]. This way, supercritical electron densities can be reached in bulk material as well as at surfaces.

A comparison of the present results obtained at large *NA* (Table 3) with literature data for smaller focusing angles shows a dramatic increase of plasma energy density, temperature and pressure with increasing *NA* [114,121,122,135,136,208,235,252,258]. Jukna et al. investigated plasma formation in water by weakly focused (*NA* = 0.0035) femtosecond pulses both experimentally and theoretically. They showed that an energetic 290-mJ, 800-nm pulse of 500 fs duration produces multiple filaments but no bubble and only weak thermo-elastic acoustic waves. Simulations predicted a peak plasma energy density of 0.030 kJ/cm$^3$, corresponding to a temperature rise of only 7 K [252]. When 35-µJ, 580-nm pulses of 100-fs duration were focused into water at moderate $NA$ = 0.185, evaluation of shadowgraphs and transmission measurements yielded $U_{avg} \approx 1$ kJ/cm$^3$ [114], slightly above the thermoelastic bubble threshold at 0.6 kJ/cm$^3$. Correspondingly, the conversion rate into bubble energy remained small ( ≈ 3 %), even well above threshold [114]. For tighter focusing at $NA$ = 0.31, modeling of breakdown in fused silica by 2.5-µJ, 800-nm pulses of 80 fs duration predicted a peak energy density around 2.5 kJ/cm$^3$ with subcritical electron density ($\rho_c/\rho_{crit} < 0.2$) [136]. The length of the plasma region was here similar to the geometrical length of the laser pulse, which led to some energy depletion by nonlinear absorption during pulse propagation accompanied by pronounced scattering. For breakdown in water and fused silica at $NA$ = 0.5 and pulse energies of up to 30 times above breakdown threshold, Potemkin and coworkers measured volumetric energy densities between 4 kJ/cm$^3$ and 14 kJ/cm$^3$ but electron densities remained below the critical density [218,235]. The present investigations for breakdown in water at $NA$ = 0.8 yield plasma energy density values up to ≈ 40 kJ/cm$^3$, which for visible and IR wavelengths are associated with supercritical electron densities. The comparison of our results for large and moderate NA with literature data reveals a strong growth of $U$ with increasing $NA$, which is due to a decreasing influence of nonlinear propagation effects, a transition from downstream to an upstream movement of the breakdown wave, and a shortening of the plasma length. This picture is consistent with model predictions for breakdown in water at very large $NA$ [102,130,131].

What are the upper limits of fs plasma energy density? Since it increases with numerical aperture, energy densities produced by focusing at $NA$ > 1 may exceed the values listed in Table 3 for $NA$ = 0.8. However, it is unlikely that they will be two orders of magnitude larger as has been stated for focusing of femtosecond pulses into sapphire at $NA$ = 1.35 [121,122,259]. In those reports, it was assumed that the plasma zone in bulk sapphire is as thin as the plasma skin layer evolving during breakdown at material surfaces [137,138], namely 65 nm. That assumption led to the conclusion that fs pulses of 20 to 120 nJ energy were absorbed in a volume of only 0.01 µm$^3$ with an energy density of several MJ/cm$^3$ and that cavities observed in the target material were produced by shock-wave-induced displacement of *cold* material surrounding the plasma [121,122]. A zone of amorphous material observed around the cavity was attributed to shock-wave-induced modifications of the target crystal, with peak pressure in the order of several TPa for a 100-nJ pulse [121,122]. This interpretation is questioned by our measurement results for water that has a band gap of ≈ 9.5 eV [36], close to the values for sapphire and fused silica of ≈ 9 eV [7,38]. The plasma photographs in Fig. 7(a) show luminescent plasma of 4.8 μm$^3$ volume at $E_L$ = 95 nJ and 11.9 μm$^3$ volume at $E_L$ = 200 nJ (Fig. 9a), corresponding to an average plasma energy density of ≈ 10 kJ/cm$^3$, more than hundred times smaller than the extreme state of matter reported in [121,122]. These results for breakdown in water at $NA$ = 0.8 are consistent with modeling results of plasma and cavity formation in fused silica that yielded an energy density of 20 kJ/cm$^3$ for a 10-nJ, 800-nm, 100-fs pulse focused at $NA$ = 1.35 [215]. With increasing pulse energy, the energy density predicted by the simulations did not increase but remained approximately constant due to the growing volume of the energy deposition zone, in agreement with our experimental observations. The predicted size of the energy deposition zone was ≈ 0.5 μm$^3$, similar to the size of the amorphous region reported in [121]. This suggests that the cavities in sapphire are formed in a region within the plasma that is softened by nonthermal or thermal melting [136,179,260], and the amorphous zone is due to rapid cooling that prevents re-crystallization. This interpretation is supported by time-resolved investigations of the electron density distribution during void creation in fused silica reported in Refs. [156,261]. Altogether, these results strongly suggest that the plasma energy density in bulk dielectrics at large $NA$ is limited by the movement of the breakdown front during the laser pulse, which shields the beam waist and increases the plasma size for growing $E_L$ [108,114,134,223].

In comparing plasma pressure data for laser-induced breakdown in water and solid dielectrics, one needs to consider the different equations of state of both materials. The pressure produced by a certain energy density is linked to the mass density and sound velocity in the medium [215,262], and both values are much larger in sapphire or fused silica than in water. Therefore, an energy density of 20 kJ/cm$^3$ leading to cavity formation in fused silica [215] corresponds to a peak pressure of ≈ 250 GPa in $SiO_2$ but to only 6.5 GPa in water (Fig. 3).

While there is a clear trend towards higher plasma energy density with increasing $NA$, there is no monotonous increase of $U$ with growing pulse energy at large $NA$. Above threshold, $U$ first increases rapidly and then stays approximately constant in an energy interval up to about 30 times $E_{th}$ (Fig. 9 and Refs. [218,235]). However, for pulse energies far above the breakdown threshold, $U$ decreases again [114,219,253]. The decrease starts, when the length of the region, in which plasma can be formed at the given pulse energy is longer than the laser pulse. The breakdown wave then changes direction and its downstream movement with the pulse is accompanied by energy depletion. Rumiantsev et al. reported an average plasma energy density merely 0.35 kJ/cm$^3$ for breakdown in water by a powerful 325-µJ, 1240-nm, 170-fs pulse focused at $NA = 0.5$ [219]. Due to the pulse energy far above threshold, the plasma length amounted to 210 µm, while the pulse itself was only 40 µm long. Therefore, the breakdown wave moved with the pulse and its energy was depleted during propagation even at large $NA$.

### 5.4. Plasma structure, energy density and pressure in nanosecond breakdown

Luminescent ns plasmas are characterized by an intricate substructure within the luminescent region and by a large size going far beyond the laser cone angle at large $NA$ (Figs. 7 and 8). We will see that both features are linked by radiative energy transport from the regions of primary energy deposition and reabsorption in the surrounding liquid, which inflates the plasma and reduces the average energy density.

The finding that the average energy density in high-density ns and fs plasmas is similar even though the pulse energies in ns breakdown are much larger (Table 3) is surprising at first sight. Femtosecond breakdown produces merely one set of free electrons, with thermalization after the laser pulse. By contrast, the cycle of free-electron generation and thermalization is repeated many times during a ns pulse. Therefore, one would expect much larger energy densities in ns breakdown than with fs pulses. However, fs plasmas are confined to the cone angle of the pump laser beam, whereas luminescent plasmas produced by tightly focused ns pulses extend well beyond the irradiated cone (Fig. 7). They feature an inhomogeneous core region of primary energy deposition that is surrounded by a diffuse luminescent halo. The region of primary energy deposition by 532-nm ns pulses is demarcated by branched plasma strings that scatter the incoming laser light [Figs. 7(c),(e),(f)]. While the energy density averaged over the entire diffuse plasma luminescence is 14 kJcm$^{-3}$, the average over primary energy deposition zone within the laser cone angle is 81 kJ/cm$^3$. Peak values are likely much higher because the strings occupy less than 1/5 of the core volume. This line of reasoning suggests transient local energy densities of up to 400 kJ/cm$^3$. Similar inhomogeneities as those produced by 532-nm pulses likely exist also in UV and IR ns-laser-produced plasmas but are not visible on the photographs because the digital camera used in the experiments does not record the scattered UV and IR laser radiation.

What causes the inhomogeneous energy deposition in nanosecond breakdown leading to strings in the region of primary energy deposition? The inhomogeneities cannot originate from nonlinear beam propagation because at $E_L = 670$ µJ [Figs. 7(e) and (f)] the laser peak power remains more than 26 times below the threshold for catastrophic self-focusing in water, which is ≈2×10$^6$W at 532 nm [263]. Instead, they result from an inherent instability of avalanche ionization that is supported by the large number of doubling sequences in ns breakdown. Since the avalanche ionization rate is proportional to irradiance and to the local free-electron density $\rho_c$, the avalanche will grow faster at locations where $\rho_c$ is already larger than elsewhere. Small intensity fluctuations across the pump laser beam result in larger fluctuations of the seed electron density produced by PI, which are then further enhanced by AI $\propto \rho_c$. Strings are produced, when the zones of high electron density grow upstream as the breakdown wave propagates towards the incoming laser beam [108,133,134]. The speed of the upstream moving breakdown wave during ns breakdown is several orders of magnitude slower than with fs pulses [the average speed is ≈ 9.1 km/s for the 733-µJ ns pulse in Fig. 7(c), compared to 28570 km/s for the 0.57-µJ fs pulse in Fig. 7(a). That provides sufficient time for a large number of doubling sequences in the ionization avalanche.

The small diameter (≈ 1 µm) and the large surface of the branched filamentous regions allow for efficient laser-plasma coupling [138,151,236,238,264]. Above $\rho_{crit}$, about 30% of the pump light are absorbed within less than 1 µm, and the rest is reflected or backscattered [236,257]. The increase of plasma absorptivity and reflectivity, when $\rho_c$ exceeds $\rho_{crit}$, [151,256], enhances the instabilities promoting string formation. The locally increased absorptivity results in a runaway towards very high ionization levels followed by rapid heating of the free electrons, and the increased reflectivity promotes the upstream growth of the string tips. Local inhomogeneities may lead to branching, such that a fractal high-density plasma structure evolves.

The strings in Figs. 7(e) and (f) resemble streamers in DC breakdown in gases [265], water [266-271], and lightning [272] – but on a much smaller spatial scale. We use the name “string” rather than “streamer” to emphasize the difference between AC and DC breakdown. Nevertheless, there are similarities too. The strong DC electric field emanating from streamer tips also accelerates electrons and drives AI, which then results in streamer growth [266-268,273]. For positive streamers, seed electrons are produced through photoionization by energetic photons emitted from the tips [265]. Their propagation velocity is > 20 km/s [266,267], similar to the speed of the breakdown front in nanosecond optical breakdown. However, since a high ionization level is reached in the highly absorbing strings, the energy density is much higher than in DC streamers before spark formation. Both strings and streamers exhibit a tree-like fractal structure [265,274,275]. The exploration of branching mechanisms and of the fractal character of branching trees is an active field of research [276-278].

String formation arising from an intrinsic AI instability in conjunction with the upstream movement of the optical breakdown wave differs from multiple filament formation arising from instabilities of nonlinear light propagation during the downstream movement of an energetic laser pulse [204,252,279]. Both string and filament formation occur during single shots, unlike laser-induced periodic structures at surfaces and periodic structures inside bulk media that arise from a cumulative interplay between electric field and thermomechanical effects during multiple exposures [280-283].

Electron diffusion and energy transport by energetic photons are potential mechanisms for the generation of the large luminescent halo within and around the region of primary energy deposition. Figure 14 compares their possible range as a function of electron or photon energy, respectively.

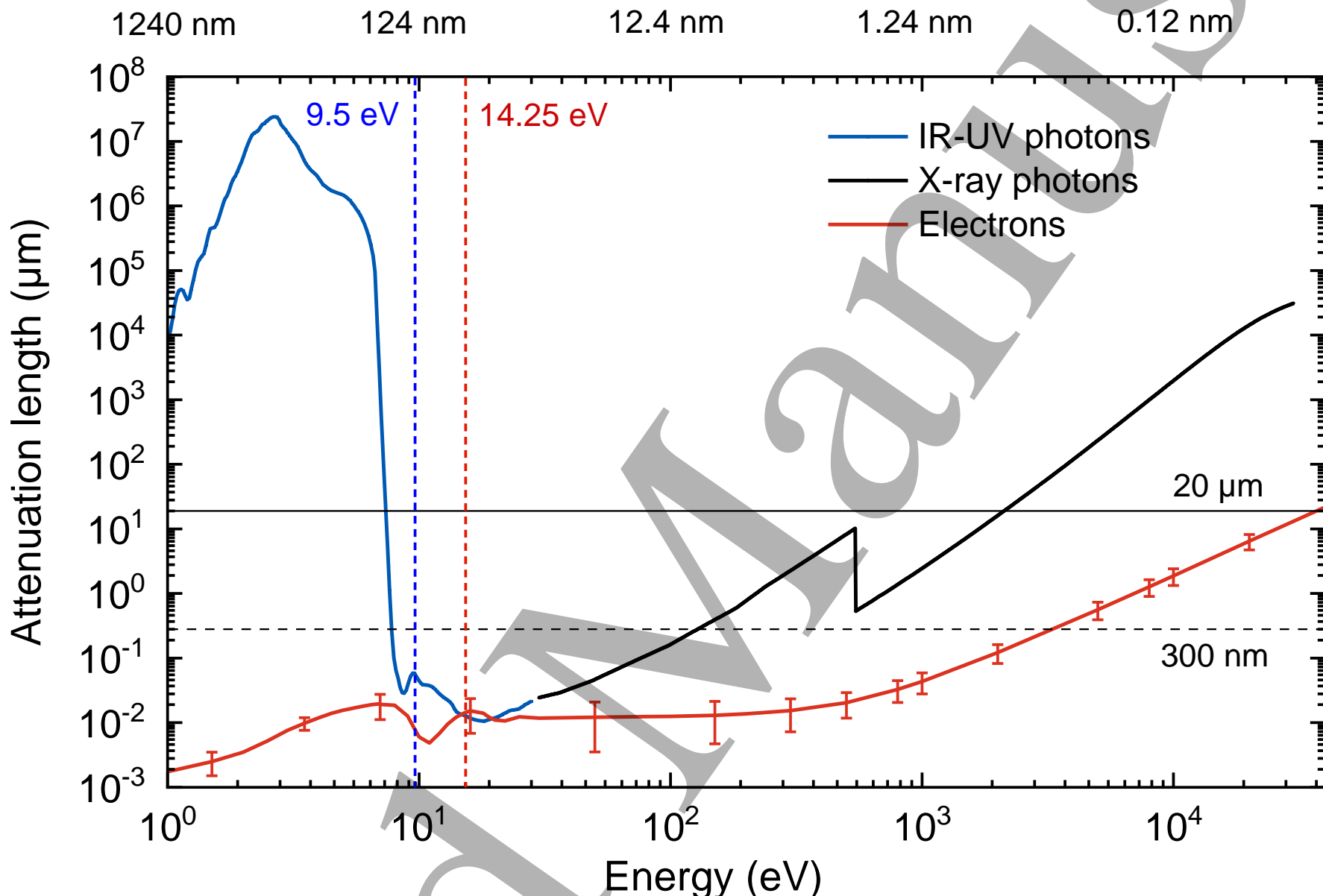

**Figure 14** Energy dependence of the attenuation length of photons and electrons in water. Data are taken from [284,285] for photons of IR to UV wavelengths, from [286,287] for x-rays, and from [288] for the attenuation length of electron beams. The latter values were obtained through Monte Carlo simulations, and the vertical bars indicate 95% confidence interval [288]. Photons can produce CB electrons by one-photon ionization, when their energy is larger than the band gap of water ($E_{gap}$ = 9.5 eV), and electrons can produce CB electrons, when their energy is larger than the impact ionization energy (1.5×$E_{gap}$ = 14.25 eV) [37]. These energy levels are marked by vertical dashed lines. The penetration depth of an electron beam provides an upper limit for the diffusion length of electrons from a plasma, where their motility is limited by the attraction from positive charges. The dashed horizontal line demarcates an attenuation length of 300 nm for which photons emitted from high-density strings of about 1 µm diameter are reabsorbed within the string region. This applies to photons in the energy range between 7.7 eV and 128 eV. The transport distance needed to form a luminescent halo extending 20 µm beyond the laser cone angle is marked by a solid line. Photons with energies between 128 eV and 2.16 keV have penetration depths between 300 nm and 20 µm; large enough to form the observed halo. The drop of penetration depth at 532 eV is due to the K-absorption edge of oxygen [286].

An energy transport over a distance of several micrometers cannot rely on electron diffusion because for energies up to 500 eV the penetration depth of an electron beam in water is below 30 nm [288], and the diffusion length for ambipolar diffusion of electrons and ions is even smaller. Electrons are thus largely confined to the region in which they have been produced, which explains the small lateral dimension of the strings that grow in length with the moving breakdown wave. Other than electrons, 500-eV photons in the water-window have a penetration range comparable to the extent of the halo [286]. Such energetic photons may arise from bremsstrahlung emitted by hot electrons in the strings. When a very high ionization level is reached, the ongoing laser energy deposition quickly raises the average electron temperature, and the photon energy of bremsstrahlung emitted from the electrons increases [289]. Initially merely UV bremsstrahlung is emitted, which is re-absorbed within a very short distance because the penetration depth for linear absorption in water is below 300 nm for photon energies between 7.7 eV and 128 eV [286]. This way, incident laser energy remains trapped in the strings, while electrons gain further energy. Finally, electrons are sufficiently hot for the generation of X-ray bremsstrahlung that can leave the region of initial energy deposition and produce plasma elsewhere. The penetration depth of the soft X-rays increases to 10 µm when the photon energy grows to 500 eV ($\lambda$ = 2.5 nm) and further to 30 µm at 2.5 keV ($\lambda$ = 0.5 nm). Such penetration depths are comparable to the extent of the luminescent halo beyond the laser cone angle. The possibility of soft X-ray emission in ns breakdown in water is corroborated by earlier experiments in which photons in the 500-eV region were emitted from ns-laser-produced plasmas on a thin water jet passing through a vacuum chamber [238,239].

Why can X-ray photons escape the high-density plasma strings? For extreme UV radiation and soft X-rays, inverse bremsstrahlung absorption is the dominant absorption mechanism both in native water (Fig. 14) and in plasma [37,290]. Due to the increasing attenuation length with decreasing wavelength (or increasing photon energy, respectively), X-rays are only weakly reabsorbed in the plasma. Even for $\rho_c = 6.68 \times 10^{22}$ cm$^{-3}$ corresponding to full ionization of the $1b_1$ valence band orbital, the frequency of X-rays with wavelengths shorter than 129 nm exceeds the plasma frequency. Therefore, no ‘plasma mirror’ evolves and X-rays can readily penetrate into the surrounding lower-density plasma or water.

The elongation of the halo in the propagation direction of the laser beam seen in Figs. 7(c) and (e) indicates that X-ray emission occurs preferentially in string direction. A possible explanation for the emission anisotropy is a field gradient from string tip towards stem favoring bremsstrahlung emission along the string direction, which evolves during string growth. Simulations of streamer growth in gases showed that the different motility of electrons and ions results in a charge separation [265,275], and we assume that the same applies for the strings observed in our experiments. Electrons and ions are simultaneously generated at the string tips, and here the local charge is neutral. During string growth, electrons in its stem move towards the periphery because equal charges repel each other, while the heavier positive ions remain in the center. This way, a potential difference evolves along the string axis from the neutral tip towards the positive core of the stem. The resulting electric field may accelerate electrons preferentially towards the stem, i.e. towards the laser beam waist, which would promote the generation of energetic bremsstrahlung in this direction.

X-ray absorption in the halo region creates a cascade of secondary low-energy electrons [160]. When the energy of these electrons drops below the ionization threshold of water, they produce no more electrons by impact ionization and are thermalized by collisions, hydration, and recombination. Bremsstrahlung emitted by electrons with energies of a few eV has an emission peak in the UV but the long-wavelength tail lies in the visible region of the optical spectrum [290]. This tail is transmitted by water, can leave the breakdown region and is visible on the photographs. Therefore, the plasma luminescence has a similar bluish appearance everywhere in the diffuse halo region, regardless of spatial variations of the initial kinetic energy distribution before thermalization. The radiation energy leaving the water cell amounts to less than 0.1 % of the absorbed laser energy [114] but a much larger fraction of the absorbed laser energy must be involved in radiative energy transport within the plasma region itself, where electron temperatures are higher and the conversion into bremsstrahlung is more efficient [291,292].

Due to the limited spectral range of our digital camera, an experimental exploration of the wavelength dependence of string formation was not possible and remains a task for future research. Since strings arise from an intrinsic spatial instability of AI and because AI becomes ever stronger with increasing wavelength, the phenomenon is expected to be a prominent feature also in ns IR breakdown but weak in UV plasmas. Moreover, the critical electron density is larger for shorter wavelengths, which promotes 'regular" nonlinear absorption and delays the onset of the absorption/reflection runaway at $\rho_c \approx \rho_{crit}$. Both factors together will tend to level out small-scale inhomogeneities at 355 nm compared to 532 nm. Nevertheless, UV plasmas produced at large *NA* inflate beyond the pump laser cone angle as seen in Fig. 7(b).

Let us now look at the plasma growth and radiative energy transport for ns plasmas produced at moderate numerical aperture, which is shown in Fig. 8 for *NA* = 0.3 and $\lambda_L$ = 355 nm. Here we do not see a region of primary energy deposition surrounded by a diffuse halo but rather a sharply delineated, mottled region extending from the beam waist in upstream direction, with a bright spot at the upper end. The bright spot is particularly pronounced and large at $E_L$ = 575 µJ, where it has a blurred, reddish periphery. The upstream movement of the breakdown wave is here faster than at large *NA*s (the average speed is 27 km/s for $E_L$ = 575 µJ), and the rate of avalanche ionization is smaller than for breakdown at 532 nm, *NA* = 0.8 discussed above. Therefore, the average plasma energy density reached during the initial upstream movement of the breakdown wave is smaller. Nevertheless, a very high ionization level may be reached in the hot spot, where the energy is deposited during the second half of the laser pulse. For $E_L$ = 575 µJ, we deduced a peak plasma density of 260 kJ/cm$^3$ in the hot spot. This value is almost as large as the estimate of 400 kJ/cm$^3$ for the high-density strings observed at large *NA*. How can such large energy densities be produced in a relatively large volume? The increase of critical electron density with decreasing wavelength [Eq. (14)] already implies the possibility of reaching large plasma energy densities. If additionally $\rho_c$ exceeds $\rho_{crit}$, the resulting drop in absorption depth can lead to an even more efficient heating of this plasma part by the ongoing laser irradiation in the second half of the laser pulse. The location, where $\rho_{crit}$ is first exceeded, depends on a balance between geometrical focusing of the pump laser beam and energy depletion by its absorption within already existing plasma. This location will lie somewhere in the upstream part of the plasma. From there, a front with increased absorption/reflection will move towards the incoming laser beam following the original breakdown wave. The reflected part of the incident laser energy is absorbed within the already existing plasma, close to the reflecting layer, and bremsstrahlung emitted by the highly absorbing region is largely re-absorbed within the cone angle of the laser beam too. This way, a large hot spot is formed.

Plasma radiation promotes plasma growth in axial direction: the plasma length is 50 µm already at the BPL threshold for *NA* = 0.3, which is much larger than the Rayleigh length of the focal region (18.5 µm). Present findings on plasma prolongation and inflation are consistent with previous observations on IR ns breakdown at moderate *NA* [108,293], which suggested that plasma radiation lowers the BPL threshold in the vicinity of the plasma and promotes the formation of large plasmas already at and close to the breakdown threshold.

The *NA* dependence of plasma energy density in ns breakdown differs from that in fs breakdown, where $U_{avg}$ drops dramatically with decreasing NA. For slm ns pulses, $U_{avg}$ remains approximately constant between *NA* = 0.8 and *NA* = 0.3 (Table 3). This similarity relies on two counteracting trends: 1. The optical breakdown wave moves faster at moderate *NA*s than for very tight focusing, which reduces $U_{avg}$. 2. On the other hand, plasma radiation is at moderate *NA*s re-absorbed within the region of primary energy deposition, which prevents a lowering of $U_{avg}$ by radiative transport. Plasma energy density starts to drop only, when the *NA* is reduced below 0.2 and the increase of plasma volume due to the faster upstream movement of the breakdown wave starts to dominate [108,114]. Correspondingly, both plasma energy density and conversion efficiency into bubble energy decrease slowly with decreasing focusing angle for $NA < 0.2$ [114]. Nevertheless, even for $NA = 0.06$, shock wave emission was observed in breakdown by 10-ns, 532-nm pulses and simulations predicted a peak energy density of ≈ 6 kJcm$^{-3}$, while fs breakdown did not even result in bubble formation [254].

The average plasma pressure values derived from $U_{avg}$ using EOS data lie between 3.6 and 11 GPa (Table 3). They agree well with previous results for ns breakdown at moderate *NA* obtained by measuring the shock wave velocity in the immediate plasma vicinity, which lie between 7 and 12 GPa [40,110,228]. Both methods do not provide peak pressure values within the plasma - the values in hot spots and strings within ns plasmas are certainly much larger. According to the EOS data in Fig. 3, the pressure at 260 kJcm$^{-3}$ energy density (as estimated for the hot spot in the plasma at E = 574 µJ in Fig. 8) should by far exceed 100 GPa (1 Mbar). An experimental validation of this conjecture is still needed. It requires ultrafast time-resolved imaging capable of tracking the formation of shock waves in larger hot spots and their propagation around the thin high-density strings (see section 6.2).

## 5.5. Bremsstrahlung and blackbody characteristics of femtosecond and nanosecond plasmas

Understanding the bremsstrahlung and blackbody characteristics of plasma luminescence in dependence on laser parameters and plasma properties enables to draw meaningful conclusions on breakdown dynamics and thermodynamic plasma properties. Spatially resolved time-integrated spectral information on fs and ns plasma luminescence in the wavelength range between 410 and 690 nm is presented in Figs. 7 and 8, and quantitative spectroscopic data on spatially integrated luminescence from ps and ns breakdown in water are available in the literature [224-226,246].

There are three sources of plasma radiation involving free-free, free-bound, and bound-bound electronic transitions. The first one involves electron-ion and electron-neutral collisions, in which a photon is emitted. It results in continuum emission and is usually referred to as bremsstrahlung [238,240,289,290,294-296]. The second involves capture of a free electron into a bound state (recombination). It is the inverse of photoionization and results in a continuum below a cut-off wave length, i.e. $\lambda < \lambda_{th}$, where $\lambda_{th}$ is the photoionization edge for the particular bound state formed. The third radiation process involves electronic transitions between bound states of an atom or ion and results in a line spectrum. For hot luminescent high-density plasmas of atoms at low atomic number (z number), bremsstrahlung dominates [294] but the luminescence is often interpreted as blackbody radiation [76,78,224-226,297]. In the spectral range, where radiation can escape the water cell and becomes photographically or spectroscopically detectable, we see merely the long-wavelength tail of the plasma radiation. Here, the blackbody spectrum corresponding to a certain thermodynamic temperature is similar to the thermal bremsstrahlung spectrum at a higher electron temperature [291,295]. Therefore, the luminescence colour on photographs and the shape of the accessible spectrum alone do not allow a clear identification of the nature of the plasma radiation. To facilitate such identification, it is useful to recall that free-free interactions between charged particles always produce bremsstrahlung but this emission assumes the properties of blackbody radiation only under specific conditions [183,291,298]. When the ensemble of CB electrons is in thermal equilibrium, thermal bremsstrahlung is emitted. Equilibrium is not yet reached during optical breakdown by a 100-fs pulse, where the kinetic energy spectrum of CB electrons changes during the pulse and approaches an asymptotic shape only towards its end [149]. However, it will be reached during longer pulses. Part of the bremsstrahlung is always self-absorbed by the plasma, whereby the absorption takes place preferentially at low frequencies because the absorption length equals the inverse-bremsstrahlung decay length [290]. With increasing density of emitting and absorbing particles, the spectrum is self-absorbed up to ever larger frequencies. When all spectral components are self-absorbed (opaque plasma) *and* the particles feature a Maxwell-Boltzmann distribution ( thermal equilibrium), we have a blackbody [291]. This condition is fulfilled if the electron density is sufficiently high and the plasma size is sufficiently large, and the pulse is sufficiently long. For a micrometer-sized plasma, the emission of a blackbody-like spectrum requires a supercritical electron density $\rho_c > 10^{21}$ cm$^{-3}$ [236,290,297] but for larger plasmas, subcritical electron densities are sufficient.

We will now use the above criteria to classify the plasma radiation visible on the photographs in Figs. 7 and 8. The plasma radiation produced by 350-fs pulses shown in Fig. 7(a) is likely thermal bremsstrahlung because the electron ensemble will be in thermal equilibrium during most of the pulse. The luminescence colour of the emitted radiation is not only linked to the kinetic energy spectrum of CB electrons evolving during the laser pulse but also to the red-shift and loss of spectral intensity during thermalization [149]. The thermalization process is fast, when the free electron density is low enough to keep the molecular and band structure of water intact. In this case, CB electrons rapidly lose energy through inelastic collisions with water molecules and hydrate within less than 300 fs [165-167,170,290]. However, when at higher irradiance and electron density the band structure dissolves and most water molecules dissociate, electron-ion recombination becomes the predominant thermalization pathway, with thermalization times of 20-30 ps [123,158,174]. The change of the thermalization process with increasing electron density will likely influence the spectral emission characteristics of the plasma and is probably responsible for the colour shift from beam waist towards upstream direction in the fs plasma images of Fig. 7(a).

The *total* duration of plasma luminescence is always longer than the laser pulse duration. However, the duration of *local* plasma luminescence may be shorter, when the thermalization time is below $\tau_L$. For ns pulses well above the breakdown threshold, the upstream moving ionization wave shields downstream regions such that thermalization can progress and bremsstrahlung is quenched in its wake. This seems to be the case for ns breakdown at moderate *NA* shown in Fig. 8. Here the downstream region of the plasma between beam waist and hot spot has sharp contours and features a mottled substructure. This indicates that the duration of the luminescence is too short to be blurred by the hydrodynamic plasma expansion leading to bubble formation. The initial bubble wall velocity in ns breakdown is well above 1000 m/s (i.e. 1 µm/ns) for a 1-mJ pulse [40]. Thus, the lack of observable blur indicates a local luminescence duration below 1 ns, which is consistent with thermal bremsstrahlung from CB electrons that ceases during their thermalization. The temperature of the thermalized electron-ion plasma is lower than the initial electron temperature. Its emission is likely weaker such that the colour of the photographs arises mainly from the initial bremsstrahlung. Numerical simulations showed that the CB electron energy spectrum in water is constant over a large irradiance range from below bubble threshold to nearly full ionization of the $1b_1$ valence band level, with average energies of 6.5-7.0 eV [149]. This finding explains the bluish colour and the homogeneity of the luminescence in the downstream plasma region.

The luminescence in and around the hot spot in the upstream plasma region has a very different appearance: it is blurred, whitish, and the colour exhibits a red shift towards the plasma periphery. At $E_L$ = 574 µJ, where the energy density is estimated to be as large as 260 kJ/cm$^3$, the luminescence extends beyond the laser cone angle. These observations indicate that the plasma emission remains strong, when local thermal equilibrium (LTE) between electron and ion populations is reached and the plasma starts to expand. The duration of plasma luminescence is longer than in the wake of the moving breakdown wave further downstream because laser energy is absorbed during the entire second half of the laser pulse and the resulting thermodynamic temperature is much larger. Due to the large size (10-20 µm) of the fully ionized hot spot, the plasma is optically thick over a large wavelength range. Hence, all criteria for blackbody emission are fulfilled. Adiabatic cooling during the plasma expansion leads to the red shift in the periphery of the blurred region around the hot spot.

During ns breakdown at large *NA*, a hot plasma core is formed within the laser cone angle, which emits energetic photons producing a cascade of secondary electrons in a larger volume. These electrons then emit the diffuse blue luminescence visible on the photographs in Figs. 7(b) and (d). The luminescent halo exhibits no red shift in its periphery and is distributed non-isotropically around the plasma core, with largest extent in the propagation direction of the laser beam. Therefore, it cannot be related to the hydrodynamic plasma expansion. The primary radiation emitted by the high-density strings in the plasma core is thermal bremsstrahlung since the interplay of AI ($\propto \rho_c$), recombination ($\propto \rho_c^2$) and thermal ionization [$\propto exp(-E_{gap}/2k_B T)$] rapidly establishes a Maxwell Boltzmann distribution typical for LTE. However, it is not blackbody radiation because the strings are optically thin. By contrast, the diffuse luminescence of the inflated plasma fulfils the blackbody criteria, because it emerges from a fairly large region with high electron density such that the plasma is optically thick.

In this paper, we have determined the thermodynamic plasma temperature and the plasma pressure through an evaluation of the measured plasma energy density based on EOS data (Table 3). Alternatively, the plasma temperature can be determined by fitting the bremsstrahlung or blackbody emission spectrum to a Planckian distribution. This has been performed for brightly luminescent ns and ps plasmas [224-226]. In ns plasmas, peak plasma temperature values in strings and hot spots will significantly exceed the average values. However, such transient local temperature variations cannot be assessed by spatially and time-integrated spectroscopic measurements at $\lambda > 300$ nm, which detect the long-wavelength tail of the diffuse luminescence from the entire halo [224-226]. Fits of a Planckian distribution to plasma emission spectra from IR breakdown in water yielded blackbody temperatures of 6200 K for 80-ps, 1-mJ pulses, 10000 K for 5-ns, 4 mJ pulses [224,246], and values of 15000 to 16000 K at much larger pulse energies ≥ 100 mJ [225,226]. A value of 11500 K was reported for ns breakdown in fused silica [297]. Since the energy density of thermal radiation is proportional to $T^4$ [291,294], spectroscopic results will be close to the peak temperature within the depth from which radiation exits the luminescent halo. In inhomogeneous plasmas exhibiting a hot spot, they represent the temperature of the hot region rather than the average temperature over the entire plasma volume. This explains, why blackbody temperatures reported in the literature are higher than the average values in Table 3. These intricacies need to be considered, when plasma temperature and temperature data are used as input for hydrodynamics calculations of shock wave emission and cavitation bubble generation.

Recently, Tian et al. determined the temperature evolution during the early plasma and bubble expansion phase after optical breakdown in bulk water by 1064-nm, 10-ns pulses of 20 mJ energy during a time interval from 75 ns to 700 ns after the laser pulse [103]. For this purpose, line emission from OH radicals and Ca tracers dissolved in the water was measured and temperatures determined by fitting experimental OH spectra with synthetic spectra and through Saha-Boltzmann plots of calcium lines [299]. Temperature values at $t$ = 75 ns were 12700 K from the Saha-Boltzmann plot and 11000 K from OH line spectra. The subsequent temperature decay from 75 ns to 700 ns matched well with adiabatic conditions during the early expansion phase, as expected from hydrodynamic bubble models [42,63]. However, the finding of a plasma temperature at $t$ = 75 ns similar to the blackbody temperature right after the laser pulse in [224-226] is very remarkable. Presumably, a high temperature is maintained during the early plasma expansion and decay through energy release by electron-ion recombination and reformation of radicals and molecules from hydrogen and oxygen atoms. Such energy release resembles the release of latent heat during condensation which keeps the temperature constant during the phase change. However, the ionization and molecular dissociation energies are much larger than the latent heat of vaporization per water molecule (0.422 eV) and the effect is, hence, also much larger.

### 5.6. Energy partitioning in dependence on plasma energy density

The energy balance of optical breakdown comprises the fraction of incident laser energy that is absorbed and the partition of $E_{abs}$ into $E_v$, $E_B$, and $E_{SW}$. The energy of the plasma radiation emitted from the diffusely luminescent plasma is here neglected because it amounts to less than 1% of $E_{abs}$ [114,297] even though the radiative transport within the plasma involves a larger energy fraction. While the dependence of nonlinear energy deposition on pulse duration, wavelength, focusing angle and pulse energy is very complex as discussed in the previous sections, the partitioning of $E_{abs}$ relies mainly on average plasma energy density and is governed by more general rules. In the following, we discuss the changes of partitioning with growing $U_{avg}$.

Below bubble threshold, all energy goes into free-electron mediated chemical effects and heating [22]. Numerical simulations for plasma formation slightly above the bubble threshold ($R_{max}$ = 230 nm) yielded the result that 0.46% of the absorbed energy is transformed into the acoustic energy of the laser-produced thermoelastic transient, and only 0.03% goes into bubble energy [22]. This is consistent with the conversion efficiency data of Fig. 10, whereby these $E_B/E_L$ data relate to the incident laser energy while modeling referred to $E_B/E_{abs}$.

There is an information gap on energy partitioning for $U_{avg}$-values close to threshold, where plasma luminescence is too weak to delineate the plasma shape. The plasma volume could be inferred from shadowgraphs or Schlieren pictures taken immediately after breakdown, or via third harmonic (THG) imaging [218,220] but such techniques were not applied in the present study. Furthermore, the determination of $E_{abs}$ merely from transmission measurements through Eq. (9) becomes inaccurate, and one would need to measure also the amount of scattered light. Determination of $V_{plasma}$ and $E_{abs}$ close to threshold remains a task for future work.

Just above threshold, absorbed energy will partition mainly into vaporization energy and bubble energy, while little energy goes into shock wave emission. Our investigations do not cover this regime because energy partitioning can be evaluated only when plasma luminescence starts to be detectable on photographs, which in our setup is the case for $U_{avg} \geq 8.5$ kJ/cm$^3$ (the limit could be shifted by integrating over more images or use of an ICCD camera). Here, 30% of $E_{abs}$ go into vaporization, almost 20% is converted into bubble energy, and about 50% goes into shock wave energy (Fig. 11). The work fractions needed to overcome surface tension and viscosity are negligible because luminescent-plasmas produce relatively large bubbles, for which the respective pressure terms scaling with $1/R$ are very small [42]. With further increase of plasma energy density, a decreasing part of $E_{abs}$ is needed for vaporization and an ever-larger fraction can be converted into mechanical energy. The increase of the conversion into mechanical energy is rapid up to about 15 kJ/cm$^3$ and becomes more gradual for larger plasma energy densities. The fraction $E_V/E_{abs}$ drops to only 7% at $\bar{\varepsilon}_{avg} = 40$ kJ/cm$^3$, while $E_B/E_{abs}$ stays approximately constant near 20% and $E_{SW}/E_{abs}$ increases to 78%.

The ratio of shock wave energy and bubble energy, $E_{SW}/E_B$, increases from a value of ≈ 2.8 at $U_{avg} = 10$ kJ/cm$^3$ to ≈ 5.2 at 40 kJ/cm$^3$. This trend is due to the rise of the plasma expansion velocity with increasing energy density. The initial plasma expansion velocity, particle velocity behind the shock front and bubble wall velocity are identical [[40,42]. Previous experiments on ps and ns breakdown showed that the initial bubble wall velocity increases from subsonic values (650 m/s) at $U_{avg} \approx 10$ kJ/cm$^3$ to supersonic values (2100 m/s) at $U_{avg} \approx 40$ kJ/cm$^3$ [40]. With supersonic plasma expansion, a shock front is formed almost instantaneously, and $E_{SW}/E_B$ becomes very large.

It is interesting to note that the peak plasma energy densities observed in this paper for laser pulses focused at $NA = 0.8$ ($U_{avg} \approx 40$ kJ/cm$^3$) are five times higher than the volumetric energy density of TNT (7.54 kJ/cm$^3$). For $NA = 0.3$, an even higher peak energy density of 260 kJ/cm$^3$ was deduced from the photographs in Fig 8. Correspondingly, the fraction of absorbed energy that is converted into shock wave energy is much larger for laser-induced high-density plasmas than for underwater explosions of TNT charges, where $E_{SW}/E_B \approx 1.5$ [58,300].

With growing energy density, the $E_{SW}/E_{abs}$ ratio increases at the cost of the fraction going into vaporization until more than ¾ of the absorbed energy is converted into shock wave energy – but then it does not increase further and also $E_{SW}/E_B$ remains constant. Most of the shock wave energy is dissipated as heat already during the early propagation phase within a few charge or plasma radii, respectively, [42,59,114]. The heat dissipation reduces the amount of mechanical energy available for structural deformation in the plasma vicinity. Moreover, the energy transfer by shock propagation and dissipation at the shock front influences the subsequent cavitation bubble dynamics by lowering surface tension and viscosity [42] and reducing the rate of heat and mass transfer by condensation and heat conduction.

The peak $E_{SW}/E_B$ ratio found in the present study is higher than the value of 2.3 in previous investigations for breakdown at $NA \leq 0.25$ and $U_{avg} \approx 40$ kJ/cm$^3$ [114]. Experimentally determined $E_{SW}/E_B$ ratios are generally higher than the values obtained in hydrodynamic simulations, in which a homogeneous energy density distribution as starting condition is assumed [42,66]. The neglect of hot spots within the luminescent plasma region leads to an underestimation of the conversion efficiency into shock wave energy, which strongly increases with peak pressure [59,114,300].

Although we have observed that the conversion into shock wave energy increases with average plasma energy density, we are not yet able to assess and quantify the influence of hot spots within the plasma on shock wave generation and energy partitioning. Further insights can be gained by time-resolved photography with ultrahigh spatiotemporal resolution that portrays not only the shock wave emission from the periphery of the luminescent region but also the launching of the shock wave from the hot spot and its propagation through the diffusely luminescent plasma region. Complementary information could be gained through finite volume modeling of hydrodynamics events considering measured absorbed energy and energy distribution in the breakdown volume. Both approaches require further developments of experimental and modeling tools as discussed in section VI.

### 5.7. Consequences for laser surgery and material processing in bulk dielectrics

We demonstrated the existence of three basic scenarios for the energy dependence of breakdown effects and enlarges the parameter space in which nanoeffects can be produced. Material processing within the framework of scenarios 1 (fs breakdown) and 2 (IR breakdown at long pulse durations) is already well established. Use of fs breakdown close to threshold allows for precise nanomorphing, cell surgery and dissection [1,13,22,24,27,301-304], while high-density plasmas produced by IR ns pulses enable to create more vigorous laser effects that are, for example, used for membrane disruption in intraocular microsurgery [12]. However, the scientific community has not been aware about the details of scenario 3 for ns breakdown at UV/VIS wavelengths, which imply that nanoeffects can be produced with slm pulses from cost-effective diode pumped Q-switched solid state lasers. Although it has been reported that cell surgery can be performed using visible and UV-A ns laser pulses [124,192,305], it remained a puzzle how nanoeffects could be produced by means of ns pulses as long as the threshold of ns breakdown was identified with plasma luminescence [19]. This question has now been resolved by the discovery and explanation of the two-step behavior of UV and VIS nanosecond breakdown that involves a small-bubble regime without bright plasma luminescence [306,307].

*Laser surgery.* We demonstrated the usefulness of single-mode UV ns pulses on corneal dissections relevant for refractive surgery in Figs. 12(a) – (c), and first results of animal experiments have been presented by Trost et al. [16]. A potential disadvantage of using UV ns laser pulses for nanosurgery on cells or tissue is the small conversion efficiency of laser energy into mechanical energy close to the bubble threshold (Fig. 9). The applied UV dose is a potential source of photochemical side effects and the mechanical effects associated with plasma-mediated dissection may lower the photochemical damage threshold. However, the photochemical damage potential is at 350 nm four orders of magnitude lower than in the UV-C region below 280 nm [17,308]. Furthermore, the use of vortex beams rather than Gaussian beams for corneal flap dissection diminishes cavitation and shock wave effects, which reduces the aggravation of UV-mediated cell stress by mechanical stress [17,309]. It remains to be investigated under which circumstances photochemical side-effects really counter-indicate the use of 355-nm sub-ns pulses in biophotonics.

Potential applications of UV ns pulses are most advantageous for transparent tissues and cells – for applications in scattering tissues ultrashort laser pulses at IR laser wavelengths remain the best choice because their penetration depth is larger due to a pronounced decrease of scattering with increasing wavelength [188]. Such applications include the induction of skin rejuvenation by subsurface optical breakdown in skin [310,311], scar treatment in vocal folds [312] and the attenuation of seizure propagation in the brain by circumscribing cuts around epileptic foci [313]. Linz et al. investigated the required pulse energy for brain interventions as a function of laser wavelength and cutting depth by numerical simulations [37].

*Material processing.* Generally, UV laser pulses offer a high precision for material processing because of their short wavelength. With ns pulses, nanoeffects can be produced in a broader energy range than with fs pulses, which goes along with a better tuning sensitivity of nonlinear energy deposition (Fig. 5c). The lack of nonlinear propagation effects enables to create filamentation-free localized energy deposition with large volumetric energy densities even at moderate *NA*s. However, this advantage is compromised by the discontinuity of the tuning curve at the transition from the low-density into the high-density plasma regime that may cause problems in some applications.

Figures 12(d) and (e) show microeffects in glass produced in both regimes by 0.56-ns UV laser pulses. With high-density plasmas, tiny cavities and cracks are observed [Fig. 12(d)], similar to those created by tightly focused 100-fs pulses at 780 nm wavelength [194]. The disruptive effects are useful for micro-marking [120]. By contrast, low-density plasmas produce the refractive index changes visible in Fig. 12(e). They may be useful for the writing of optical waveguides [2-4] if glitches into the high-density regime with cavity formation can be avoided. Glitches visible in Fig. 12(e) are mostly located in corners of the scanning pattern, where a larger number of laser pulses were applied. They can possibly be prevented by using higher pulse repetition rates than in the present experiments, which would allow for heat accumulation in the focal region and thermal annealing [314,315]. Another approach could be the use of longer pulses of several ns duration, which feature a broader energy range with nanoeffects than the 0.56-ns pulses used in the present experiment (see Fig. 5c).

The abrupt transition between low- and high-density plasma regimes compromises the reproducibility of laser effects in biological targets featuring inhomogeneous nonlinear absorption properties due to centers of reduced excitation energy for multiphoton ionization. For cells and some cell culture media, a fairly broad transition zone has been observed in which both regimes overlap such that either very small or much larger bubbles are produced [193,242]. Nevertheless, close to the bubble threshold, low-density plasma effects could reliably be create in intestinal epithelium using 355-nm, 560-ps pulses [242]. The discontinuity of the $R_{max}(E_L)$ curve does not matter for applications in homogeneous breakdown media in which the tuning range of interest belongs completely to either the low- or the high-density plasma regime. In some cases, easy switching between both modes may even be desired.

Altogether, our findings show that for various applications mode-locked femtosecond lasers may be replaceable by compact, cost-effective microchip lasers emitting single-mode sub-ns pulses in the UV/VIS wavelength range.

### 5.8. Consequences for laser ablation and breakdown spectroscopy in liquids

A deeper mechanistic understanding of optical breakdown in bulk liquid also helps to understand the rules governing plasma growth and hydrodynamic phenomena in laser ablation in liquids (LAL) for nanoparticle generation [71-81] and in underwater laser-induced breakdown spectroscopy (LIBS) for remote analysis of target constituents [82-89]. LAL is an ecofriendly and versatile approach that reduces the need for additional purification steps. It has been successfully employed to synthesize nanoparticles from metals, oxides semiconductors and organic materials directly in water and organic liquid media, making it attractive for applications requiring high-purity colloidal nanoparticles [72,316]. Underwater LIBS is of major importance for deep-sea exploration of minerals and hydrothermal vent fluid chemistry [85,86,88,89].

*Plasma formation and target shielding.* In both cases, plasma is formed at a linear absorbing solid target immersed in water (or another liquid). For discussing the ablation of linear absorbing materials, we will in the following refer to the radiant exposure threshold rather than to the irradiance threshold. The surrounding liquid features little linear absorption at the laser wavelength, and the optical breakdown threshold of the liquid, $F_{th,liquid}$, lies above the ablation threshold of the target, $F_{th,target}$. Thus, for pulse energies close to the ablation threshold of the target material, plasma formation is restricted to the target surface (Scenario I). However, at higher pulse energies, ionizing radiation from plasma formed at the target surface may ignite breakdown in the liquid, which then shields the target, as observed previously for breakdown in air ignited by plasma at a water/air surface [317] (Scenario II). Pure shielding would impair ablation in LAL and reduce the spectroscopic signal from the analyte in LIBS. However, the plasma formed in the liquid can, in conjunction with the target plasma also heat and melt the target and eject relatively large droplets [69,70,76]. Both scenarios are more strongly separated in LAL of metallic targets, where $F_{th,target} \ll F_{th,liquid}$ than in LAL of oxides, semiconductors and organic materials as well as in LIBS, where ablation thresholds may be closer to $F_{th,liquid}$.

For process optimization, it is useful to know the thresholds for plasma-mediated target ablation and liquid breakdown as a function of pulse duration, wavelength and mode of laser operation. However, this knowledge alone is not sufficient to understand the conditions for ignition of liquid breakdown at the target surface and its propagation into the liquid during the laser pulse. Electrons are ejected from solid targets during plasma-mediated pulsed laser ablation [71,318-320] but the penetration depth of electrons into water is below 30 nm for energies up to 500 eV and the attenuation length of UV radiation remains below 20 nm for photon energies up to 27 eV (Fig. 14). For direct bremsstrahlung emission beyond 20 nm, the kinetic energy of the plasma electrons would have to exceed 27 eV, which would require electron temperatures > 300,000 K [321-324]. Such high electron temperatures can be reached under conditions, where a plasma skin layer forms [178,325] but this will not occur in pulsed laser ablation of metals for pulse energies below the liquid breakdown threshold. We conclude that for $F_L < F_{th,liquid}$ an ignition of liquid breakdown resembling the halo formation outside the laser cone angle in Fig. 7 (b)-(f) is not possible.

Inside the region traversed by the laser beam, another mechanism for breakdown ignition may come into play, which relies on radiation-induced generation of seed electrons for AI. The band structure of water features an intermediate state at the energy level of solvated electrons around 6.5 eV that lies below the lower edge of the conduction band at around 9.5 eV. Therefore, plasma initiation can proceed through excitation of water molecules above 6.5 eV with subsequent electron abstraction and upconversion into the CB [36,37]. Electron emission from the target plasma will first initiate liquid breakdown in a several nanometer thick layer next to the target. The interplay of IBA and impact ionization produce CB electrons in this layer with a kinetic energy spectrum reaching up to ≈ 14.5 eV (the energy required for impact ionization), with average kinetic energy around 6.8 eV [149]. Bremsstrahlung from these electrons with photon energies larger than 6.5 eV can excite water molecules and initiate plasma formation in adjacent liquid. Figure 14 shows that the optical penetration depth is as large as 10 cm for 6.5-eV photons and drops rapidly to 30 nm when the photon energy increases to 8.5 eV. Thus, the penetration depth suffices to drive a radiation-ionization wave propagating through creation of seed electrons for AI in upstream regions, from which then again UV radiation is emitted. Our hypothesis is corroborated by the relatively large extent of nanosecond plasmas at the BPL threshold in bulk liquid (Fig. 8) [108]. Once plasma is formed, the breakdown process becomes here independent of the creation of initial electrons by multiphoton ionization, and the threshold is lowered. Therefore, the plasma grows further than predicted by the moving breakdown model, where the plasma extent is defined by iso-intensity lines at $I_{th}$. Nanosecond pulse durations favor the formation of a radiation-ionization wave as they give time for an AI-driven buildup of ionization strong enough for significant bremsstrahlung emission providing seed electrons in new liquid regions. The possible depth of the liquid breakdown region depends on the AI rate, which grows with irradiance and wavelength (Section 2.1).

For $F_{th,target} < F_L < F_{th,liquid}$, it remains to be explored under which conditions laser energy is merely absorbed by the target and an expanding target plasma (Scenario I) [75,76,78], and when plasma grows also by ignition of liquid breakdown through ionizing radiation from the primary plasma (Scenario II).

To date, most experimental studies on LAL and LIBS present plasma images only for times ≥ 100 ns and with limited spatial resolution [78,90,91,326-329]. An exception with regard to temporal resolution is the study by Song et al. on LAL of a metallic glass with image series taken at 100 Mio frames/s [330]. For a pulse energy well below the breakdown threshold of bulk water, the authors observe a thin luminescent plasma layer at the target surface and point-like plasmas in the liquid within the 3-mm laser beam that may be formed at nanoparticles produced earlier. In the context of LIBS, the early plasma phase and initial bubble expansion has been investigated in free liquid [88,331] but not yet at solid targets. With large pulse energies, the plasma volume seen on photographs at $t \geq 100$ ns is often very large but it is unclear whether this is due to an expansion of the target plasma or whether liquid breakdown is involved. High-resolution plasma photography in side view could elucidate details of plasma formation and the possible movement of a radiation-ionization wave into the liquid. Time-gated photography with a fast ICCD camera can further elucidate the temporal evolution – but without the spectral information available from CCD or CMOS color sensors.

*Ablation rate and efficiency.* Without plasma shielding, the ablation depth with ultrashort laser pulses up to ≈ 20 ps exhibits an approximately logarithmic dependence on pulse energy, $d_{abl} \propto \ln E_L$ [177,187,332]. The formation of a highly absorptive plasma skin layer in the target reduces the ablation rate at high radiant exposures [177,178]. Under conditions, for which the $d_{abl} \propto \ln E_L$ law holds, the ablation efficiency (ablated mass per unit energy) has a maximum at 2.7 times the ablation threshold [75,188]. This favorable condition is likely still part of the ablation regime, where liquid breakdown can be avoided. In nanosecond ablation, the energy dependence of the ablation rate is more complex, and ablation efficiency increases with the transition from surface vaporization to phase explosion regime, which goes along with substantial ionization of vaporized metal atoms [322]. The freed electrons will favor the ignition of nearby liquid breakdown.

*Connection between ablation and bubble dynamics.* The thermodynamic conditions during cavitation bubble initiation, expansion, and collapse influences the particle size distribution in LAL [74,75,81,90-94,333-336]. Valuable mechanistic insights into the initial ablation and bubble formation phase have been obtained through atomistic simulations for metallic targets [81,94,337]. Three distinct nanoparticle generation mechanisms were identified in the simulations: (1) the formation of a thin transient metal layer at the interface between the ablation plume and water environment followed by its decomposition into large molten nanoparticles, (2) nucleation, growth, and rapid cooling/solidification of small nanoparticles at the very front of the emerging cavitation bubble, above the transient interfacial metal layer, and (3) spinodal decomposition of a part of the ablation plume located below the transient interfacial layer, leading to the formation of a large population of nanoparticles growing in a high-temperature environment through inter-particle collisions and coalescence. The atomistic simulations start with a certain energy density distribution in the target material and do not consider the plasma formation dynamics in the target and the surrounding liquid. Therefore, they are most valuable for the regime in which energy deposition occurs through IBA by electrons in the metal without ionization and without a contribution of liquid breakdown. Furthermore, because of their numerical expense, such simulations are restricted to selected parameters. A systematic exploration of the parameter dependence of the entire sequence of LAL processes is still lacking. The $R_{max}(E_L)$ curves in Figs. 5 and 6 resulting from such systematic exploration helped to delineate the regimes of scenarios 1 to 3 for plasma formation in bulk liquid. Measurement of $R_{max}(E_L)$ for the approximately hemispherical bubbles produced at solid targets may help to delineate the regimes and transitions between scenarios I and II in LAL. Reich et al. already showed that scaling laws for LAL ablation rates and bubble size are related [92], and this approach can be further sophisticated through a larger and more detailed data base.

An energy balance of LAL and LIBS processes requires knowledge of the energy partitioning into phase transitions and the mechanical pathways of shock wave emission and bubble formation. The $R_{max}(E_L)$ data yield the bubble energy, and plasma pressure and shock wave energy can be assessed by optical measurements of the initial shock wave propagation [40,114,228]. The shock speed $u_S(t)$ derived from measured $r_S(t)$ data together with the Hugoniot relations linking shock speed and pressure then yields $p_s(r)$ [40,330]. Since most shock wave energy is dissipated already in close plasma vicinity, its approximate amount can be obtained by integrating over the energy dissipation losses $\varepsilon_{SW}(r)$, which is accessible by evaluation of $p_s(r)$ [114]. The energy remaining in the far-field can be obtained by measuring and evaluating the far-field pressure profile [229], and the sum of both contributions yields the total energy [114]. This approach together with $R_{max}(E_L)$ data provides important elements for the creation of an *energy balance for LAL and LIBS processes*. The balance can be completed if the absorbed (non-reflected) fraction of the incident laser energy and its partitioning into vaporization energies of target and liquid are also known. Such data may come from measurements of the ablated target volume and photographs with high spatiotemporal resolution from which the vaporized liquid volume can be estimated.

*Optimization strategies for LAL suggested by the findings of the present study*: For optimum LAL and LIBS performance the plasma energy density should remain at a moderate level to maximize ablation efficiency and avoid shielding of the target by plasma formation in the liquid in front of the target. The use of laser pulses with reproducible shape (slm pulses) largely improves the reproducibility of ns optical breakdown and will likely facilitate to establish a clear gap between breakdown threshold at the target and in the liquid to minimize shielding and support a mono-disperse particle size distribution. This is due to the increased sharpness and the higher breakdown threshold with slm pulses compared to mlm pulses (Table 2, Figs. 4 and 5). Furthermore, ns pulses at IR wavelengths provide a better separation of $F_{\text{th,liquid}}$ from $F_{\text{th,target}}$ than shorter wavelengths because their breakdown threshold in water is higher due to the larger hurdle for multiphoton-induced seed electron generation with smaller photon energies. Use of a top hat beam instead of a Gaussian beam consolidates the separation of $F_{\text{th,liquid}}$ from $F_{\text{th,target}}$ because in a Gaussian beam, liquid breakdown may be ignited in the center, while in its periphery plasma is formed only at the target. Spatially multimode beams often have an approximately top hat profile. Otherwise, the beam profile of a Gaussian beam can be somewhat homogenized by insertion of a helical phase plate that transforms it into a propagation-invariant vortex beam with ring-shaped cross section [309]. This approach avoids hot spots in the liquid in front of the target that may lead to unwanted plasma formation in the liquid. Finally, the laser beam should be focused behind the target to avoid an intensity peak in the liquid, and the spot size should be adjusted to provide radiant exposure values around 2.7 times ablation threshold, which yield maximum ablation efficiency in blow-off processes [75,188].

*Target shielding during pulse series in LAL.* In LAL, maximization of the ablated nanoparticle mass and the optimization of their size distribution and reactive properties are a major goal. Therefore, pulse series with large average power are often employed [73,338], which leads to the problem that bubble remnants and bubbles formed at nanoparticles shield the target during consecutive ablation events [91,93,336,339]. Shielding by bubble remnants can be avoided by lateral pulse separation through fast scanning [73,338], by pulse sequences, which favorably influence the bubble dynamics to clear the ablation site [316,340], and by flow configurations [341,342]. For envisioning strategies to minimize bubble formation at nanoparticles, one needs to look at the mechanisms of bubble formation, which depend on the laser wavelength used for ablation. In metallic nanoparticles, light-driven collective oscillations of free electrons induced by electromagnetic waves form plasmons. Plasmon oscillations heat the particle, and they enhance the external electric field through charge concentration at the polar points, i.e. along the polarization direction of the field [343,344]. Near fields are evanescent, and the field-enhanced region is thin [10-20 nm for a gold NP with 50 nm radius, as shown in Fig. 15(a)].

Absorption and heating peak at resonance wavelengths in the visible wavelength range, where the frequency of incident light matches with that of the plasmons, and drop off rapidly for longer wavelengths [Fig. 15(b)]. The field enhancement peaks at the resonance wavelength but is still larger than 10-fold at IR wavelengths [Fig. 15(c)]. As a consequence, bubble formation near resonance is driven by heat diffusion from the heated nanoparticle into the surrounding liquid, while off-resonance plasma formation may occur in the liquid surrounding the particle, if the laser irradiance in the field-enhanced regions exceeds the breakdown threshold [345-347]. This results in a more vigorous effect driven by a volumetric phase transition of the liquid in the plasma even though the particle temperature is lower.

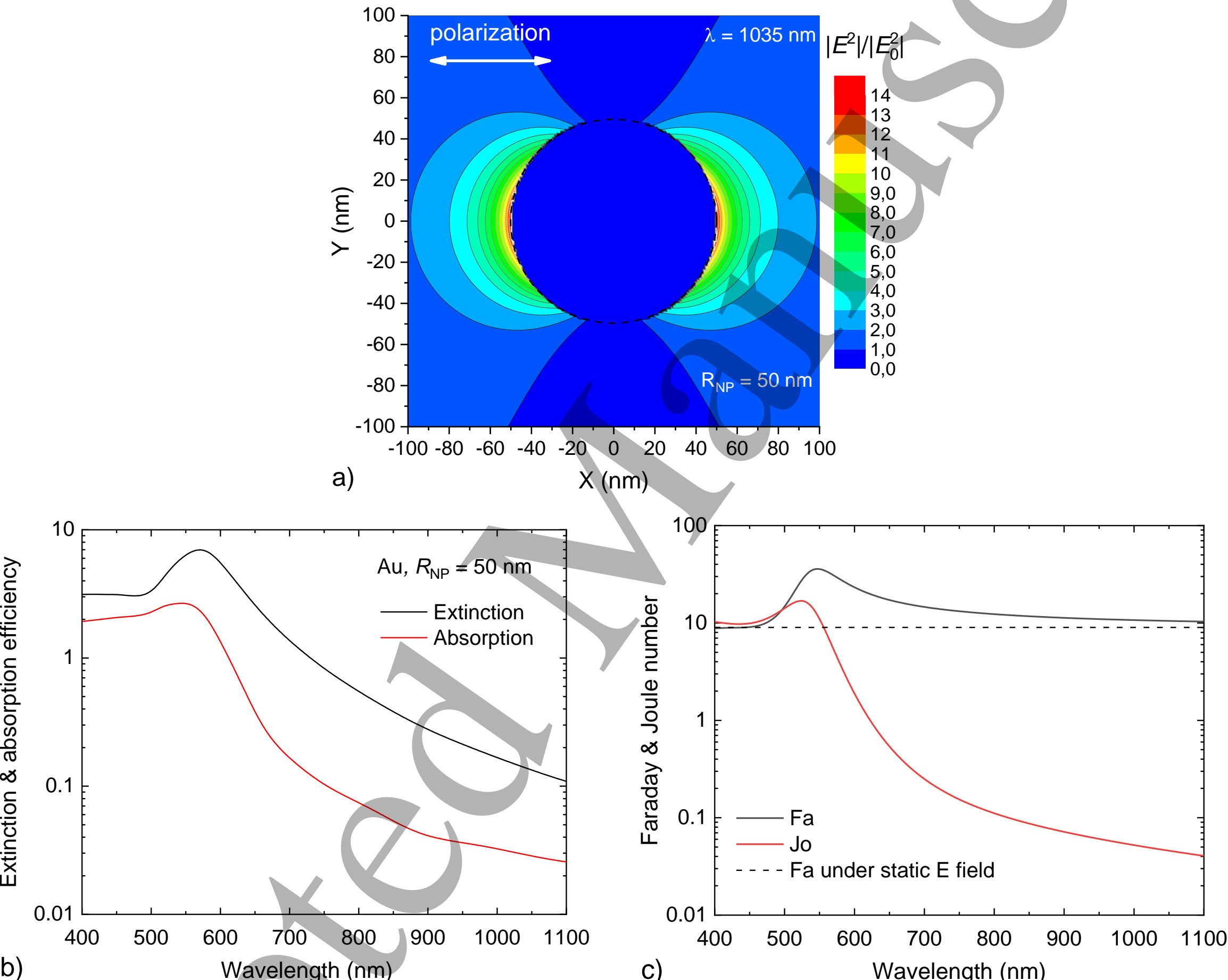

**Figure 15** Wavelength dependence of nanoparticle properties illustrated on a gold nanoparticle with 50 nm radius. (a) Near-field enhancement around the nanoparticle at 1035 nm wavelength, calculated using the Matlab tool box *SPlaC v1.0* from Ref. [344]. For this purpose, the original tool box enabling 2-D calculations was extended to allow for 3-D calculations. Maximum enhancement of field intensities is achieved along the *x*-axis in the polarization direction of the external field. (b) Wavelength dependence of plasmonic absorption and extinction (absorption + scattering) in a gold nanoparticle of 50 nm radius, calculated using Ref. [348]. (c) Wavelength dependence of Faraday number (Fa) and Joule number (Jo) characterizing the near-field enhancement outside the nanoparticle and the ability of the electric field inside a metallic nanoparticle to generate heat, respectively [349,350]. The Faraday number is the maximum value of the near-field enhancement factor outside the NP, and the Joule number characterizes the ability of a metal to generate heat in an electric field.

Plasma-driven bubble generation at IR wavelengths has been demonstrated for 45-fs, 800-nm laser pulses [346,351,352] but the results of Fig. 15(c) suggest that it will be possible also at wavelengths > 1000 nm, which are often employed in LAL, and at longer pulse durations. The boundary of the ($\tau_L$, $\lambda_L$, $E_L$) parameter space, in which they play a role, remains to be explored. It will be influenced by the temporal structure of the laser pulses. Intensity spikes characteristic for mlm Nd:YAG laser pulses facilitate plasma ignition in field-enhanced regions, and the threshold irradiance averaged over the spikes of mlm pulses is lower than for temporally smooth slm pulses (Fig. 4). Therefore, bubble formation at nanoparticles could likely be inhibited by using IR slm ns pulses – but once breakdown occurs, bubbles are larger (Fig. 5).

*Laser-induced particle fragmentation.* Nanoparticle generation by LAL often leads to an undesirably broad size distribution. Secondary laser irradiation can be used to fragment large free nanoparticles for homogenization of their size and modification of their crystalline structure [95,96,98,100]. Atomistic simulations and X-ray scattering methods have provided deep insight in the fragmentation mechanisms [97,99,353,354]. The simulations assume energy deposition into the particle, which is typical for laser wavelengths near the plasmon resonance. Stress-confined energy deposition by ultrashort laser pulses favors the generation of thermoelastic tensile and (in nonspherical particles) shear stress, which in combination with heat-diffusion-mediated bubble formation results in fragmentation. Rapid cooling yields a crystalline structure boosting the reactivity and catalytic efficiency of the fragments.

The use of off-resonance IR laser pulses with plasma-mediated bubble generation around the nanoparticle modulates the interplay between internal thermoelastic stress generation and external forces. At low radiant exposures close to the bubble threshold, nanoparticle integrity is well maintained by the relatively small temperature rise inside the particle that avoids melting. Targeted nanosurgeries on cells such as photoporation for gene modification or gene silencing with antibody conjugated nanoparticles [25,355] can be performed with good transfection efficiency and nanoparticle integrity [347,356]. However, at large radiant exposure values, where events become more vigorous, the high plasma pressure in the field-enhanced region and the high pressure at bubble collapse may induce fragmentation, or at least support the action of the internal thermoelastic stress. The concentration of plasma pressure in two field-enhanced regions [Fig. 15(a)] leads to shear stresses that are more favorable for fragmentation than homogeneous compressive stress. In particle clouds, the interaction between adjacent bubbles leads to aspherical collapse events with liquid jet impact, which also produces shear stress [63,74,357]. Hence, tuning the laser wavelength could alter the fragmentation mechanism for metallic nanoparticles, which would affect also the physicochemical properties of the fragments. Nevertheless, fragmentation mediated through energy deposition into an absorbing particle remains the more universal mechanism. Here, the use of longer wavelengths can help to maintain its chemical integrity by reducing the occurrence of multiphoton-mediated alterations [342,358].

*Plasma-mediated chemistry.* Laser ablation and particle fragmentation in liquids are considered physical synthesis techniques but accompanied by chemical changes even if they are performed in water [72,75,359-361]. Many changes are either directly produced by free-electrons in the laser plasma or indirectly via free-electron-generated reactive oxygen species as reviewed in [98]. Hydrogen, $O_2$ and $H_2O_2$ are produced through ns breakdown at nanoparticles [362,363] and during LAL for metallic nanoparticle generation [339]. Direct free-electron-mediated disintegration of water dominates for noble metals such as gold and platinum, while redox reactions at the particle surface dominate for less noble metals like aluminum. They lead to massive gas evolution, which forms persistent bubbles disturbing the LAL process, as discussed further above. For gold and platinum, the plasma liquid chemistry provides a large $H_2O_2$ yield, with platinum being the best catalyst [339]. Laser-induced water splitting works also without nanoparticle support [364]. Yan et al. demonstrated efficient $H_2$ and $H_2O_2$ generation using energetic 532-nm ns pulses weakly focused in pure water, which yields a large interaction volume for plasma-mediated chemistry [365]. In high-density water plasmas, free-electron-mediated molecular disintegration goes along with thermal dissociation [172,362,366] and it is not easy to separate the contributions of both factors. Water splitting and other free-electron-mediated chemical reactions can also occur in cold plasmas created by series of fs pulses below the cavitation threshold (section 6.3). Their relative importance increases more and more with decreasing temperature. In pure water, they will dominate below $\approx$ 3000 K, above which thermal dissociation sets in [172].

A generalization of the combination of plasma-mediated and surface chemistry in LAL and laser fragmentation is reactive laser synthesis, in which different solvents or adding electrolytes/ligands allows control of the nanoparticles' surface chemistry and elemental composition [98,100]. The ability to produce nanoparticles with controlled surface ligands, and tunable composition has broadened the range of LAL-synthesized nanoparticle applications to catalysis [98,100], biotechnology[341], photovoltaics[367], and nanomedicine [368-370].

*Application of multiple pulses in LIBS.* In LIBS, plasma-mediated ablation is not the primary goal but a way for obtaining a characteristic spectroscopic signal from the atomized and excited target material after the initial continuum emission from the plasma has ceased [84,88,101-103]. The laser beam is usually scanned to explore different regions of the target, and pulse sequences with low or moderate repetition rate but large pulse energy are employed to maximize the spectroscopic signal. Therefore, shielding of the target by bubbles from previous pulses is of little concern but shielding by plasma formation in the liquid is a major problem. One could avoid this by adjusting the spot size such that the radiant exposure at the target remains close to the ablation threshold. However, this approach is difficult to realize in many practical applications. A way out is the application of double or even triple laser pulses during a time interval shorter than the cavitation bubble oscillation time. Application of double pulses has been introduced already in the early work on LIBS in liquids [82] and is a topic of active research [87,371-374]. A sequence of more than two pulses (pre-pulse for void creation and up to 6 ablation pulses) has been used to improve the efficiency of pulsed laser tissue ablation in confined liquid environment while reducing mechanical side effects [375]. For LIBS in a liquid environment, triple pulses could be particularly useful: The cavity created by the first pulse frees the optical path through the target. An energetic second pulse can then ablate target material without shielding by plasma formation in the liquid. A third pulse further atomizes and excites the ablated material inside the bubble to improve the spectroscopic signal. For good coupling of laser energy into the bubble, pulses 2 and 3 should be applied when the wall of the expanding bubble is still smooth. Separation of the three tasks into three steps with appropriate timing and pulse energy could provide an even better overall performance than double pulses.

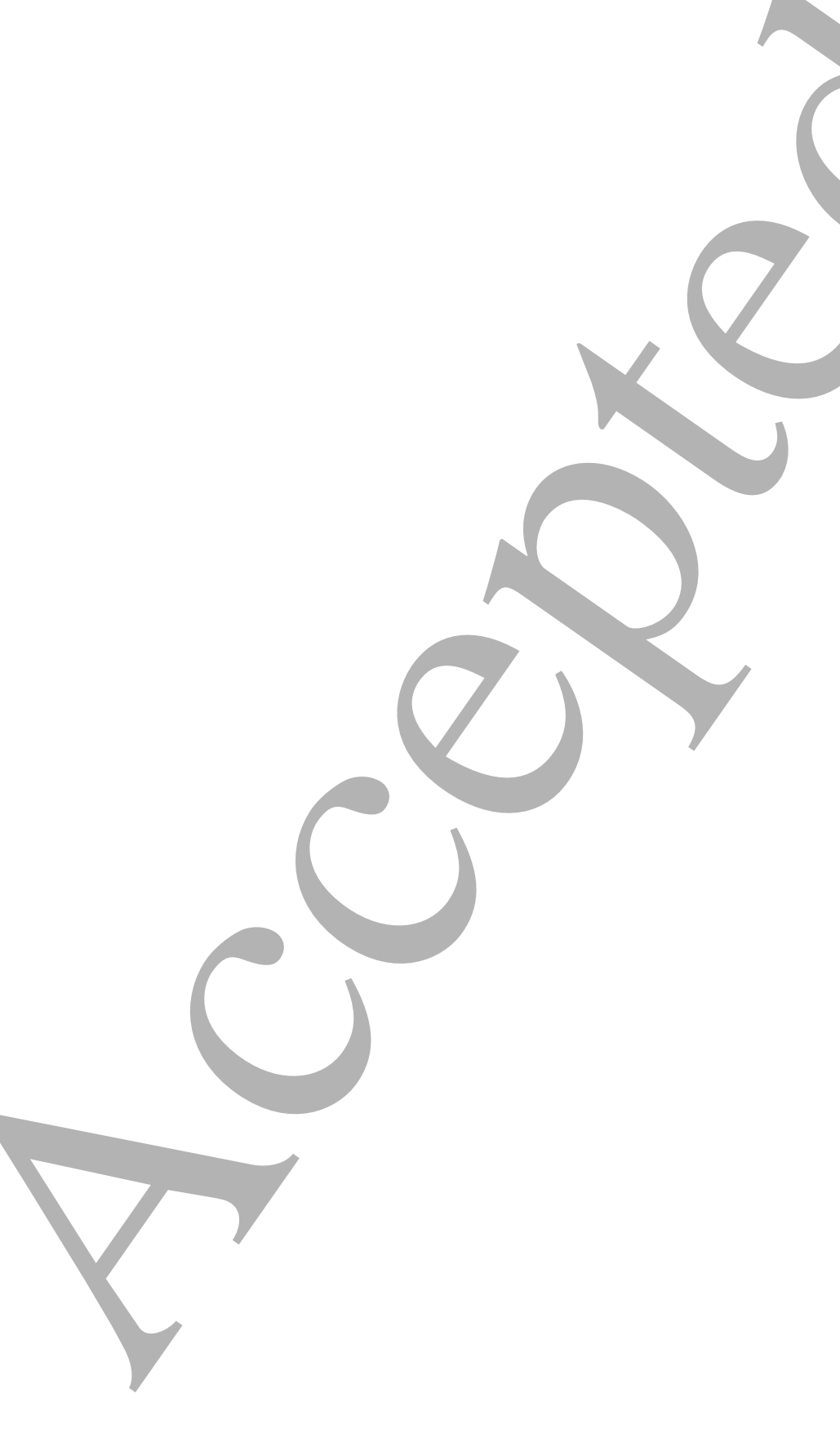

## 6. Conclusions and outlook

We performed comprehensive investigations of the energy dependence of aberration-free plasma generation in water by tightly focused laser pulses that cover a wide range of pulse durations (fs to ns), wavelengths (UV to IR), and numerical apertures. Deterministic behavior was observed not only in fs breakdown but also, when UV- or VIS ns pulses with smooth, reproducible temporal shape are tightly focused into pure dielectrics. This extends the basis for parameter selection in laser surgery and material processing.

We established a conceptual framework of nonlinear energy deposition in ($\tau_L$, $\lambda_L$, $NA$, $E_L$) parameter space by analyzing the interplay between PI, AI, recombination, and thermal ionization. On this basis, we identified three different scenarios of nonlinear energy deposition and confirmed their existence through experimental investigations. The scenarios are: 1. *Tunable energy deposition by ultrashort laser pulses of all wavelengths, with continuous transition from nanoeffects to extreme energy densities and disruptive effects*. The possibility of nanoeffects relies on the abundance of multiphoton-mediated seed electrons for AI already at the breakdown threshold, and the smooth tunability is due to time constraints of AI. 2. *Big effects already at threshold in IR ns breakdown.* Here, the generation of seed electrons by MPI is the critical hurdle for breakdown and, due to the high order of multiphoton effects, this hurdle is overcome only at high irradiance values, where AI largely overshoots the bubble threshold. 3. *Stepwise transition from nanoeffects close to threshold to extreme energy densities well above threshold in UV and VIS ns breakdown*. Here, seed electrons are readily available due to the low order of MPI, and AI is close to the bubble threshold inhibited by recombination, which provides fine-tunability of energy deposition. Above a second threshold, the temperature in the focal volume is so high that thermal ionization sets in during the laser pulse. Now, AI and TI jointly overcome the inhibition by recombination and suddenly high degrees of ionization are reached.

On a mesoscopic level, ionization involves an interplay between downstream or upstream movement of the breakdown wave, spatiotemporal instabilities of free-electron multiplication by AI, and X-ray-mediated radiative energy transport from high-density strings or hot spots into their surroundings. The combination of interplays on a molecular and a mesoscopic level results in a higher complexity of optical breakdown events, especially for ns plasmas, than previously envisioned.

The plasma energy density in bulk media is limited by the movement of the breakdown front during the laser pulse. The downstream movement of the breakdown wave in fs breakdown at small to moderate

*NA* in conjunction with nonlinear beam propagation results in energy depletion during pulse propagation and low electron and energy density. By contrast, high energy densities are produced at large *NA*, when the plasma is shorter than the laser pulse and the breakdown front moves upstream. The largest average plasma energy densities reach up to $U_{avg} \approx 40$ kJ/cm$^3$ at $NA = 0.8$ both for fs and ns breakdown. This corresponds to peak pressures above 10 GPa. The local energy density of ns plasmas can reach much higher values. String-like regions of primary energy deposition during optical breakdown at large *NA* have an estimated energy density of up to 400 kJ/cm$^3$, and the estimated peak value in the hot spot observed in the upstream part of the plasma at $NA = 0.3$ is 260 kJ/cm$^3$, corresponding to a pressure above 100 GPa.

The $R_{max}(E_L)$ curves obtained at different pulse durations, wavelengths, and *NA*s, characterize the different scenarios and delineate regimes with low and high plasma energy density. The $R_{max}(E_L)$ curves also show the existence of a general scaling law for high-density plasmas, regardless of pulse duration and wavelength. In this regime, well above the respective breakdown thresholds, the bubble volume and energy are proportional to the incident laser pulse energy. The partitioning of absorbed laser energy into vaporization-, bubble-, and shock wave energy was explored as a function of plasma energy density in a range between 8.5 kJ/cm$^3$ and 40 kJ/cm$^3$. With increasing energy density, the conversion into mechanical energy grows rapidly up to about 15 kJ/cm$^3$ and more gradually for larger plasma energy densities. While the fraction going into bubble energy stays approximately constant near 20%, the fraction needed for vaporization drops to only 7% at $U_{avg} = 40$ kJ/cm$^3$, while more than ¾ of the absorbed energy is converted into shock wave energy.

The discovery that precisely tunable energy deposition is possible in larger regions of the ($\tau_L$, $\lambda_L$) space than previously thought provides new possibilities for cost-effective and compact solutions adapted to specific applications. We demonstrated the utility of tunable nonlinear energy deposition with UV ns pulses on the examples of refractive corneal surgery as well as on micro-marking and refractive index modifications in glass. These examples indicate that results for water obtained in this study are representative for breakdown features in many transparent dielectrics.

The identification of different scenarios through systematic investigation of $R_{max}(E_L)$ and the exploration of energy partitioning of absorbed laser energy into bubble and shock wave energy demonstrated in this paper on breakdown in bulk liquid can be useful also for delineating regimes with different working mechanisms in LAL and LIBS. We found that scenarios 1-3 are most clearly separated, when nanosecond laser pulses with smooth temporal shape are employed rather than the common multi-longitudinal-mode pulses exhibiting intensity spikes from mode beating. Therefore, it seems worthwhile to investigate the influence of the mode of laser operation also in LAL and LIBS.

## 6.1. Roadmap for future experimental work

The experimental results in this paper were obtained by probe beam scattering that enabled to collect a large amount of $R_{max}(E_L)$ data in the ($\tau_L$, $\lambda$, $NA$) parameter space, and by time-integrated high-resolution plasma photography in side view together with transmission measurements. Evaluation of plasma volume and transmission yielded the plasma energy density. The color photographs made is possible to distinguish between regions of primary energy deposition and radiative plasma enlargement and helped to discriminate bremsstrahlung and blackbody radiation. Information on the breakdown dynamics was inferred with the help of the moving breakdown model. An extension of high-resolution color photography to plasma formation at solid targets may provide a deeper understanding also for the mechanisms of LAL and LIBS.

Further exploration of the breakdown dynamics both in bulk media and at solid targets requires time resolved investigations of the sequence of breakdown events with high spatiotemporal resolution. Time-resolved ICCD imaging with ultrashort gating times can directly portray the plasma growth during nanosecond breakdown and visualize the formation of branched high-density strings and of large hot spots during the movement of the breakdown wave. Shadowgraph and schlieren photography can provide insights into the pressure distribution inside and around the plasma and the resulting shock wave formation. Presumably, the local hydrodynamic expansion of string regions and hot spots featuring high energy density and pressure will be faster than the expansion of the larger region emitting diffuse plasma luminescence in which energy density and pressure are lower. The difference has to date been obscured because tracking of acoustic transients and shock wave propagation was restricted to the plasma surroundings. With sufficiently bright narrowband light sources, picosecond exposure times and sub-micrometer image resolution, the propagation of shock waves emanating from hot spots or strings can possibly be tracked also inside the diffusely luminescent region.

Time-resolved spectroscopy can provide information on the transition from the initial plasma state to a vapor bubble during the initial expansion phase. Blackbody temperatures of the breakdown plasma are available [224-226], and spectroscopic temperatures from line spectra are also available for $t \geq 75$ ns [103] but the transition from continuum emission to line emission has not yet been continuously tracked. Coverage of this phase with spectroscopic measurements of plasma temperature and electron density will help to understand the energy conversions going along with the change from plasma to the thermodynamic states of supercritical water and hot vapor in the expanding bubble.

Sub-micrometer spatial resolution of shock wave emission and early bubble expansion requires novel techniques for laser-based speckle-free stroboscopic photography with picosecond exposure times, which are presently under development in our group. Highly reproducible breakdown events can be recorded by stroboscopic photography [40], whereas multi-exposure single-frame photography enables to capture also events with lower reproducibility. This approach was pioneered by Alcock et al. [376] (with high-quality prints in [377]) and has recently been sophisticated using illumination pulse bursts from a fiber laser with time separations down to 5 ns [378,379]. Availability of bursts consisting of about 10 pulses with ≈ 1 ns time separation will enable precise single-shot recordings of the early shock wave formation and emission. As described in section 5.8, the pressure evolution $p_s(r)$ can be derived from measured $r_s(t)$ data, and the shock wave energy can be obtained by integrating over the energy dissipation, which is encoded in the $p_S(r)$ curves [114].

Confocal probe beam scattering is a convenient method for single-shot measurements of cavitation bubble dynamics with high temporal resolution. While transmission measurements merely yield the bubble oscillation time $T_{osc}$, we have demonstrated that also radius time curves can be acquired by recording the back reflected scattering signal. Interference of reflections from the front and rear bubble walls yields a signal from which $R(t)$ can be deduced [242].

Availability of $R(t)$ and $p_s(r)$ data from single shot recordings in combination with spectroscopic information on the evolution of plasma temperature and electron density will provide a data base that can advance the development of numerical models on plasma formation and early hydrodynamic events. Single-shot methods for $R(t)$ and $r_s(t)$ also open new avenues for the investigation of optical breakdown and cavitation events in viscoelastic materials in general and transparent biological tissues and cells in particular [380,381]. They are of great interest for cavitation rheology that to date relies on stroboscopic photography [382,383] or high-speed videography [384,385].

For a visualization of processes inside the cavitation bubble at a solid target during LAL and LIBS, a good view into the bubble with high spatiotemporal resolution is of major importance. For optical imaging, this can be achieved by Koehler illumination of the breakdown region at large numerical aperture [105,386]. Phase contrast X-ray imaging also offers a view through bubbles [91,387,388], while time-resolved small-angle X-ray scattering provides additionally the particle distribution inside the bubble [74,91,389].

### 6.2. Roadmap for multiscale modelling of plasma formation and hydrodynamic events

The conceptual framework and experimental results provided in this paper motivate to establish a unified model of plasma formation, shock wave emission, and bubble generation in water. Such comprehensive model should portray the interplay of MPI, AI, recombination and TI, and link it to the modeling of hydrodynamic events in a way that the experimental results of Figs. 4-6 with its different breakdown scenarios are reproduced. Combining a breakdown model with simulations of spherical bubble generation in a compressible liquid [42,390] is a promising approach for achieving this goal because the relative simplicity of such models facilitates parametric studies in ($\tau_L$, $\lambda_L$, $E_L$) space.

On a higher level of sophistication, modeling should consider the spatiotemporal evolution of the energy density distribution in the breakdown region as starting point for the hydrodynamic part of the modeling chain including shock wave emission and bubble oscillations. Advanced models of femtosecond optical breakdown in bulk media covering the interplay of beam nonlinear propagation and nonlinear absorption are already available [130,131,136,140,141,289]. However, modeling approaches covering the features of strings, hot spots, and radiative energy transport relevant for nanosecond breakdown (Figs. 7, 8) are still lacking.

Remarkably, our heuristic considerations based on generic equations assuming a constant band structure of a transparent dielectric could delineate different breakdown scenarios in water that were experimentally observed in a large range of pulse energies and plasma energy densities. However, future modeling must consider that this approach is fully appropriate only in the low-energy density regime close to the optical breakdown threshold and becomes insufficient when the target material disintegrates and an electron-ion plasma forms.

For cavitation produced by the expansion of an electron-ion plasma, a dual EOS approach assuming a gaseous domain inside the plasma/bubble volume and a liquid EOS outside, which is used in most models, does not cover the real complexity of events. Recently, the use of a Noble-Abel stiffened gas EOS for bubble interior and exterior has become popular, in which a covolume is introduced that allows to account for repulsive short-distance effects linked to molecular motion [391]. This EOS resolves the temperature inaccuracy associated with the commonly employed Tait EOS for liquids [392], and allows to express also a real-gas EOS by appropriate parameter choice [43]. However, it does not overcome the dual EOS approach itself. Particle-based approaches and incorporation of chemical reactions [296,393] promise more flexibility and accuracy for modeling the plasma-vapor transition - at the cost of increased complexity. While chemical processes driven by high temperature and pressure in electron-ion plasmas have been considered in the modeling of bubble collapse in SBSL and sonochemistry [183,394-396], their integration in the modeling of laser-induced plasma formation and bubble generation is still in its infancy [393].

Information on the spatiotemporal dynamics of shock wave emission and energy partitioning arising from an inhomogeneous energy distribution in the plasma can be gained through ***finite volume modeling*** of the hydrodynamic events [64,67,397-401]. They are also a versatile tool for modeling laser-induced cavitation under asymmetric boundary conditions at solid targets such as in LAL and LIBS. At present, they often assume isotropic energy distribution into a spherical start volume and neglect heat diffusion as well as phase transitions at the bubble wall but the integration of spatiotemporal features of energy deposition and heat and mass transfer is a field of active research. Information about the spatiotemporal evolution of the laser-induced energy density distribution that define the starting conditions for refined finite volume modeling can be obtained either from high-resolution plasma and stroboscopic photographs, or through numerical simulations of plasma formation. Calculations of the deposited energy distribution have been performed for fs breakdown [136,215,252] but still pose a challenge for ns breakdown, because neither the spatiotemporal instabilities of avalanche ionization leading to string formation nor radiative energy transport are yet included in current models.

Dehghani et al. [402] and Preso et al. [403] recently reviewed perspectives and challenges in 'green' applications of sonochemistry. Sonochemical reactions rely on the high pressure and temperature developed during bubble collapse [74,394,396]. Like single bubble sonoluminescence [183], also laser-induced cavitation is a model system for detailed investigations of bubble collapse under controlled conditions. Collapse behavior is tunable through variation of breakdown plasma features as they govern the chemical reactions taking place upon breakdown, which determine the amount and composition of permanent gas during the collapse phase. A large gas content cushions the collapse and reduces the collapse pressure, while pure vapor bubbles collapse more vigorously. It was recently shown that a large energy density upon breakdown goes along with lower collapse pressure than a moderate energy density – likely due to an increased rate of molecular dissociation and permanent gas production in hotter plasmas [42,66].

Generally, understanding the physicochemical interactions connected with laser-induced breakdown and cavitation is essential for many applications and deserves more attention in the Physics community. Tracking the chemical reaction chain from optical breakdown to collapse with its connection to hydrodynamic and thermodynamic processes is also a rewarding task for validating numerical simulations in fundamental research and for guiding applications.

The complexity of the sequence of breakdown-induced events in different parameter regimes and under different boundary conditions imply a similar complexity of modeling tasks. First attempts have been made to use artificial intelligence for physical model-guided machine learning for accelerating laser induced plasma-mediated micro-machining process optimization [404]. Based on a modular system of different physical modeling tools and on data collected in systematic experiments and during material processing, artificial intelligence could help to select and combine appropriate modules for different regions of the parameter space.

### 6.3. Exploration of low-density plasma effects below the breakdown threshold

In the present paper, we have focused attention to nonlinear energy deposition by individual laser pulses with energies above the breakdown threshold. However, when series of tightly focused pulses with small energies are applied at high repetition rates, free electrons can produce material modifications also below the cavitation bubble threshold [22]. Here, in the low-density plasma regime, the laser action changes from mechanical to chemical effects. Figure 16 presents the irradiance range and the corresponding range of free-electron densities defining the low-density plasma regime. It covers seven orders of magnitude from the irradiance at which free electrons start to produce photodamage in nonlinear microscopy [405-409] up to the cavitation bubble threshold with single pulses.

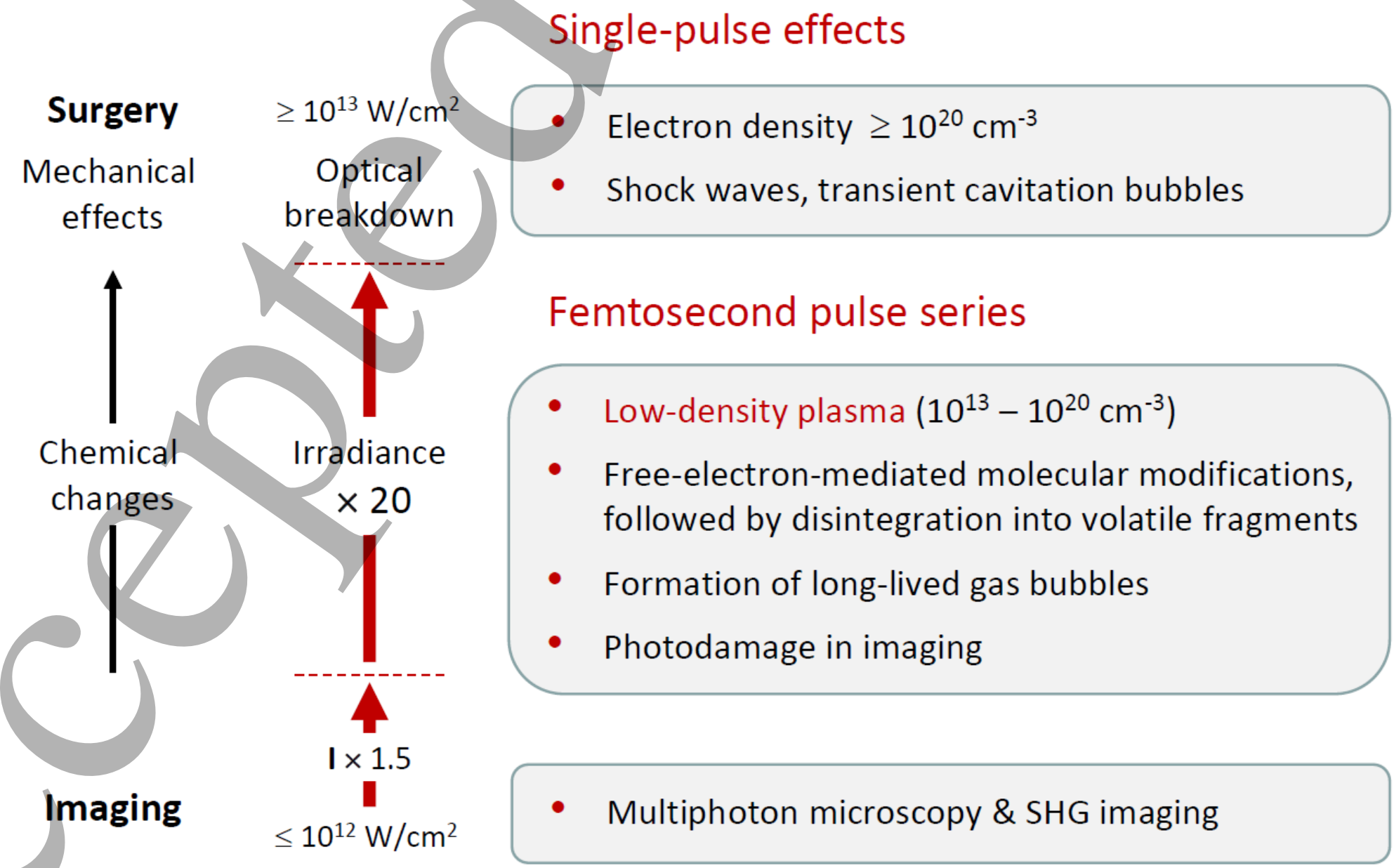

**Figure 16** The realm of low-density plasmas between irradiance values employed for nonlinear imaging with fs laser pulse series and the optical breakdown threshold above which cavitation-associated laser surgery is performed.

Nonlinear microscopy is performed at irradiance values between $10^{11}$ W/cm$^2$ and $10^{12}$ W/cm$^2$ [410]. Photodamage sets in already ≈ 1.5 times above the irradiance level that is regularly used for multiphoton imaging [407]. Here, the free-electron-yield becomes comparable to the nonlinear photon yield [406]. At about 20 times larger irradiance, a vapor bubble can be produced by a single laser pulse [37]. In the low-density plasma regime, the dominant laser-material interactions are free-electron-mediated chemical changes. In the transition regime just below the cavitation threshold, thermoelastic stresses still play a role for tissue and cell surgery, in conjunction with free-electron-mediated bond breaks [17,22]. However, at lower irradiance, the heating by individual pulses or pulse trains is far too small to cause any thermal changes [411]. Here, direct chemical bond breaking by free electrons goes hand in hand with indirect effects via dissociation of water molecules forming •OH radicals, which then interact with biomolecules [198]. Molecular fragmentation produces long-lived bubbles of noncondensable gas. Gas bubble formation is usually preceded by hyperfluorescence of photoproducts and weak plasma luminescence through Bremsstrahlung from CB electrons. All three markers can be used as physical real-time indicators for free-electron-mediated chemical changes [407,409].

A mechanistic understanding of such low-density plasma effects can be achieved by numerical simulations of plasma formation that consider the plasma initiation channels constituted by the specific band structure of water and by biomolecules with low ionization energy [36,37,196-198]. A key feature governing the interaction between electrons and biomolecules in the low-density plasma regime is the kinetic energy spectrum of the laser-induced conduction band electrons [149], which depends strongly on irradiance and laser wavelength [198]. An essential prerequisite for its calculation is the consideration of the stepwise gain of kinetic energy of CB electrons by inverse bremsstrahlung absorption, which can be achieved by solving a multi-rate equation [149,198]. Knowledge of the kinetic energy spectrum enables to tune the biomolecular effects by appropriate choice of the laser parameters, if the action spectra for different chemical changes as a function of kinetic electron energy are known [198,203,412,413].

Free-electron-mediated chemical changes with control of the kinetic energy spectrum of CB electrons through appropriate choice of laser parameters make it possible to induce specific modifications of various types of biomolecules. Examples are the induction of different damage types in DNA for studying repair processes in living cells [198,203,414], gentle fragmentation of large protein segments for improving resolution in mass spectroscopy [415,416], and the induction of local refractive index changes in the cornea for the correction of myopia or presbyopia [17,417]. Life-cell experiments with tailored electron energy spectra can provide a better understanding of LEE-induced reaction pathways in targeted laser nanosurgery and radiation therapy [198].

Free-electron-mediated molecular disintegration and chemical changes play a role also in the transition zone from optical breakdown involving phase transitions by single pulses to low-density plasma formation with gas bubble formation by multiple pulses. Micro-and nanosurgical applications in the transition zone include corneal dissection by low-energy fs pulses applied at MHz repetition rate [17] and gentle optoporation of cells mediated by molecularly targeted gold NP that are irradiated by series of off-resonance fs pulses [418-420]. Reactive LAL in organic solvents for NP synthesis and tailoring is also a mixture of thermodynamic events, surface chemistry, and free-electron-mediated chemical effects [98,100]. The same applies for plasma formation by electric discharge at liquid surfaces and within liquids [271,421-423].

Altogether, the energy dependence of plasma formation in dielectrics comprises a very large range of laser pulse energies and plasma energy densities, and the laser-induced effects exhibit an enormous variety. The results and the framework for the analysis presented in this paper may help to guide future mechanistic analysis and stimulate the exploration of novel applications reaching from free-electron-mediated molecular modifications to the generation and characterization of extreme states of matter. A better understanding of the complete range of plasma-induced processes from the thermomechanical effects to free-electron-mediated changes opens new pathways for laser surgery, material processing, material research, various sustainable technologies, and can even improve nonlinear microscopy.

## Acknowledgements

This work was supported by U.S. Air Force Office of Scientific Research, grants FA8655-05-1-3010, FA9550-15-1-0326, FA9550-18-1-0521, FA9550-22-1-0289, and by the German Federal Ministry of Economics and Technology, grant ZIM KF2341201FR9.

We thank Dr. Thomas Mattsson (Sandia National Laboratories, U.S.A.) for providing QMD and SESAME high-density EOS data of water, and we thank Jean-Paul Jay-Gerin and Jintana Meesungnoen (Université de Sherbrooke, Québec, Canada) for providing data of the attenuation length of electron beams.

## Author declarations

**Conflict of Interest**

The authors have no conflicts to disclose.

## Author contributions

**Norbert Linz:** Conceptualization (supporting), methodology (equal), experiments (equal), data interpretation (equal), writing (supporting), **Sebastian Freidank:** Methodology (equal), experiments (equal), data interpretation (supporting), **Xiao-Xuan Liang:** Methodology (equal), data interpretation (equal), **Alfred Vogel:** Conceptualization (lead), fund raising, methodology (equal), data interpretation (equal), writing (lead).

## Data availability

Most data supporting the findings of this study are available within the article. The raw data used for determining the dependence of energy partitioning of absorbed laser energy as a function of average plasma energy density (Fig. 11) are available on request from the authors.

## ORCID IDs:

| | |
|---|---|
| Norbert Linz | https://orcid.org/0000-0002-7843-5966 |
| Sebastian Freidank | https://orcid.org/0000-0002-5453-269X |
| Xiao-Xuan Liang | https://orcid.org/0000-0002-8325-1627 |
| Alfred Vogel | https://orcid.org/0000-0002-4371-9037 |

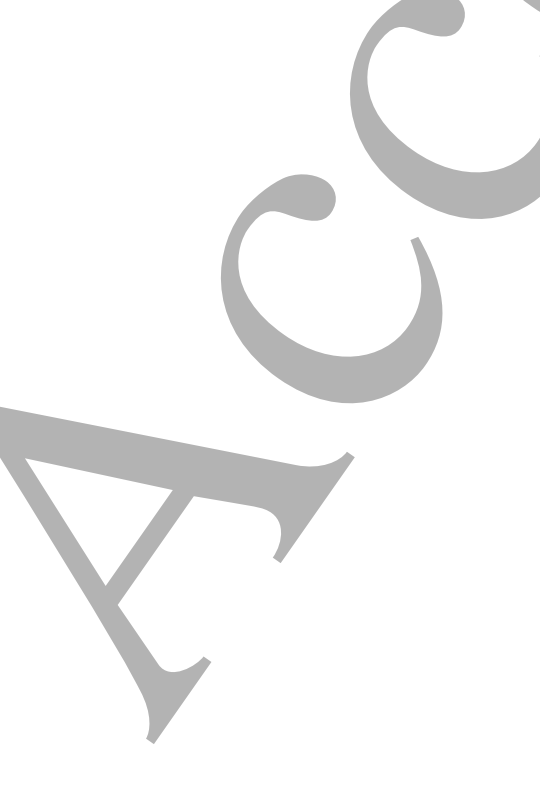

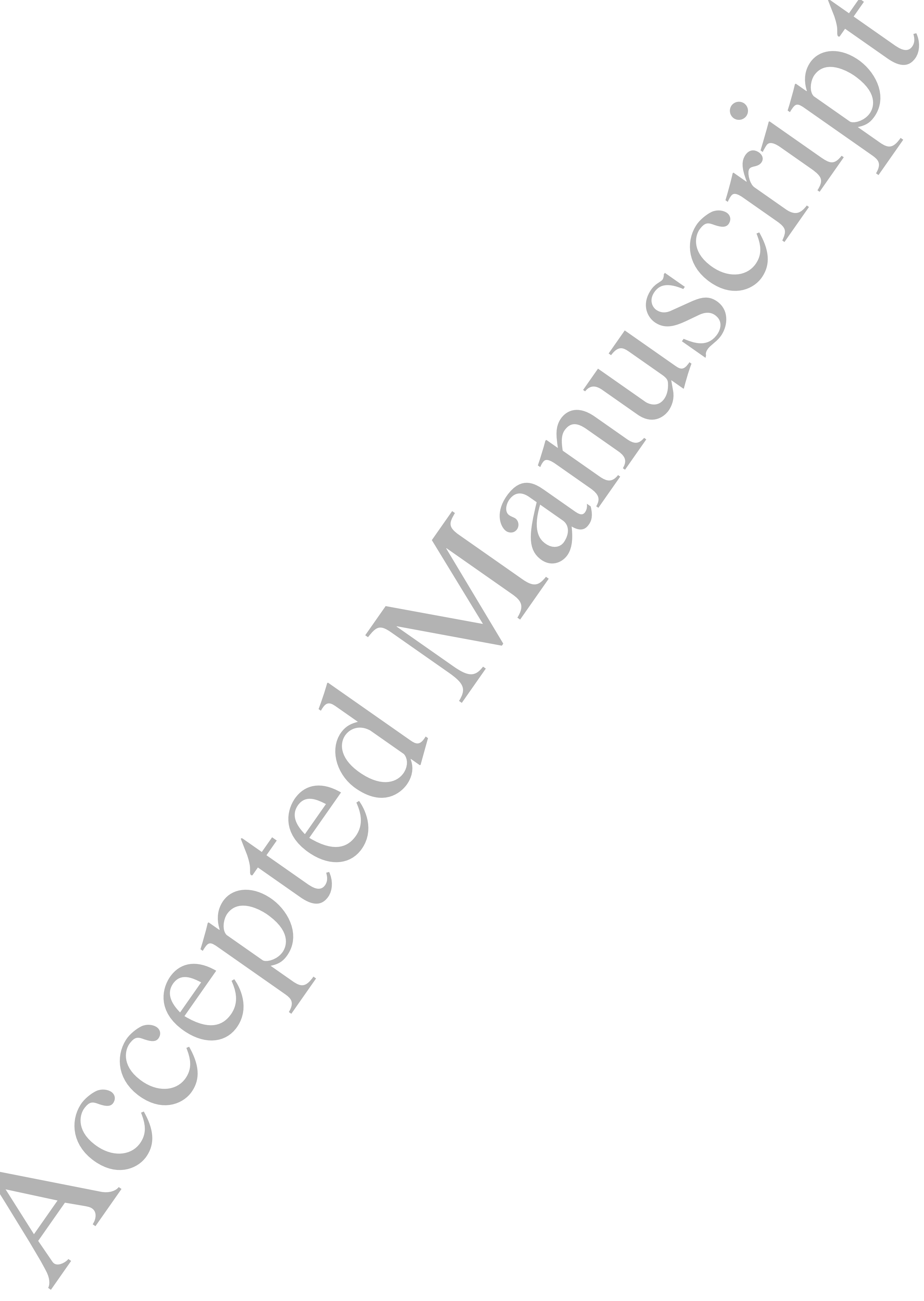